\documentclass [preprint,12pt] {elsarticle}
\usepackage {graphicx}
\usepackage {amssymb}
\usepackage {amsmath}
\usepackage {xcolor}
\usepackage {breakcites}
\usepackage {lscape}
\usepackage {rotating}
\usepackage {longtable}
\usepackage {ifthen}
\usepackage {listings} 
\usepackage {gnuplottex}

\newcommand{\Res}{\textnormal{Res}}

\makeatletter 
\renewcommand{\@biblabel}[1]{} 
\makeatother

\biboptions{round,semicolon} 

\begin{document}
\title{Angular distribution of Bremsstrahlung photons \\ and of positrons for
calculations of \\ terrestrial gamma-ray flashes and positron beams}
\author{Christoph K\"ohn$^a$ and Ute Ebert$^{a,b}$}
\ead{koehn@cwi.nl, ebert@cwi.nl}
\address{$^a$ CWI, P.O.Box 94079, 1090GB Amsterdam, The Netherlands, and\\
$^b$ Eindhoven University of Technology, P.O.Box 513, 5600MB Eindhoven, The Netherlands.}

\begin{abstract}
Within thunderstorms electrons can gain energies of up to hundred(s) of MeV. These electrons can create X-rays and gamma-rays as Bremsstrahlung when they collide with air molecules. Here we calculate the distribution of angles between incident electrons and emitted photons as a function of electron and photon energy. We derive these doubly differential cross-sections by integrating analytically over the triply differential cross-sections derived by Bethe and Heitler; this is appropriate for light atoms like nitrogen and oxygen
(Z=7,8) if the energy of incident and emitted electron is larger than 1~keV.
We compare our results with the approximations and cross section
used by other authors.
We also discuss some simplifying limit cases, and we derive some simple approximation for the most probable scattering angle.

We also provide cross sections for the production of electron positron pairs from energetic photons when they interact with air molecules. This process is related to the Bremsstrahlung process by some physical symmetry. Therefore the results above can be transferred to predictions on the angles between incident photon and emitted positron, again as a function of photon and positron energy. We present the distribution of angles and again a simple approximation for the most probable scattering angle.

Our results are given as analytical expressions as well as in the form of a C++ code that can be directly be implemented into Monte Carlo codes.
\end{abstract}


\maketitle

\section {Introduction}

\subsection {  Flashes
 of gamma-rays, electrons and positrons above thunderclouds}

Terrestrial gamma ray flashes (TGFs) were first observed above thunderclouds by the Burst and Transient Source Experiment
(BATSE) (Fishman et al., 1994). It was soon understood that these energetic photons were generated by the Bremsstrahlung process when energetic electrons collide with air
molecules \cite{fishman,emission}; these electrons were accelerated by some mechanism within the thunderstorm. Since then, measurements of TGF's were extended and largely refined by the Reuven Ramaty Energy Solar Spectroscopic Imager (RHESSI)
(Cummer et al., 2005; Smith et al., 2005, Grefenstette et al., 2009, Smith
et al., 2010)
, by the Fermi Gamma-ray Space Telescope~\cite{fermi}, by the Astrorivelatore Gamma a Immagini Leggero (AGILE) satellite which recently measured TGFs with quantum energies of up to 100 MeV
\cite{AGILE_1,AGILE_2}, and by the Gamma-Ray Observation of Winter Thunderclouds (GROWTH) experiment \cite{GROWTH}.

Hard radiation was also measured from approaching lightning
leaders (Moore et al., 2001; Dwyer et al., 2005); and there are also a number of laboratory experiments where very energetic photons were generated during the streamer-leader stage of discharges in open air
(Stankevich and Kalinin, 1967; Dwyer et al., 2005b; Kostyrya et al., 2006; Dwyer et
al., 2008b; Nguyen et al., 2008; Rahmen et al., 2008; Rep'ev and Repin, 2008;
Nguyen et al., 2010; March and Montany\`{a}, 2010; Shao et al., 2011).

Next to gamma-ray flashes, flashes of energetic electrons have been detected above thunderstorms~\cite{electron}; they are distinguished from gamma-ray flashes by their dispersion and their location relative to the cloud - as charged particles in sufficiently thin air follow the geomagnetic field lines. In December 2009 NASA's Fermi satellite detected a substantial amount of positrons within these electron beams \cite{positron}. It is now generally assumed that these positrons come from electron positron pairs that are generated when gamma-rays collide with air molecules.

Two different mechanisms for creating large amounts of energetic electrons in thunderclouds are presently discussed in the literature. The older suggestion is a relativistic run-away process in a rather homogeneous electric field inside the cloud
(Wilson, 1925; Gurevich, 1961; Gurevich et al., 1992; Gurevich,
2001 Dwyer, 2003, 2007; Milikh and Roussel-Dupr\'{e}, 2010).

More recently research focusses on electron acceleration in the streamer-leader process with its strong local field enhancement (Moss et al., 2006; Li et al., 2007; Chanrion and Neubert, 2008; Li et al., 2009; Carlson et al., 2010; Celestin and Pasko, 2011; Li et al., 2012).

\subsection{The need for doubly differential cross-sections}

Whatever the mechanism of electron acceleration in thunderstorms is, ultimately one needs to calculate the energy spectrum and angular distribution of the emitted Bremsstrahlung photons. As the electrons at the source form a rather directed beam pointing against the direction of the local field, the electron energy distribution together with the angles and energies of the emitted photons determine the photon energy spectrum measured by some remote detector. The energy resolved photon
scattering angles are determined by so-called doubly differential cross-sections that resolve simultaneously energy $\hbar\omega$ and scattering angle $\Theta_i$ of the photons for given energy $E_i$ of the incident electrons. The data is required for scattering on the light elements nitrogen and oxygen with atomic numbers $Z=7$ and $Z=8$, while much research in the past has focussed on metals with large atomic numbers $Z$. The energy range up to 1 GeV is relevant for TGF's; we here will provide data valid for energies above 1 keV.

As illustrated by Fig.~1, the full scattering problem is characterized by three angles. The two additional angles $\Theta_f$ and $\Phi$ determine the direction of the
scattered electron relative to the incident electron and the emitted photon. The full angular and energy dependence of this process is determined by so-called triply differential cross-sections. A main result of the present paper is the analytical integration over the angles $\Theta_f$ and $\Phi$ to determine the doubly differential cross-sections relevant for TGF's. 

As the cross-sections for the production of electron positron pairs from photons in the field of some nucleus are related by some physical symmetry to the Bremsstrahlung process, we study these processes as well; we provide doubly differential cross-sections for scattering angle $\Theta_+$ and energy
$E_+$ of the emitted positrons for given incident photon energy $\hbar\omega$ and atomic number
$Z$.\\
With the doubly differential cross sections for Bremsstrahlung and
pair production a feedback model can be constructed tracing
Bremsstrahlung photons and positrons as a possible explanation of TGFs (Dwyer, 2012).

\begin {figure}
\includegraphics [scale=0.38] {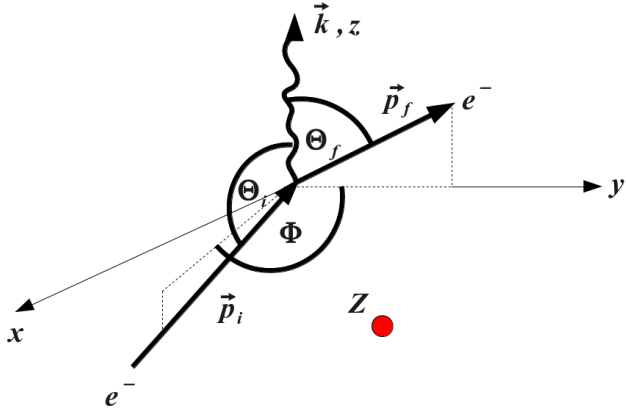} 
\caption {Parametrization of the Bremsstrahlung process: Momenta of incident electron $\mathbf{p}_i$, scattered electron $\mathbf{p}_f$ and emitted photon $\hbar\mathbf{k}$ form the angles $\Theta_i=\sphericalangle(\mathbf{p}_i, \mathbf{k})$ and $\Theta_f=\sphericalangle(\mathbf{p}_f,\mathbf{k})$, and $\Phi$ is the angle between the planes spanned by the vector pairs $(\mathbf{p}_i, \mathbf{k})$ and $(\mathbf{p}_f,\mathbf{k})$. The scattering nucleus has atomic number $Z$.}\label{start_fig.1}
\end {figure}

\subsection {Available cross-sections for Bremsstrahlung}

Our present understanding of Bremsstrahlung and pair production was largely developed in the first half of the 20th century.
It was first calculated by Bethe and Heitler (1934). Important older reviews are by
Heitler (1944), by Hough (1948), and by Koch and Motz (1959). We also used some recent text books (Greiner and Reinhardt, 1995; Peskin and Schroeder, 1995); together with Heitler (1944) and Hough (1948), they provide a good introduction into the
quantum field theoretical description of Bremsstrahlung and pair production.
The calculation of these two processes is related through some physical symmetry
as will be explained in chapter \ref{pair}.

As drawn in Fig.~1, when an electron scatters at a nucleus, a photon with frequency
$\omega$ can be emitted. The geometry of this process is described by the three
angles $\Theta_i$, $\Theta_f$ and $\Phi$. Cross sections can be total or differential. Total cross sections determine whether a
collision takes place for given incident electron energy, singly differential cross sections give additional
information on the photon energy or on the angle between incident electron and emitted photon, and doubly differential cross sections contain both.
Triply differential cross sections additionally contain the angle at which the electron is scattered. As two angles are required to characterize the direction of the scattered electron, one could argue that this cross section should actually be called quadruply differential, but the standard terminology for the process is triply differential.

Koch and Motz (1959) review many different expressions for different limiting cases, but without derivations. Moreover, some experimental results are discussed and compared with the presented theory. Bethe and Heitler (1934), Heitler (1944), Hough (1948), Koch
and Motz (1959), Peskin and Schroeder
(1995), Greiner and Reinhardt (1955) use the Born
approximation to derive and describe Bremsstrahlung and pair production
cross sections. 

Several years later new ansatzes were made to describe
Bremsstrahlung. Elwert and Haug (1969) use approximate Sommerfeld-Maue eigenfunctions to derive
cross sections for Brems\-strah\-lung under the assumption of a pure Coulomb
field. They derive a triply differential cross section and beyond that also
numerically a doubly differential cross section. Furthermore they
compare with results obtained by using the Born approximation. They show
that there is a small discrepancy for high atomic numbers between the Bethe-Heitler
theory and
experimental data, and they provide a correcting factor to fit the
Bethe-Heitler approximation better to experimental data for large $Z$. However, they only
investigate properties of Bremsstrahlung for $Z=13$ (aluminum) and $Z=79$ (gold). 

Tseng and
Pratt (1971) and Fink and Pratt (1973) use exact numerical calculations using Coulomb screened
potentials and Furry-Sommerfeld-Maue wave functions, respectively.
They investigate Bremsstrahlung and pair production for $Z=13$ and for $Z=79$
and show that their results with more accurate wave functions
do not fit with the Bethe Heitler cross section exactly. 
This is not surprising as the Bethe-Heitler approximation is developed for low atomic
numbers $Z$ and for $Z$ dependent electron energies as discussed in
section~\ref{valid}.\\ \indent
Shaffer et al.~(1996) review the Bethe Heitler and the Elwert Haug theory. They discuss
that the Bethe Heitler approach is good for small atomic numbers and give a
limit of $Z>29$ for experiments to deviate from theory. For
$Z<29$ the theory of Bethe and Heitler, however, is stated to be in good
agreement with experiments for energies above the keV range. They calculate
triply differential cross sections using partial-wave and multipole
expansions in a screened potential numerically for $Z=47$ (silver) and $Z=79$ and
compare their results with experimental data. Actually their results are
close to the Elwert Haug theory which fits the experimental data better than
their theory.

Shaffer and Pratt (1997) also discuss the theory of Elwert and Haug (1969) and compare it with the Bethe Heitler theory and, additionally, with the Bethe Heitler results multiplied with the Elwert factor and with the exact partial wave method.
They show that all theories agree within a factor 10 in the keV energy
range, and that the Elwert-Haug theory fits the exact partial wave method
best. However, they only investigate Bremsstrahlung for atomic nuclei with $Z=47$, 53 (iodine), 60 (neodymium), 68 (erbium) and 79, but not for small atomic numbers $Z=7$ and 8 as relevant in air.  In summary,
Elwert and Haug (1969), Tseng and Pratt (1971), Fink and Pratt (1973), Shaffer
et al. (1996) and Shaffer and Pratt (1997) 
calculate cross sections for Bremsstrahlung and pair production for
atomic numbers $Z=13$ and $Z>47$ numerically,
but not analytically, and they do not provide any formula or data which can be
used to simulate discharges in air.

The EEDL database consists
mainly of experimental data which have been adjusted to nuclear model
calculations. For the low energy range Geant4 takes over this data and gives
a fit formula. The singly differential cross section related to $\omega$
which is used in the Geant4 toolkit is valid in an energy range from 1 keV to 10 GeV and taken from Seltzer and Berger (1985). The singly differential cross section
related to $\Theta_i$ is based on the doubly differential cross section
by \cite {Tsai_1,Tsai_2} and valid for very high energies, i.e., well above
$(1-10)$ MeV. But in the preimplemented cross sections of
Geant4 the dependence on the photon
energy is neglected in this case so that it is actually a singly
differential cross section describing $\Theta_i$.

Table \ref{brems.tab} gives an overview of the available literature and data for total or singly, doubly or triply differential Bremsstrahlung cross sections; parameterized angles or photon energies are given, as well as the different energy ranges of the incident electron.
Furthermore, the table shows the atomic number $Z$ investigated and includes some further
remarks.

For calculating the angularly resolved photon energy spectrum of TGF's, we need a
doubly differential cross section resolving both energy and emission angle of the photons; we need it in the energy range between 1 keV and 1 GeV for the small atomic numbers $Z=7$ and 8. Therefore most of the literature reviewed here is not applicable.
However, the Bethe-Heitler approximation is valid for atomic numbers $Z<29$ and for electron energies above 1~keV~\cite{shaffer_tong}. 
How the range of validity depends on the atomic number $Z$ is discussed in
section \ref{valid}. We therefore will use the triply differential cross
section derived by Bethe and Heitler (1934) to determine the correct doubly differential
cross.

\begin {landscape}
\small
\begin{table}
 \begin{tabular}{|c|c|c|c|c|}


  \hline
  Data/Paper & Information & Energy range & Atomic Number $Z$ & Remarks\\
  \hline
  Bethe and & $\omega$ & 1 keV - 1 GeV &
  7,8 & energy range depends on $Z$ \\
  Heitler (1934,1944)& $\omega,\Theta_i,\Theta_f,\Phi$ & & & \\
  \hline
   & Total & different lower & depends on
  the &  \\
  Koch and & $\omega$ & bounds, no & used
  formulae & \\
  Motz (1959) & $\omega,\Theta_i,\Phi$ & upper bounds & & \\
   & $\omega,\Theta_i,\Theta_f,\Phi$ & & & \\
  \hline
  Aiginger (1966) & $\omega,\Theta_i$
  & 180, 380 keV & 79, Al$_2$O$_3$ & experimental\\
  \hline
  Elwert and & $\omega,\Theta_i$
  & keV range & 13,79 & \\
  Haug (1969)& $\omega,\Theta_i,\Theta_f,\Phi$ & & & \\
  \hline
  Penczynski and  & $\omega,\Theta_i$ & $(300\pm10)$ keV & 82 & experimental\\
  Wehner (1970) & & & & \\
  \hline
  Tseng and & $\omega,\Theta_i,\Theta_f,\Phi$
  & keV, MeV range& 13,79 &\\
  Pratt (1971) & & & & \\
  \hline
  Fink and & $\omega$ & keV, MeV range & 6,13,79,92 & also for pair production\\
  Pratt (1973) & $\omega,\Theta_i,\Theta_f,\Phi$ & & &\\
  \hline
  Tsai (1974,1977) & $\omega,\Theta_i$
  & $>$ few 10 MeV & all & \\
  \hline
  Seltzer and & $\omega$ & 1 keV - 10 GeV &
  Z=6,13,29,47,74,92 & \\
  Berger (1985) & & & &\\
  \hline
 EEDL (1991) & Total & 5 eV - 1 TeV & all & see \cite{EEDL} \\
  \hline
  Nackel (1994) & $\omega,\Theta_i$ & keV & 6,29,47,79 & only twodimensional
  description\\
  \hline
  Schaffer & $\omega,\Theta_i,\Phi$ & keV range & 6,13,29,47,74,92 & \\
  et al. (1996) & & & & \\
  \hline
  Schaffer and & $\omega,\Theta_i,\Theta_f,\Phi$ & keV
  range & 47,53,60,68,79 & \\
  Pratt (1997) & $\omega,\Theta_i$ & & & \\
  \hline
  Lehtinen (2000) & $\omega,\Theta_i$
  & 1 keV - 1 GeV & 7,8 & Simple product ansatz for\\
  & & & & angular and frequency part\\
  \hline
  & Total & 5 eV - 1 TeV & all & based on EEDL \\
  Geant 4 (2003) & $\omega$ & 1 keV - 10 GeV &
  6,13,29,47,74,92 & based on Seltzer and Berger (1985) \\
  & $\Theta_i$ & $>$ few 10 MeV & all & based on
  Tsai (1974, 1977)\\
  \hline
 \end{tabular}
 \caption {Available data for Bremsstrahlung cross sections. Besides the
 available information on total or singly, doubly or triply differential cross-sections, the range of validity of the incident electron energy and of
 the atomic number is given. If not stated
 otherwise, these are theoretical expressions.} \label{brems.tab}
 \end{table}
\end {landscape}

I\subsection{Bremsstrahlung data used by other TGF researchers}

Carlson et al.~(2009, 2010) use Geant 4, a library of sotware
tools with a preimplemented  database to
simulate the production of Terrestrial Gamma-Ray Flashes. But Geant 4 does not supply an energy resolved angular distribution as it does not contain a doubly differential cross section, parameterizing both energy and emission angle of the Bremsstrahlung photons (see Table \ref{brems.tab}).
Furthermore, it is designed for high electron energies. It also includes the
Landau-Pomeranchuk-Migdal (LPM) (Landau and Pomeranchuk,\ 1953) effect and dielectric suppression
(Ter-Mikaelian,\ 1954) which do
not contribute in the keV and MeV range.
We will briefly discuss the cross sections and effects implemented in Geant 4 in 
\ref{app_geant4}.

Lehtinen has suggested a doubly differential cross section in his PhD thesis
(Lehtlinen, 2000) that is also used by Xu et al. (2012). Lehtinen's ansatz is a heuristic approach based on factorization into two factors. The first factor is the singly differential cross section of Bethe and Heitler (1934) that resolves only electron and photon energies, but no angles. The second factor is due to Jackson
(1975, p. 712 et seq.), it depends on the variable $(1-\beta^2)\;[(1-\beta\cos\Theta_i)^2 +
(\cos\Theta_i-\beta)^2]\;/\; (1-\beta\cos\Theta_i)^4$, where $\beta = |{\bf} v_i|/c$ measures the incident electron velocity on the relativistic scale. However, this factor derived in Jackson
(1975, p. 712 et seq.) is calculated in the classical and not quantum mechanical
case, and it is valid only if the photon energy is much smaller than the total energy of the incident electron.
We will compare this ansatz with our results in \ref{app_leht}.

Dwyer (2007) chooses to use the triply differential cross section by Bethe and Heitler (1934), but with an additional form factor parameterizing the structure of the nucleus~\cite{brems_theory}.
We will show in \ref{app_form} that this form factor, however, does not contribute for
energies above 1~keV. This cross section depends on all three angles as shown in Fig.~1.
If one is only interested in the angle $\Theta_i$ between incident electron
and emitted Bremsstrahlung photon, the angles $\Theta_f$ and $\Phi$ have to
be integrated out --- either numerically, or the analytical results derived
in the present paper can be used.

\subsection {Organization of the paper}

In chapter \ref{brems} we introduce the triply differential cross
section derived by Bethe and Heitler 
Then we integrate over the two angles $\Theta_f$ and $\Phi$
to obtain the doubly differential cross section which gives a correlation
between the energy of the emitted photon and its direction relative to the
incident electron. Furthermore, we investigate the limit
of very small or very large angles and of high photon energies; this also serves as a consistency check for the correct integration of the full expression.

In chapter \ref{pair} we perform the same calculations for pair production, i.e., when an incident photon interacts with an atomic
nucleus and creates a
positron electron pair. As we explain, this process is actually related by some physical symmetry to Bremsstrahlung, therefore results can be transferred from Bremsstrahlung to pair production. We get a doubly differential cross section for energy and emission angle of the created positron relative to direction and energy of the incident photon.

The physical interpretation and implications of our analytical results is discussed in chapter \ref{disc}. Energies and emission angles of the created photons and positrons are described in the particular case of scattering on nitrogen nuclei. For electron energies below 100~keV, the emission of Bremsstrahlung photons in different directions varies typically by not more than an order of magnitude, while for higher electron energies the photons are mainly
emitted in forward direction. For this case, we derive an analytical approximation for the most likely emission angle of Bremsstrahlung photons and positrons for given particle energies. 

In chapter \ref{concl} we will briefly summarize the results of our calculations.

Details of our calculations can be found in
\ref{app_res} - \ref{app_theta}. Beyond that we provide a C++ code 
The C++ code can be downloaded directly from the website of the journal.

\newpage

\section {Bremsstrahlung} \label{brems}

\subsection {Definition of the process} \label{brems_def}

If an electron with momentum $\mathbf{p}_i$ approaches the nucleus of an atom, it can change its direction due to Coulomb interaction with the
nucleus; the electron acceleration creates a Bremsstrahlung photon with momentum $\mathbf{k}$ that can be emitted at an angle $\Theta_i$ relative to the initial direction of the electron. The new direction of the electron
forms an angle $\Theta_f$ with the direction of the photon. The angle $\Phi$ is the angle between the planes spanned by the vector pairs $(\mathbf{p}_i, \mathbf{k})$ and $(\mathbf{p}_f,\mathbf{k})$. This process is shown in figure 1. A virtual photon (allowed by Heisenberg's uncertainty principle) transfers a momentum ${\bf q}$ between the electron and the nucleus. Therefore both energy and momentum are conserved in the scattering process.

The corresponding triply differential cross section 
was derived by Bethe and Heitler (1934):
\begin {eqnarray}
d^4\sigma &=&
\frac{Z^2\alpha_{fine}^3\hbar^2}{(2\pi)^2}\frac{|\mathbf{p}_f|}{|\mathbf{p}_i|}
\frac{d\omega}{\omega}\frac{d\Omega_i d\Omega_f d\Phi}{|\mathbf{q}|^4}\times
\nonumber \\
&\times&\left[
\frac{\mathbf{p}_f^2\sin^2\Theta_f}{(E_f-c|\mathbf{p}_f|\cos\Theta_f)^2}\left
(4E_i^2-c^2\mathbf{q}^2\right)\right. \nonumber\\
&+&\frac{\mathbf{p}_i^2\sin^2\Theta_i}{(E_i-c|\mathbf{p}_i|\cos\Theta_i)^2}\left
(4E_f^2-c^2\mathbf{q}^2\right) \nonumber \\
&+&2\hbar^2\omega^2\frac{\mathbf{p}_i^2\sin^2\Theta_i+\mathbf{p}_f^2\sin^2\Theta_f}{(E_f-c|\mathbf{p}_f|\cos\Theta_f)(E_i-c|\mathbf{p}_i|\cos\Theta_i)}
\nonumber\\
&-&2\left.\frac{|\mathbf{p}_i||\mathbf{p}_f|\sin\Theta_i\sin\Theta_f\cos\Phi}{(E_f-c|\mathbf{p}_f|\cos\Theta_f)(E_i-c|\mathbf{p}_i|\cos\Theta_i)}\left(2E_i^2+2E_f^2-c^2\mathbf{q}^2\right)\right].
\label{bcs.1}
\end {eqnarray}
Here $Z$ is the atomic number of the nulceus, $\alpha_{fine}\approx {1}/{137}$
is the fine structure constant, $h\approx 6.63\cdot 10^{-34}$~Js is Planck's
constant, $\hbar=h/{2\pi}$ and $c\approx
3\cdot 10^{8}$ m/s is the speed of light. The kinetic energy
$E_{kin,i/f}$ of the electron in the initial and final state is 
related to its total energy and momentum as
\begin {eqnarray}
E_{i/f}=E_{kin,i/f}+m_e c^2=\sqrt{m_e^2c^4+\mathbf{p}_{i/f}^2c^2} \label{bcs.2}
\end {eqnarray}
where $m_e\approx 9.1\cdot 10^{-31}$ kg is the electron mass. The
conservation of energy implies
\begin {eqnarray}
E_f=E_i-\hbar\omega   \label{bcs.2b}
\end {eqnarray}
  which determines $E_f$ as a function
of $E_i$ and $\hbar\omega$.
The directions of the emitted photon with energy $\hbar\omega$ and of the scattered electron are parameterized by the three angles (see Fig.~1)
\begin {eqnarray}
\Theta_i&=&\sphericalangle (\mathbf{p}_i,\mathbf{k}),  \label{bcs.3} \\
\Theta_f&=&\sphericalangle (\mathbf{p}_f,\mathbf{k}),  \label{bcs.4}\\
\Phi&=&\textnormal{Angle between the planes $(\mathbf{p}_i,\mathbf{k})$ and
$(\mathbf{p}_f,\mathbf{k})$}. \label{bcs.5}
\end {eqnarray}
The differentials are
\begin {eqnarray}
d\Omega_i &=& \sin\Theta_i \ d\Theta_i,\\
d\Omega_f &=& \sin\Theta_f \ d\Theta_f.
\label{bcs.6}
\end {eqnarray}
Furthermore one can get an expression for the absolute value of the virtual
photon $\mathbf{q}$ with the help of the momenta, the photon
energy $\hbar\omega$ and the angles (\ref{bcs.3}) - (\ref{bcs.5}). Its value
is
\begin {eqnarray}
-\mathbf{q}^2&=&-|\mathbf{p}_i|^2-|\mathbf{p}_f|^2-\left(\frac{\hbar}{c}\omega\right)^2+2|\mathbf{p}_i|\frac{\hbar}{c}
\omega\cos\Theta_i-2|\mathbf{p}_f|\frac{\hbar}{c} \omega\cos\Theta_f
\nonumber\\
&+&2|\mathbf{p}_i||\mathbf{p}_f|(\cos\Theta_f\cos\Theta_i+\sin\Theta_f\sin\Theta_i\cos\Phi).
\label {bcs.7}
\end {eqnarray}

\subsection {Validity of the cross sections of Bethe and Heitler} \label{valid}

The cross sections of Bethe and Heitler (\ref{bcs.1}) are valid if the Born approximation~\cite {bethe} holds
\begin {eqnarray}
v\gg\frac{Zc}{137} \label{born.1}
\end {eqnarray}
For nitrogen with $Z=7$ and for oxygen with $Z=8$, this holds for electron velocities 
$|\mathbf{v}_{Z=7}|\gg 15\cdot 10^6$ m/s and
$|\mathbf{v}_{Z=8}|\gg 18\cdot 10^6$ m/s; this 
is equivalent to a kinetic energy of
\begin {eqnarray}
E_{kin}=\frac{m_e c^2}{\sqrt{1-\frac{v^2}{c^2}}}-m_ec^2 \gg \left\{ \begin
{array}{c}670 \ \textnormal{eV},\ Z=7 \\
875 \ \textnormal{eV}, \ Z=8 \end {array}\right. .
\label{born.2}
\end {eqnarray}
This means that incident electron energies above 1 keV can be treated with Eq.~\ref{bcs.1}, for lower energies, one cannot calculate with free electron waves anymore, but has to use Coulomb waves
(Heitler, 1944; Greiner und Reinhardt, 1995); in this case one cannot derive cross sections like
(\ref{bcs.1}) analytically any more. Thus the Bethe Heitler
cross section and our results must not be used for energies of the electron
in the initial and final state smaller than 1 keV. However, for higher
energies of the electron in the initial and final state, the approximation
by Bethe and Heitler becomes more accurate; thus this approximation is
better if $E_{kin}\ge 10$ keV.

\subsection {Integration over $\Phi$} \label {sec_phi}

The easiest way is to integrate over the angle $\Phi$ between the
scattering planes first   (see Fig. \ref{start_fig.1})  . For this purpose
it is useful to redefine some
quantities in the following way; therefore (\ref{bcs.1}) can be written much
more simply:
\begin {eqnarray}
\alpha &:=& 2|\mathbf{p}_i||\mathbf{p}_f|\sin\Theta_i\sin\Theta_f, \label{int.1}\\
\beta &:=&
-\mathbf{p}_i^2-\mathbf{p}_f^2-\left(\frac{\hbar}{c}\omega\right)^2-2\frac{\hbar}{c}\omega|\mathbf{p}_f|\cos\Theta_f+
2 \frac{\hbar}{c}\omega|\mathbf{p}_i|\cos\Theta_i \nonumber\\
&+& 2|\mathbf{p}_i||\mathbf{p}_f|\cos\Theta_i\cos\Theta_f \label{int.2}\\
A &:=& \frac{Z^2\alpha_{fine}^3}{(2\pi)^2}\frac{|\mathbf{p}_f|}{|\mathbf{p}_i|}
\frac{\hbar^2}{\omega}, \label{int.3} \\
a_1&:=&\left(\frac{|\mathbf{p}_f|^2c^2\sin^2\Theta_f}{(E_f-|\mathbf{p}_f|c\cos\Theta_f)^2}+\frac{|\mathbf{p}_i|^2c^2\sin^2\Theta_i}{(E_i-|\mathbf{p}_i|c\cos\Theta_i)^2}\right)\cdot
A, \label{int.4} \\
a_2&:=&\left(-\frac{2|\mathbf{p}_i||\mathbf{p}_f|c^2\sin\Theta_i\sin\Theta_f}{(E_i-|\mathbf{p}_i|c\cos\Theta_i)(E_f-|\mathbf{p}_f|c\cos\Theta_f)}
\right)\cdot A, \label{int.5}\\
a_3&:=&\left( \frac{4E_i^2
|\mathbf{p}_f|^2\sin^2\Theta_f}{(E_f-|\mathbf{p}_f|c\cos\Theta_f)^2}+\frac{4E_f^2
|\mathbf{p}_i|^2\sin^2\Theta_i}{(E_i-|\mathbf{p}_i|c\cos\Theta_i)^2}
\right. \nonumber \\
&+&\left.\frac{2\hbar^2\omega^2(|\mathbf{p}_i|^2\sin\Theta_i+|\mathbf{p}_f|^2\sin^2\Theta_f)}{(E_i-|\mathbf{p}_i|c\cos\Theta_i)(E_f-|\mathbf{p}_f|c\cos\Theta_f)}\right)
\cdot A, \label{int.6} \\
a_4&:=&\left(-\frac{|\mathbf{p}_i||\mathbf{p}_f|\sin\Theta_i\sin\Theta_f(4E_i^2+4E_f^2)}{(E_i-|\mathbf{p}_i|c\cos\Theta_i)(E_f-|\mathbf{p}_f|c\cos\Theta_f)}
\right)\cdot A. \label{int.7}
\end {eqnarray}
With (\ref{int.1}) - (\ref{int.7}), Eq. (\ref{bcs.1}) can be written as:
\begin {eqnarray}
\frac{d^4\sigma}{d\omega \Omega_i d\Omega_f d\Phi} &=&
\frac{a_1}{\alpha\cos\Phi+\beta}
+\frac{a_2\cos\Phi}{\alpha\cos\Phi+\beta} \nonumber\\
&+&\frac{a_3}{(\alpha\cos\Phi+\beta)^2}
+\frac{a_4\cos\Phi}{(\alpha\cos\Phi+\beta)^2}; \label{int.8}
\end {eqnarray}
thus the integration over $\Phi$ simply reads
\begin {eqnarray}
\frac{d^3\sigma}{d\omega d\Omega_i d\Omega_f} &=&
\int\limits_{0}^{2\pi}d\Phi\left[\frac{a_1}{\alpha\cos\Phi+\beta} +
\frac{a_2\cos\Phi}{\alpha\cos\Phi+\beta} \right.\nonumber\\
&+&\left.\frac{a_3}{(\alpha\cos\Phi+\beta)^2}
+\frac{a_4\cos\Phi}{(\alpha\cos\Phi+\beta)^2}\right] \label{int.9}
\end {eqnarray}
where $a_i,\ i\in\{1,\ldots,4\},\alpha$ and $\beta$ still depend on $\Theta_f$
and $\Theta_i$.
These integrals can be calculated with the help of the residue theorem which
is reviewed briefly in \ref{app_res}. If $R(x,y):\mathbb{R}^2\rightarrow\mathbb{R}$ is
a rational function without poles on the unit circle $x^2+y^2=1$, then
\begin {eqnarray}
\int\limits_{0}^{2\pi} R(\cos\Phi,\sin\Phi) d\Phi  = 2\pi i \sum\limits_
{|z|<1} \Res(f,z) \label{int.10}
\end {eqnarray}
where $f$ is a complex function which is defined as
\begin {eqnarray}
f(z):=\frac{1}{iz} R\left(\frac{1}{2}\left(z+\frac{1}{z}\right),
\frac{1}{2i} \left(z-\frac{1}{z}\right)\right). \label {int.11}
\end {eqnarray}
  The residuum of a pole $z_j$ of order $n$ is defined as
\begin {eqnarray}
\Res(f,z_j)=\frac{1}{(n-1)!}\lim_{z\rightarrow z_j}\frac{d^{n-1}}{dz^{n-1}}
\Big[(z-z_j)f(z)\Big].
\end {eqnarray}
To integrate the functions in Eq. (\ref{int.9}), we write
\begin {eqnarray}
R_1(x,y)&:=&\frac{a_1}{\alpha x+\beta}, \label{int.12}\\
R_2(x,y)&:=&\frac{a_2x}{\alpha x+\beta}, \label{int.13}\\
R_3(x,y)&:=&\frac{a_3}{(\alpha x+\beta)^2}, \label{int.14}\\
R_4(x,y)&:=&\frac{a_4x}{(\alpha x+\beta)^2}; \label{int.15}
\end {eqnarray}
and   get from (\ref{int.11})
\begin {eqnarray}
f_1(z)&=&\frac{2a_1}{i(\alpha z^2+2\beta z+\alpha)}, \label{int.16}\\
f_2(z)&=&\frac{a_2z^2+a_2}{zi(\alpha z^2+2\beta z+\alpha)}, \label{int.17}
\\
f_3(z)&=&\frac{4a_3z}{i\left(\alpha z^2+2\beta z+\alpha
\right)^2}, \label{int.18}\\
f_4(z)&=&\frac{2a_4(z^2+1)}{i\left(\alpha z^2+2\beta z+
\alpha\right)^2}. \label{int.19}
\end {eqnarray}
  The   poles of   the functions $f_i(z)$
in   (\ref{int.16}) - (\ref{int.19}) are given by
\begin {eqnarray}
z_{1,2}=-\frac{\beta}{\alpha}\pm\sqrt{\left(\frac{\beta}{\alpha}\right)^2-1}.
\label {int.20}
\end {eqnarray}
  For   $f_{1,2}$ poles are of order one
and for   $f_{3,4}$ of order two. In addition $f_2$ has a pole at
\begin {eqnarray}
z_3=0. \label {int.21}
\end {eqnarray}
According to (\ref{int.10}) one needs poles with $|z_i|<1$. For $z_3$ it
is quite clear that $|z_3|=0<1$.
  As the angles $\Theta_i$ and $\Theta_f$ are between $0$ and
$\pi$, the expression $\alpha>0$ in Eq. (\ref{int.1}).
Furthermore $\cos\Theta_f>-1$, $\cos\Theta_i<1$ and $|\mathbf{p_i}|> \hbar/c
\ \omega$. Hence
\begin {eqnarray}
&&\beta \\
&=&-\mathbf{p}_i^2-\mathbf{p}_f^2-\left(\frac{\hbar}{c}\omega\right)^2-2\frac{\hbar}{c}
\omega|\mathbf{p}_f|\cos\Theta_f+2\frac{\hbar}{c}\omega|\mathbf{p}_i|
\cos\Theta_i \nonumber \\
&+&2|\mathbf{p}_i||\mathbf{p}_f|\cos\Theta_i\cos\Theta_f \\
&=&-\mathbf{p}_i^2-\mathbf{p}_f^2-\left(\frac{\hbar}{c}\omega\right)^2
+2\frac{\hbar}{c}\omega|\mathbf{p}_i| \cos\Theta_i\nonumber\\
&+&2|\mathbf{p}_f|\cos\Theta_f\left(-\frac{\hbar}{c}\omega+|\mathbf{p}_i|\cos\Theta_i\right)\\
&<&-\mathbf{p}_i^2-\mathbf{p}_f^2-\left(\frac{\hbar}{c}\omega\right)^2
+2\frac{\hbar}{c}\omega|\mathbf{p}_i|
+2|\mathbf{p}_f|\cos\Theta_f\left(-\frac{\hbar}{c}\omega+|\mathbf{p}_i|\right)\\
&<&-\mathbf{p}_i^2-\mathbf{p}_f^2-\left(\frac{\hbar}{c}\omega\right)^2
+2\frac{\hbar}{c}\omega|\mathbf{p}_i|
+2|\mathbf{p}_f|\left(-\frac{\hbar}{c}\omega+|\mathbf{p}_i|\right)\\
&=&-\mathbf{p}_i^2-\mathbf{p}_f^2-\left(\frac{\hbar}{c}\omega\right)^2
+2\frac{\hbar}{c}\omega|\mathbf{p}_i|
-2|\mathbf{p}_f|\frac{\hbar}{c}\omega+2|\mathbf{p}_f||\mathbf{p}_i|\\
&=&-\left(|\mathbf{p}_i|-|\mathbf{p}_f|-\frac{\hbar}{c}\omega\right)^2 < 0
\end {eqnarray}
Therefore $\beta/\alpha$ in Eq. (\ref{int.20}) is a negative
real number. Furthermore $\sin\Theta_i<1$ and $\sin\Theta_f<1$.
Thus
\begin {eqnarray}
&&-\beta-\alpha \\
&=&\mathbf{p}_i^2+\mathbf{p}_f^2+\left(\frac{\hbar}{c}\omega\right)^2+2\frac{\hbar}{c}
\omega|\mathbf{p}_f|\cos\Theta_f-2\frac{\hbar}{c}\omega|\mathbf{p}_i|
\cos\Theta_i \nonumber \\
&-&2|\mathbf{p}_i||\mathbf{p}_f|\cos\Theta_i\cos\Theta_f-2|\mathbf{p}_i|
|\mathbf{p}_f|\sin\Theta_i\sin\Theta_f \\
&>& \mathbf{p}_i^2+\mathbf{p}_f^2+\left(\frac{\hbar}{c}\omega\right)^2-2\frac{\hbar}{c}
\omega|\mathbf{p}_f|-2\frac{\hbar}{c}\omega|\mathbf{p}_i|-4|\mathbf{p}_i||\mathbf{p}_f|
\\
&>& \mathbf{p}_i^2+\mathbf{p}_f^2+\left(\frac{\hbar}{c}\omega\right)^2 > 0\\
&\Rightarrow& -\frac{\beta}{\alpha} > 1
\end {eqnarray}
It follows immediately that $|z_1|>1$ and $|z_2|<1$.
For all residua one obtains
\begin {eqnarray}
\Res(f_1,z_2)&=&-\frac{a_1}{i}\frac{1}{\sqrt{\beta^2-\alpha^2}},
\label{int.22}\\
\Res(f_2,z_2)&=&\frac{a_2\beta}{\alpha i}\frac{1}{\sqrt{\beta^2-\alpha^2}},
\label{int.23}\\
\Res(f_2,z_3)&=&\frac{a_2}{\alpha i}, \label{int.24}\\
\Res(f_3,z_2)&=&-\frac{a_3\beta}{i}\frac{1}{(\sqrt{\beta^2-\alpha^2})^3},
\label{int.25}\\
\Res(f_4,z_2)&=&a_4\alpha\frac{1}{(\sqrt{\beta^2-\alpha^2})^3}.
\label{int.26}
\end {eqnarray}
With the knowledge of these residua and using (\ref{int.10}), the integral in
(\ref{int.9}) can be calculated elementarily
\begin {eqnarray}
\frac{d^3\sigma}{d\omega d\Omega_i d\Omega_f} =
\frac{2\pi a_2}{\alpha}&+&\frac{2\pi}{\sqrt{\beta^2-\alpha^2}} \left[-a_1+
\frac{a_2\beta}{\alpha} \right. \nonumber \\
&-&\left.\frac{a_3\beta}{|\beta^2-\alpha^2|}+\frac{a_4
\alpha}{|\beta^2-\alpha^2|}\right]. \label{int.27}
\end {eqnarray}

\subsection {Integration over $\Theta_f$} \label{theta_int}

After having obtained an expression for the  ``triply'' \footnote {
  Here ``triply'' really means the dependence on the photon frequency and two angles.
 }   differential cross
section, there is still the integration over $\Theta_f$ left. This
calculation is mainly straight forward, but rather tedious.  
Using expression (\ref{int.27}), it is
\begin {eqnarray}
\frac{d^2\sigma}{d\omega d\Omega_i} = \int\limits_{0}^{\pi}
d\Theta_f\left[\frac{2\pi a_2}{\alpha}\right.&+&\frac{2\pi}{\sqrt{\beta^2-\alpha^2}}
\left(-a_1+\frac{a_2\beta}{\alpha} \right. \nonumber \\
&-&\left.\left.\frac{a_3\beta}{|\beta^2-\alpha^2|}+\frac{a_4
\alpha}{|\beta^2-\alpha^2|}\right)\right]\sin\Theta_f. \label{theta.1}
\end {eqnarray}
Let's now consider the first integral of (\ref{theta.1}). If one inserts
(\ref{int.1}) and (\ref{int.5}), it becomes
\begin {eqnarray}
\int\limits_0^{\pi}d\Theta_f\frac{2\pi a_2}{\alpha}\sin\Theta_f &=&
-\frac{2\pi Ac^2}{E_i-cp_i\cos\Theta_i}\int\limits_0^{\pi}d\Theta_f
\frac{\sin\Theta_f}{E_f-cp_f\cos\Theta_f} \label{theta.2} \\
&=&-\frac{2\pi Ac^2}{E_i-cp_i\cos\Theta_i}\int\limits_{-1}^{+1}dx
\frac{1}{E_f-cp_fx} \label{theta.3}
\end {eqnarray}
where the substitution $x:=\cos\Theta_f$ was
made in the second step. (\ref{theta.3}) is rather simple and yields
\begin {eqnarray}
\int\limits_0^{\pi}d\Theta_f\frac{2\pi a_2}{\alpha}\sin\Theta_f = -
\frac{2\pi Ac}{(E_i-cp_i\cos\Theta_i)p_f}\ln\left(\frac{E_f+p_fc}
{E_f-p_fc}\right). \label{theta.4}
\end {eqnarray}
This was a quite simple calculation. All the other integrals can be
calculated similarly, but with more effort. As another example let's consider
the last integral. Before inserting (\ref{int.1}), (\ref{int.2}) and
(\ref{int.7}) one can define   for simplicity
\begin {eqnarray}
\Delta_1&:=& -\mathbf{p}_i^2-\mathbf{p}_f^2-\left(\frac{\hbar}{c}\omega\right)^2+2\frac{\hbar}{c}\omega|\mathbf{p}_i|\cos\Theta_i,
\label {theta.5} \\
\Delta_2&:=& -2\frac{\hbar}{c}\omega|\mathbf{p}_f|+2|\mathbf{p}_i||\mathbf{p}_f|\cos\Theta_i. \label {theta.6}
\end {eqnarray}
  The expression $\beta$ from Eq. (\ref{int.2}) is then
\begin {eqnarray}
\beta=\Delta_1+\Delta_2\cos\Theta_f.   \label{theta.7}
\end {eqnarray}
Thus the regularly appearing term $\beta^2-\alpha^2$ can be written as
\begin {eqnarray}
\beta^2-\alpha^2&=&(\Delta_2^2+4p_i^2p_f^2\sin^2\Theta_i)\cos^2\Theta_f
+2\Delta_1\Delta_2\cos\Theta_f \nonumber \\
&+&(\Delta_1^2-4p_i^2p_f^2\sin^2\Theta_i) \label{theta.8} \\
&=& \Box_1^2\cos^2\Theta_f+2\Delta_1\Delta_2\cos\Theta_f+\Box_2^2
\label{theta.9}
\end {eqnarray}
where the definitions
\begin {eqnarray}
\Box_1^2&:=& \Delta_2^2+4p_i^2p_f^2\sin^2\Theta_i, \label{theta.10} \\
\Box_2^2&:=& \Delta_1^2-4p_i^2p_f^2\sin^2\Theta_i \label{theta.11}
\end {eqnarray}
have been introduced.

By using (\ref{int.1}), (\ref{int.7}) and (\ref{theta.9}), the last integral
of (\ref{int.27}) becomes
\begin {eqnarray}
&&\int\limits_0^{\pi} d\Theta_f \frac{2\pi
a_4\alpha}{\sqrt{(\beta^2-\alpha^2)^3}} \sin\Theta_f
=-\frac{16\pi A p_i^2p_f^2\sin^2\Theta_i(E_i^2+E_f^2)}{E_i-cp_i\cos\Theta_i}
\nonumber\\
&\times&\int\limits_0^{\pi} d\Theta_f\frac{\sin^2\Theta_f}{\sqrt{(\Box_1^2\cos\Theta_f^2
+2\Delta_1\Delta_2\cos\Theta_f+\Box_2^2)^3}(E_f-cp_f\cos\Theta_f)}\sin\Theta_f
\nonumber \\ \label{theta.12} \\
&=&-\frac{16\pi A p_i^2p_f^2\sin^2\Theta_i(E_i^2+E_f^2)}{E_i-cp_i\cos\Theta_i}
\nonumber \\
&\times&\int\limits_{-1}^{+1} dx \frac{1-x^2}{\sqrt{(\Box_1^2 x^2
+2\Delta_1\Delta_2x+\Box_2^2)^3}(E_f-cp_f x)} \label{theta.13}
\end {eqnarray}
where $x=\cos\Theta_f$ has been   substituted   again.
\\ \indent
This integration can be performed elementarily by finding   the
indefinite integral
\tiny
\begin {eqnarray}
&&\frac{\Box_1^2x^2+2\Delta_1\Delta_2x+\Box_2^2}{\sqrt{(\Box_1^2x^2+2\Delta_1\Delta_2x+\Box_2^2)^3}(\Delta_1^2\Delta_2^2-
\Box_1^2\Box_2^2)(\Box_1^2E_f^2+2\Delta_1\Delta_2E_fp_fc+\Box_2^2p_fc)}\times
\nonumber\\
&\times&\left(-\Box_1^4E_fx+\Box_2^4p_fc+2\Delta_1^2\Delta_2^2(E_fx-p_fc)+\Delta_1
\Delta_2\Box_2^2(E_f+p_fcx)\right. \nonumber\\
&-&\left.\Box_1^2(\Box_2^2(E_fx-p_fc)\Delta_1\Delta_2
(E_f+p_fcx))\right)\nonumber\\
&+&\frac{E_f^2-p_f^2c^2}{\sqrt{(\Box_1^2E_f^2+2\Delta_1\Delta_2E_fp_fc+\Box_2^2p_fc)^3}}
\ln\Bigl((E_f-p_fcx)(\Box_1^2E_fx+\Box_2^2p_fc \nonumber\\
&+&\Delta_1\Delta_2(E_f+p_fcx)+\sqrt{\Box_1^2x^2+2\Delta_1\Delta_2x+\Box_2^2}\times
\nonumber\\
&\times&\sqrt{\Box_1^2E_f^2+2\Delta_1\Delta_2E_fp_fc+\Box_2^2p_fc})\Bigr),
 \label{theta.14}
\end {eqnarray}
\normalsize
by inserting $+1$ and $-1$ as upper and lower limit, using (\ref{theta.10}) and
(\ref{theta.11}) and simplifying.   The integral in
(\ref{theta.12}) is then finally
\tiny
\begin {eqnarray}
\int\limits_0^{\pi} d\Theta_f \frac{2\pi
a_4\alpha}{\sqrt{(\beta^2-\alpha^2)^3}} \sin\Theta_f&-&\frac{16\pi Ap_i^2p_f^2\sin^2\Theta_i(E_i^2+E_f^2)}{E_i-cp_i\cos\Theta_i}\times
\nonumber\\
&\times&\left[-\frac{2(\Delta_2p_fc+\Delta_1E_f)}
{(-\Delta_2^2+\Delta_1^2-4p_i^2p_f^2\sin^2\Theta_i)((\Delta_2E_f+\Delta_1p_fc)^2+4m^2c^4p_i^2p_f^2\sin^2\Theta_i)}\right.
\nonumber\\
&+&\frac{m^2c^4}{\sqrt{((\Delta_2E_f+\Delta_1p_fc)^2+4m^2c^4p_i^2p_f^2\sin^2\Theta_i)^3}} \times
\nonumber\\
&\times&\ln\Bigg(\Big((E_f-cp_f)(4p_i^2p_f^2\sin^2\Theta_i(-E_f-p_fc)+(\Delta_1-
\Delta_2)((\Delta_2E_f+\Delta_1p_fc) \nonumber\\
&-&\sqrt{\Box_1^2E_f^2+2\Delta_1\Delta_2E_fp_fc+\Box_2^2p_fc})\Big)\Big((E_f+cp_f)(4p_i^2p_f^2\sin^2\Theta_i(+E_f-p_fc)
\nonumber \\
&+&(\Delta_1+\Delta_2)((\Delta_2E_f+\Delta_1p_fc)-\sqrt{\Box_1^2E_f^2+2\Delta_1\Delta_2E_fp_fc+\Box_2^2p_fc})\Big)^{-1}\Bigg)\Bigg].
\nonumber \\ \label{theta.15}
\end {eqnarray}
\normalsize
All the other integrals can be calculated similarly where one always has to
  substitute   $x=\cos\Theta_f$. With this technique the whole doubly
differential cross section finally becomes
\begin {eqnarray}
\frac{d^2\sigma (E_i,\omega,\Theta_i)}{d\omega d\Omega_i
}=\sum\limits_{j=1}^{6} I_j \label{theta.16}
\end {eqnarray}
with the following contributions:
\tiny
\begin {eqnarray}
I_1&=&\frac{2\pi A}{\sqrt{\Delta_2^2+4p_i^2p_f^2\sin^2\Theta_i}}
\ln\left(
\frac{\Delta_2^2+4p_i^2p_f^2\sin^2\Theta_i-\sqrt{\Delta_2^2+4p_i^2p_f^2\sin^2
\Theta_i}(\Delta_1+\Delta_2)+\Delta_1\Delta_2}{-\Delta_2^2-4p_i^2p_f^2\sin^2\Theta_i
-\sqrt{\Delta_2^2+4p_i^2p_f^2\sin^2 \Theta_i}(\Delta_1-\Delta_2)+\Delta_1\Delta_2
}\right)  \nonumber\\
&\times&\left[1+\frac{c\Delta_2}{p_f(E_i-cp_i\cos\Theta_i)}-\frac{p_i^2c^2\sin^2\Theta_i}
{(E_i-cp_i\cos\Theta_i)^2}
-\frac{2\hbar^2\omega^2p_f\Delta_2}{c(E_i-cp_i\cos
\Theta_i)(\Delta_2^2+4p_i^2p_f^2\sin^2\Theta_i)}\right], \label{theta.17}\\
I_2&=&-\frac{2\pi Ac}{p_f(E_i-cp_i\cos\Theta_i)}\ln\left(
\frac{E_f+p_fc}{E_f-p_fc}\right), \label{theta.18} \\
I_3&=&\frac{2\pi A}{\sqrt{(\Delta_2E_f+\Delta_1p_fc)^2+4m^2c^4p_i^2p_f^2\sin^2\Theta_i
}} \nonumber \\
&\times&\ln\Bigg(\Big((E_f+p_fc)(4p_i^2p_f^2\sin^2\Theta_i(E_f-p_fc)+(\Delta_1+\Delta_2)
((\Delta_2E_f+\Delta_1p_fc) \nonumber\\
&-&\sqrt{(\Delta_2E_f+\Delta_1p_fc)^2+4m^2c^4p_i^2p_f^2\sin^2\Theta_i}))\Big)\Big((E_f-p_fc)
(4p_i^2p_f^2\sin^2\Theta_i(-E_f-p_fc) \nonumber \\
&+&(\Delta_1-\Delta_2)
((\Delta_2E_f+\Delta_1p_fc)-\sqrt{(\Delta_2E_f+\Delta_1p_fc)^2+4m^2c^4p_i^2p_f^2\sin^2\Theta_i}))\Big)^{-1}
\Bigg) \nonumber\\
&\times&\left[-\frac{(\Delta_2^2+4p_i^2p_f^2\sin^2\Theta_i)(E_f^3+E_fp_f^2c^2)+p_fc(2
(\Delta_1^2-4p_i^2p_f^2\sin^2\Theta_i)E_fp_fc+\Delta_1\Delta_2(3E_f^2+p_f^2c^2))}{(\Delta_2E_f+\Delta_1p_fc)^2+4m^2c^4p_i^2p_f^2\sin^2\Theta_i}\right.\nonumber\\
&-&\frac{c(\Delta_2E_f+\Delta_1p_fc)}{p_f(E_i-cp_i\cos\Theta_i)} \nonumber\\
&-&\frac{4E_i^2p_f^2(2(\Delta_2E_f+\Delta_1p_fc)^2-4m^2c^4p_i^2p_f^2\sin^2\Theta_i)(\Delta_1E_f+\Delta_2p_fc)}{((\Delta_2E_f+\Delta_1p_fc)^2+4m^2c^4p_i^2p_f^2\sin^2\Theta_i)^2} \nonumber \\
&+&\left.\frac{8p_i^2p_f^2m^2c^4\sin^2\Theta_i(E_i^2+E_f^2)-2\hbar^2\omega^2p_i^2\sin^2\Theta_ip_fc(\Delta_2E_f+\Delta_1p_fc)+
2\hbar^2\omega^2p_f m^2c^3(\Delta_2E_f+\Delta_1p_fc)}
{(E_i-cp_i\cos\Theta_i)((\Delta_2E_f+\Delta_1p_fc)^2+4m^2c^4p_i^2p_f^2\sin^2\Theta_i)}\right],
\nonumber\\ \label{theta.19}\\
I_4&=&-\frac{4\pi Ap_fc(\Delta_2E_f+\Delta_1p_fc)}{(\Delta_2E_f+\Delta_1p_fc)^2+4m^2c^4p_i^2p_f^2\sin^2\Theta_i}
-\frac{16\pi E_i^2p_f^2
A(\Delta_2E_f+\Delta_1p_fc)^2}{((\Delta_2E_f+\Delta_1p_fc)^2+4m^2c^4p_i^2p_f^2\sin^2\Theta_i)^2}, \label{theta.20}
\\
I_5&=&\frac{4\pi A}{(-\Delta_2^2+\Delta_1^2-4p_i^2p_f^2\sin^2\Theta_i)
((\Delta_2E_f+\Delta_1p_fc)^2+4m^2c^4p_i^2p_f^2\sin^2\Theta_i)} \nonumber\\
&\times&\left[\frac{\hbar^2\omega^2p_f^2}{E_i-cp_i\cos\Theta_i}\right.\nonumber\\
&\times&\frac{E_f[2\Delta_2^2(\Delta_2^2-\Delta_1^2)+8p_i^2p_f^2\sin^2\Theta_i(\Delta_2^2+\Delta_1^2)]
+p_fc[2\Delta_1\Delta_2(\Delta_2^2-\Delta_1^2)-16\Delta_1\Delta_2p_i^2p_f^2\sin^2\Theta_i]}{\Delta_2^2+4p_i^2p_f^2\sin^2\Theta_i}
\nonumber\\
&+& \frac{2\hbar^2\omega^2 p_i^2\sin^2\Theta_i(2\Delta_1\Delta_2
p_fc+2\Delta_2^2E_f+8p_i^2p_f^2\sin^2\Theta_i E_f)}{E_i-cp_i\cos\Theta_i}\nonumber\\
&+&
\frac{2E_i^2p_f^2\{2(\Delta_2^2-\Delta_1^2)(\Delta_2E_f+\Delta_1p_fc)^2
+8p_i^2p_f^2\sin^2\Theta_i[(\Delta_1^2+\Delta_2^2)(E_f^2+p_f^2c^2)
+4\Delta_1\Delta_2E_fp_fc]\}}{((\Delta_2E_f+\Delta_1p_fc)^2+4m^2c^4p_i^2p_f^2\sin^2\Theta_i)}\nonumber\\
&+&\left.\frac{8p_i^2p_f^2\sin^2\Theta_i(E_i^2+E_f^2)(\Delta_2p_fc +\Delta_1
E_f)}{E_i-cp_i\cos\Theta_i}\right], \label{theta.21}\\
I_6&=&\frac{16\pi E_f^2p_i^2\sin^2\Theta_i A}{(E_i-cp_i\cos\Theta_i)^2
(-\Delta_2^2+\Delta_1^2-4p_i^2p_f^2\sin^2\Theta_i)}. \label{theta.22}
\end {eqnarray}
\normalsize
  Eq. (\ref{theta.16}) depends explicitly on
$E_i,\omega$ and $\Theta_i$ while $E_f$ and $p_f$ are functions of
$E_i$ and $\omega$ through (\ref{bcs.2}) and (\ref{bcs.2b}).
(\ref{theta.16}) is the final result of the integration of (\ref{bcs.1})
over $\Phi$ and $\Theta_f$ with the help of the residue theorem and some
basic calculations. Now this result can be used both as input for Monte Carlo
code and for discussing some basic properties of the behaviour of produced
Bremsstrahlung photons.\\ \indent
Actually (\ref{theta.16}) is also valid for $\Theta_i=0$, as will be shown
in the next section, but the simple way just to set $\Theta_i=0$ in
(\ref{theta.16}) will fail, especially for numerical purposes, because the logarithmic part in (\ref{theta.17})
tends to $\ln(0/0)$ for $\Theta_i\rightarrow0$ and so
fails for numerical applications. Thus we
need an additional expression for $\Theta_i=0$ which has to be consistent
with (\ref{theta.16}).

\subsection {Special limits: $\Theta_i=0,\pi$ and $\hbar\omega \rightarrow E_{kin,i}$}

For some special cases the integration of (\ref{bcs.1}) over $\Phi$ and $\Theta_f$
  is easier. This
information   can   also   be used
  to verify (\ref{theta.16}) by checking consistency and use them for Monte
Carlo codes.

\subsubsection {$\Theta_i=0$ or $\Theta_i=\pi$}

If one is only interested in forward and backward scattering, one can set
$\Theta_i=0$ or $\Theta_i=\pi$ before integrating;  then   (\ref{bcs.1}) becomes
\begin {eqnarray}
\frac{d^4\sigma}{d\omega d\Omega_i d\Omega_f d\Phi} &=&
\frac{Z^2\alpha_{fine}^3\hbar^2}{(2\pi)^2}\frac{|\mathbf{p}_f|}{|\mathbf{p}_i|}
\frac{1}{\omega}\frac{1}{|\mathbf{q}|^4}
\nonumber \\
&\times&\left(
\frac{\mathbf{p}_f^2\sin^2\Theta_f}{(E_f-c|\mathbf{p}_f|\cos\Theta_f)^2}\left
(4E_i^2-c^2\mathbf{q}^2\right) \right. \nonumber\\
&+&2\left.
\hbar^2\omega^2\frac{\mathbf{p}_f^2\sin^2\Theta_f}{(E_f-c|\mathbf{p}_f|\cos\Theta_f)(E_i\mp
c|\mathbf{p}_i|\cos\Theta_i)}
\right) \label{small.1}
\end {eqnarray}
where the momentum $\mathbf{q}$ of the virtual photon can be written as
\begin {eqnarray}
-\mathbf{q}^2=-\mathbf{p}_i^2-\mathbf{p}_f^2-\left(\frac{\hbar}{c}\omega\right)^2
-2\frac{\hbar}{c}|\mathbf{p}_f|\cos\Theta_f\pm 2\frac{\hbar}{c}\omega
|\mathbf{p}_i|\pm 2|\mathbf{p}_i||\mathbf{p}_f|\cos\Theta_f. \nonumber\\
\label{small.2}
\end {eqnarray}
  Here   the upper sign   corresponds
  to $\Theta_i=0$ and the lower one to
$\Theta_i=\pi$.\\ \indent
As (\ref{small.1}) and (\ref{small.2}) do not depend on $\Phi$ at all, the
$\Phi$ integration simply gives a factor of $2\pi$,   and
  (\ref{small.1})   becomes
\begin {eqnarray}
\frac{d^3\sigma}{d\omega d\Omega_i d\Omega_f} &=&
\frac{Z^2\alpha_{fine}^3\hbar^2}{2\pi}\frac{|\mathbf{p}_f|}{|\mathbf{p}_i|}
\frac{1}{\omega}\frac{1}{|\mathbf{q}|^4}
\nonumber \\
&\times&\left(
\frac{\mathbf{p}_f^2\sin^2\Theta_f}{(E_f-c|\mathbf{p}_f|\cos\Theta_f)^2}\left
(4E_i^2-c^2\mathbf{q}^2\right) \right. \nonumber\\
&+&2\left.
\hbar^2\omega^2\frac{\mathbf{p}_f^2\sin^2\Theta_f}{(E_f-c|\mathbf{p}_f|\cos\Theta_f)(E_i\mp
c|\mathbf{p}_i|\cos\Theta_i)}
\right). \label{small.3}
\end {eqnarray}
Finally this expression has to be integrated over $\Theta_f$ in order to
obtain the doubly differential cross section. Similarly to the total
integration of (\ref{theta.1}) it is convenient to define
\begin {eqnarray}
\tilde{\Delta}_1&:=& -\left(p_i\mp\frac{\hbar}{c}\omega\right)^2-p_f^2, \label{small.4}\\
\tilde{\Delta}_2&:=& -2\frac{\hbar}{c}\omega p_f\pm2p_ip_f\label{small.5}
\end {eqnarray}
where   $\tilde{\Delta}_{1,2}=\Delta_{1,2}(\Theta_i=0,\pi),j\in\{1,2\}$ , with
definitions (\ref{theta.5}) and (\ref{theta.6}).   Eq.   (\ref{small.2}) can   then   be rewritten as
\begin {eqnarray}
-\mathbf{q}^2=\tilde{\Delta}_1+\tilde{\Delta}_2\cos\Theta_f \label{small.6}
\end {eqnarray}
and
\begin {eqnarray}
\frac{d^2\sigma}{d\omega d\Omega_i
}&=&\frac{Z^2\alpha_{fine}^3\hbar^2}{2\pi}\frac
{|\mathbf{p}_f|}{|\mathbf{p}_i|}\frac{1}{\omega}\int\limits_{0}^{\pi}
d\Theta_f \left[
\frac{|\mathbf{p}_f|^2c^2\sin^2\Theta_f
}{(E_f-|\mathbf{p}_f|c\cos\Theta_f)^2(\tilde{\Delta}_1+\tilde{\Delta}_2\cos\Theta_f)}
\right. \nonumber\\
&+&\frac{4E_i^2
|\mathbf{p}_f|^2\sin^2\Theta_f}{(E_f-|\mathbf{p}_f|c\cos\Theta_f)^2(\tilde{\Delta}_1+\tilde{\Delta}_2\cos\Theta_f)^2}
\nonumber\\
&+&\left.\frac{2\hbar^2\omega^2|\mathbf{p}_f|^2\sin^2\Theta_f}{(E_i\mp
c|\mathbf{p}_i|)(E_f-|\mathbf{p}_f|c\cos\Theta_f)(\tilde{\Delta}_1+\tilde{\Delta}_2\cos\Theta_f)^2}\right]
\sin\Theta_f  \nonumber \\ \label{small.7}
\end {eqnarray}
where the integration is rather elementary and can be performed by
substituting $x=\cos\Theta_f$ again. Thus (\ref{small.7}) yields
\begin {eqnarray}
&\ & \frac{d^2\sigma}{d\omega d\Omega_i
}(E_i,\omega,\Theta_i=0,\pi)=\frac{Z^2\alpha^3\hbar^2}{2\pi}\frac
{|\mathbf{p}_f|}{|\mathbf{p}_i|}\frac{1}{\omega} \left[-\frac{2|\mathbf{p}_f|c}{\tilde{\Delta}_2
E_f+\tilde{\Delta}_1 |\mathbf{p}_f| c}\right. \nonumber\\
&+&\frac{|\mathbf{p}_f|^2c^2(-\tilde{\Delta}_1^2+\tilde{\Delta}_2^2)}{\tilde{\Delta}_2(\tilde{\Delta}_2E_f+\tilde{\Delta}_1
|\mathbf{p}_f|
c)^2}\ln\left(\frac{\tilde{\Delta}_1+\tilde{\Delta}_2}{\tilde{\Delta}_1-\tilde{\Delta}_2}\right) \nonumber\\
&+&\frac{2\tilde{\Delta}_1E_f|\mathbf{p}_f|c+\tilde{\Delta}_2(E_f^2+|\mathbf{p}_f|^2c^2)}{(\tilde{\Delta}_2 E_f+\tilde{\Delta}_1
|\mathbf{p}_f|c)^2}\ln\left(\frac{E_f+|\mathbf{p}_f|c}{E_f-|\mathbf{p}_f|c}\right) \nonumber\\
&-&\frac{16E_i^2|\mathbf{p}_f|^2}{(\tilde{\Delta}_2E_f+\tilde{\Delta}_1|\mathbf{p}_f|
c)^2}-\frac{4\hbar^2 |\mathbf{p}_f|^2\omega^2}{(\tilde{\Delta}_2 E_f+\tilde{\Delta}_1
|\mathbf{p}_f| c)(E_i\mp
c|\mathbf{p}_i|)\tilde{\Delta}_2} \nonumber\\
&-&\frac{8E_i^2|\mathbf{p}_f|^2
(\tilde{\Delta}_1 E_f+\tilde{\Delta}_2 |\mathbf{p}_f| c)}{(\tilde{\Delta}_2 E_f+\tilde{\Delta}_1
|\mathbf{p}_f| c)^3}\ln\left(
\frac{(\tilde{\Delta}_1-\tilde{\Delta}_2)(E_f-|\mathbf{p}_f|c)}{(\tilde{\Delta}_1+\tilde{\Delta}_2)(E_f+|\mathbf{p}_f|c)}\right)
\nonumber\\
&+&\frac{2\hbar^2|\mathbf{p}_f|^2\omega^2(2\tilde{\Delta}_1\tilde{\Delta}_2E_f
+\tilde{\Delta}_1^2|\mathbf{p}_f|c+\tilde{\Delta}_2^2|\mathbf{p}_f| c)}{(\tilde{\Delta}_2E_f+\tilde{\Delta}_1
|\mathbf{p}_f| c)^2(E_i\mp
c|\mathbf{p}_i|)\tilde{\Delta}_2^2}\ln\left(\frac{\tilde{\Delta}_1+\tilde{\Delta}_2}{\tilde{\Delta}_1-\tilde{\Delta}_2}\right)
\nonumber\\
&+&\left. \frac{2\hbar^2|\mathbf{p}_f|\omega^2(E_f^2-c^2|\mathbf{p}_f|^2)}{(\tilde{\Delta}_2
E_f+\tilde{\Delta}_1
|\mathbf{p}_f| c)^2(E_i\mp
c|\mathbf{p}_i|)c}\ln\left(\frac{E_f-|\mathbf{p}_f|c}{E_f+|\mathbf{p}_f|c}\right)\right]. \label{small.8}
\end {eqnarray}
This expression is much simpler than (\ref{theta.16}), but only valid for
$\Theta_i=0$ or $\Theta_i=\pi$. Actually this expression has also been obtained by
calculating the limit $\Theta_i\rightarrow 0$ or $\Theta_i\rightarrow\pi$ in
(\ref{theta.16});   hence the consistency check is succesful.
Details can be found in \ref{app_small}.

\subsubsection {$\hbar\omega\rightarrow E_{kin,i}$}

The other case which can be investigated easily is when almost all
kinetic energy of the incident electron is   transferred
  to the emitted photon,   i.e.,
\begin {eqnarray}
E_f=E_i-\hbar\omega=E_{kin,i}+m_ec^2-\hbar\omega \xrightarrow{\hbar\omega
\rightarrow E_{kin,i}} m_ec^2 \label{limit.1}
\end {eqnarray}
and
\begin{eqnarray}
|\mathbf{p}_f|=\sqrt{\frac{E_f^2}{c^2}-m_e^2c^2}\xrightarrow{\hbar\omega\rightarrow
E_{kin,i}}
\sqrt{\frac{m_e^2c^4}{c^2}-m_e^2c^2}\equiv 0 \label {limit.2}
\end{eqnarray}
and consequently   from Eq. (\ref{bcs.7})
\begin {eqnarray}
-\mathbf{q}^2\xrightarrow{\hbar\omega\rightarrow
E_{kin,i}} -\mathbf{p}_i^2-\left(\frac{\hbar}{c}\omega\right)^2
+2\frac{\hbar}{c}\omega |\mathbf{p}_i|\cos\Theta_i &=:& \delta \label{limit.3} \\
\Rightarrow \mathbf{q}^4 &\xrightarrow{\hbar\omega\rightarrow
E_{kin,i}}& \delta^2. \label{limit.4}
\end {eqnarray}
With these limits it follows for the triply differential cross section
(\ref{bcs.1})
\begin {eqnarray}
\frac{d^4\sigma}{d\omega d\Omega_i d\Omega_f d\Phi} &\xrightarrow{\hbar\omega\rightarrow
E_{kin,i}}&
\frac{Z^2\alpha_{fine}^3
\hbar^2}{(2\pi)^2} \frac{|\mathbf{p}_f|}{|\mathbf{p}_i|}\frac{1}{\omega}
\frac{1}{\delta^2} \left[ \frac{|\mathbf{p}_i|^2\sin^2\Theta_i}{(E_i-c
|\mathbf{p}_i|\cos
\Theta_i)^2}\right. \nonumber\\
&\times&(4E_f^2+\delta c^2)+2\left. \hbar^2\omega^2
\frac{|\mathbf{p}_i|^2\sin^2\Theta_i}{(E_i-c|\mathbf{p}_i|\cos\Theta_i)E_f}
\right].\nonumber\\ \label{limit.5}
\end {eqnarray}
Actually (\ref{limit.5}) depends neither on $\Phi$,
nor on $\Theta_f$. Therefore
\begin {eqnarray}
\int\limits_0^{2\pi}d\Phi\int\limits_0^{\pi}\sin\Theta_i d\Theta_i=4\pi
\label{limit.6}
\end {eqnarray}
which leads to a very simple expression for the doubly differential cross
section
\begin {eqnarray}
\frac{d^2\sigma}{d\omega d\Omega_i} \xrightarrow{\hbar\omega\rightarrow
E_{kin,i}}&&
\frac{Z^2\alpha^3
\hbar^2}{\pi} \frac{|\mathbf{p}_f|}{|\mathbf{p}_i|}\frac{1}{\omega}
\frac{1}{\delta^2} \left[
\frac{|\mathbf{p}_i|^2\sin^2\Theta_i}{(E_i-c|\mathbf{p}_i|\cos
\Theta_i)^2}(4E_f^2+\delta c^2)\right. \nonumber \\
&+&2\left. \hbar^2\omega^2
\frac{|\mathbf{p}_i|^2\sin^2\Theta_i}{(E_i-c|\mathbf{p}_i|\cos\Theta_i)E_f}
\right]. \label{limit.7}
\end {eqnarray}
Although taking the limit $\hbar\omega\rightarrow E_{kin,i}$
contradicts Eq. (\ref{born.2}) as the   energy of the emitted
  electron should   be larger than $1$ keV   (\ref{born.2}),
(\ref{limit.7})   can be used   for two purposes.\\ \indent
  As   (\ref{limit.7}) can be obtained,
as well,   by taking the limit $|\mathbf{p}_f|\rightarrow 0$ in
(\ref{theta.16}), the complicated   expression (\ref{theta.16})
is checked for consistency analytically.
For further details the reader   is   referred to
\ref{app_large}. Furthermore we will see in section
\ref{max_scat.sec} that the most probable scattering angle does not depend
on the photon energy for $E_{kin,i}\ge 1$ MeV. Therefore this cross section
can be used for calculating the most probable scattering
angle in this energy range.  

\newpage

\section {Pair production} \label{pair}
  Pairs of electrons and positrons can be produced
if a photon interacts with the   nucleus   of an atom.
This process   is related by some symmetry to   the production of
Bremsstrahlung photons.    Bremsstrahlung occurs when an electron is affected
by the nucleus of an atom, scattered and then emits a photon. So there are
three real particles involved: incident electron, scattered electron and
emitted photon. As the photon has no antiparticle one can change the time
direction of the photon. For antimatter it is well known that antiparticles
can be interpreted as the corresponding particles moving back in time. So
one can substitute the incident electron by an positron moving forward in
time. Thus by substituting emitted photon by incident photon and incident
electron by emitted positron (due to time reversal and changing its charge)
it is possible to describe pair production from Bremsstrahlung. Thus
  the emitted photon in the Bremsstrahlung process has to
be   substituted   by the incident photon from the nucleus and the incident electron by
the produced positron. With these two replacements one gets
the differential cross section for pair production \cite{heitler,QED}
\begin {eqnarray}
d^4\sigma &=&
\frac{Z^2\alpha_{fine}^3c^2}{(2\pi)^2\hbar}|\mathbf{p}_+||\mathbf{p}_-|
\frac{dE_+}{\omega^3}\frac{d\Omega_+ d\Omega_- d\Phi}{|\mathbf{q}|^4}\times
\nonumber \\
&\times&\left[-
\frac{\mathbf{p}_-^2\sin^2\Theta_-}{(E_--c|\mathbf{p}_-|\cos\Theta_-)^2}\left
(4E_+^2-c^2\mathbf{q}^2\right)\right. \nonumber\\
&-&\frac{\mathbf{p}_+^2\sin^2\Theta_+}{(E_+-c|\mathbf{p}_+|\cos\Theta_+)^2}\left
(4E_-^2-c^2\mathbf{q}^2\right) \nonumber \\
&+&2\hbar^2\omega^2\frac{\mathbf{p}_+^2\sin^2\Theta_++\mathbf{p}_-^2\sin^2\Theta_-}{(E_+-c|\mathbf{p}_+|\cos\Theta_+)(E_--c|\mathbf{p}_-|\cos\Theta_-)}
\nonumber\\
&+&2\left.\frac{|\mathbf{p}_+||\mathbf{p}_-|\sin\Theta_+\sin\Theta_-\cos\Phi}{(E_+-c|\mathbf{p}_+|\cos\Theta_+)(E_--c|\mathbf{p}_-|\cos\Theta_-)}\left(2E_+^2+2E_-^2-c^2\mathbf{q}^2\right)\right],
\nonumber \\ \label{pair.1}
\end {eqnarray}
where $Z$, $\alpha_{fine}$, $h$, $\hbar$ and $c$ are the same  
parameters as in Eq.   (\ref{bcs.1}). $\omega$ is the frequency of the incident
photon, $E_{\pm}$   and $p_{\pm}$ are
the total energy   and the momentum   of the positron/electron
  with
\begin {eqnarray}
E_{\pm}=\sqrt{\mathbf{p}_{\pm}^2c^2+m_e^2c^4}. \label{pair.2}
\end {eqnarray}
Similarly to (\ref{bcs.1}) there are three angles, $\Theta_{\pm}$ between
the direction of the photon and the
positron/electron direction,   $\Theta_+=\sphericalangle
(\mathbf{p}_+,\mathbf{k}), \Theta_-=\sphericalangle
(\mathbf{p}_-,\mathbf{k})$,   and $\Phi$ is the angle between the scattering
  planes $(\mathbf{p}_+,\mathbf{k})$ and
$(\mathbf{p}_-,\mathbf{k})$.
The absolute value of the momentum of the virtual photon is
\begin {eqnarray}
-\mathbf{q}^2&=&-|\mathbf{p}_+|^2-|\mathbf{p}_-|^2-\left(\frac{\hbar}{c}\omega\right)^2+2|\mathbf{p}_+|\frac{\hbar}{c}
\omega\cos\Theta_+ +2|\mathbf{p}_-|\frac{\hbar}{c} \omega\cos\Theta_-
\nonumber\\
&-&2|\mathbf{p}_+||\mathbf{p}_-|(\cos\Theta_+\cos\Theta_-+\sin\Theta_+\sin\Theta_-\cos\Phi).
\label {pair.q}
\end {eqnarray}
Algebraically one obtains (\ref{pair.1}) from (\ref{bcs.1}) by replacing
\begin {eqnarray}
E_f &\rightarrow& E_-, \label{pair.8}\\
E_i &\rightarrow& -E_+, \label{pair.9}\\
\mathbf{p}_i &\rightarrow& -\mathbf{p}_+, \label{pair.10}\\
\mathbf{p}_f &\rightarrow& \mathbf{p}_-, \label{pair.11}\\
\omega &\rightarrow& -\omega,  \label{pair.12}\\
\Theta_i &\rightarrow& \pi-\Theta_+,  \label{pair.13}\\
\Theta_f &\rightarrow& \Theta_-,  \label{pair.14}\\
\Phi &\rightarrow& \Phi-\pi  \label{pair.15}
\end {eqnarray}
where the quantities on the left hand side are   for   Bremsstrahlung,
  and those   on the right hand side   for
  pair production. At the end one has to
multiply with an additional factor to get the correct prefactor. With all
the mentioned substitutions it is
\begin {eqnarray}
d^4\sigma_{brems} &\leftarrow&
\frac{\hbar^3\omega^2}{|\mathbf{p}_+|^2c^2}\frac{d\omega}{dE_+}
d^4\sigma_{pair}, \label{pair.16}
\end {eqnarray}
Because of this symmetry   the results for pair production follow
easily from those for Bremsstrahlung.  \\ \indent
The direction of the positron   relative to the incident photon
is given by
integrating (\ref{pair.1}) over $\Phi$ and $\Theta_-$.   But this
  is the same
exercise as to integrate (\ref{bcs.1}) over $\Phi$ and $\Theta_f$. Because
of the symmetry between Bremsstrahlung and pair production one can take
(\ref{theta.16})   and substitute   (\ref{pair.8}) - (\ref{pair.15})
  to obtain   a doubly
differential cross section
\begin {eqnarray}
\frac{d^2\sigma (E_+,\omega,\Theta_+)}{dE_+d\Omega_+} =
\sum\limits_{j=1}^{6} I_j \label{pair.18}
\end {eqnarray}
with the following contributions
\tiny
\begin {eqnarray}
I_1&=&\frac{2\pi A}{\sqrt{(\Delta^{(p)}_2)^2+4p_+^2p_-^2\sin^2\Theta_i}}
\nonumber \\ &\times&
\ln\left(\frac{(\Delta^{(p)}_2)^2+4p_+^2p_-^2\sin^2\Theta_i-\sqrt{(\Delta^{(p)}_2)^2+4p_+^2p_-^2\sin^2
\Theta_i}(\Delta^{(p)}_1+\Delta^{(p)}_2)+\Delta^{(p)}_1\Delta^{(p)}_2}{-(\Delta^{(p)}_2)
^2-4p_+^2p_-^2\sin^2\Theta_i
-\sqrt{(\Delta^{(p)}_2)^2+4p_+^2p_-^2\sin^2 \Theta_i}(\Delta^{(p)}_1-\Delta^{(p)}_2)+\Delta^{(p)}_1\Delta^{(p)}_2
}\right) \nonumber\\
&\times&\left[-1-\frac{c\Delta^{(p)}_2}{p_-(E_+-cp_+\cos\Theta_i)}+\frac{p_+^2c^2\sin^2\Theta_i}
{(E_+-cp_+\cos\Theta_i)^2}
-\frac{2\hbar^2\omega^2p_-\Delta^{(p)}_2}{c(E_+-cp_+\cos
\Theta_i)((\Delta^{(p)}_2)^2+4p_+^2p_-^2\sin^2\Theta_i)}\right], \nonumber\\ \label{pair.19}\\
I_2&=&\frac{2\pi Ac}{p_-(E_+-cp_+\cos\Theta_i)}\ln\left(
\frac{E_-+p_-c}{E_--p_-c}\right), \label{pair.20} \\
I_3&=&\frac{2\pi A}{\sqrt{(\Delta^{(p)}_2E_-+\Delta^{(p)}_1p_-c)^2+4m^2c^4p_+^2p_-^2\sin^2\Theta_i
}} \nonumber \\
&\times&\ln\Bigg(\Big((E_-+p_-c)(4p_+^2p_-^2\sin^2\Theta_i(E_--p_-c)+(\Delta^{(p)}_1+\Delta^{(p)}_2)
((\Delta^{(p)}_2E_-+\Delta^{(p)}_1p_-c) \nonumber\\
&-&\sqrt{(\Delta^{(p)}_2E_-+\Delta^{(p)}_1p_-c)^2+4m^2c^4p_+^2p_-^2\sin^2\Theta_i}))\Big)\Big((E_--p_-c)
(4p_+^2p_-^2\sin^2\Theta_i(-E_--p_-c) \nonumber\\
&+&(\Delta^{(p)}_1-\Delta^{(p)}_2)
((\Delta^{(p)}_2E_-+\Delta^{(p)}_1p_-c)-\sqrt{(\Delta^{(p)}_2E_-+\Delta^{(p)}_1p_-c)^2+4m^2c^4p_+^2p_-^2\sin^2\Theta_i}))\Big)^{-1}\Bigg)  \nonumber\\
&\times&\left[\frac{c(\Delta^{(p)}_2E_-+\Delta^{(p)}_1p_-c)}{p_-(E_+-cp_+\cos\Theta_i)}\right.\nonumber\\
&+&\Big[((\Delta^{(p)}_2)^2+4p_+^2p_-^2\sin^2\Theta_i)(E_-^3+E_-p_-c)+p_-c(2
((\Delta^{(p)}_1)^2-4p_+^2p_-^2\sin^2\Theta_i)E_-p_-c \nonumber\\
&+&\Delta^{(p)}_1\Delta^{(p)}_2(3E_-^2+p_-^2c^2))\Big]\Big[(\Delta^{(p)}_2E_-+\Delta^{(p)}_1p_-c)^2+4m^2c^4p_+^2p_-^2\sin^2\Theta_i\Big]^{-1}
\nonumber\\
&+&\Big[-8p_+^2p_-^2m^2c^4\sin^2\Theta_i(E_+^2+E_-^2)-2\hbar^2\omega^2p_+^2\sin^2\Theta_ip_-c(\Delta^{(p)}_2E_-+\Delta^{(p)}_1p_-c)
\nonumber\\
&+&2\hbar^2\omega^2p_- m^2c^3(\Delta^{(p)}_2E_-+\Delta^{(p)}_1p_-c)\Big]
\Big[(E_+-cp_+\cos\Theta_i)((\Delta^{(p)}_2E_-+\Delta^{(p)}_1p_-c)^2+4m^2c^4p_+^2p_-^2\sin^2\Theta_i)\Big]^{-1} \nonumber\\
&+&\left.\frac{4E_+^2p_-^2(2(\Delta^{(p)}_2E_-+\Delta^{(p)}_1p_-c)^2-4m^2c^4p_+^2p_-^2\sin^2\Theta_i)(\Delta^{(p)}_1E_-+\Delta^{(p)}_2p_-c)}{((\Delta^{(p)}_2E_-+\Delta^{(p)}_1p_-c)^2+4m^2c^4p_+^2p_-^2\sin^2\Theta_i)^2}\right],
\label{pair.21}\\
I_4&=&\frac{4\pi Ap_-c(\Delta^{(p)}_2E_-+\Delta^{(p)}_1p_-c)}{(\Delta^{(p)}_2E_-+\Delta^{(p)}_1p_-c)^2+4m^2c^4p_+^2p_-^2\sin^2\Theta_i}
+\frac{16\pi E_+^2p_-^2
A(\Delta^{(p)}_2E_-+\Delta^{(p)}_1p_-c)^2}{((\Delta^{(p)}_2E_-+\Delta^{(p)}_1p_-c)^2+4m^2c^4p_+^2p_-^2\sin^2\Theta_i)^2},
\label{pair.22}
\\
I_5&=&\frac{4\pi A}{(-(\Delta^{(p)}_2)^2+(\Delta^{(p)}_1)^2-4p_+^2p_-^2\sin^2\Theta_i)
((\Delta^{(p)}_2E_-+\Delta^{(p)}_1p_-c)^2+4m^2c^4p_+^2p_-^2\sin^2\Theta_i)} \nonumber\\
&\times&\left[\frac{\hbar^2\omega^2p_-^2}{E_+cp_+\cos\Theta_i}
\Big[E_-[2(\Delta^{(p)}_2)^2((\Delta^{(p)}_2)^2-(\Delta^{(p)}_1)^2)+8p_+^2p_-^2\sin^2\Theta_i((\Delta^{(p)}_2)^2+(\Delta^{(p)}_1)^2)]
\right.\nonumber\\
&+&p_-c[2\Delta^{(p)}_1\Delta^{(p)}_2((\Delta^{(p)}_2)^2-(\Delta^{(p)}_1)^2)-16\Delta^{(p)}_1\Delta^{(p)}_2p_+^2p_-^2\sin^2\Theta_i]\Big]\Big[(\Delta^{(p)}_2)^2+4p_+^2p_-^2\sin^2\Theta_i\Big]^{-1}
\nonumber\\
&+& \frac{2\hbar^2\omega^2 p_{+}^2 \sin^2\Theta_i(2\Delta^{(p)}_1\Delta^{(p)}_2
p_-c+2(\Delta^{(p)}_2)^2E_-+8p_+^2p_-^2\sin^2\Theta_i E_-)}{E_+-cp_+\cos\Theta_i}\nonumber\\
&-&
\Big[2E_+^2p_-^2\{2((\Delta^{(p)}_2)^2-(\Delta^{(p)}_1)^2)(\Delta^{(p)}_2E_-+\Delta^{(p)}_1p_-c)^2
+8p_+^2p_-^2\sin^2\Theta_i[((\Delta^{(p)}_1)^2+(\Delta^{(p)}_2)^2)(E_-^2+p_-^2c^2)
\nonumber\\
&+&4\Delta^{(p)}_1\Delta^{(p)}_2E_-p_-c]\}\Big]\Big[(\Delta^{(p)}_2E_-+\Delta^{(p)}_1p_-c)^2+4m^2c^4p_+^2p_-^2\sin^2\Theta_i\Big]^{-1}\nonumber\\
&-&\left.\frac{8p_+^2p_-^2\sin^2\Theta_i(E_+^2+E_-^2)(\Delta^{(p)}_2p_-c +\Delta^{(p)}_1
E_-)}{E_+-cp_+\cos\Theta_i}\right], \label{pair.23}\\
I_6&=&-\frac{16\pi E_-^2p_+^2\sin^2\Theta_i A}{(E_+-cp_+\cos\Theta_i)^2
(-(\Delta^{(p)}_2)^2+(\Delta^{(p)}_1)^2-4p_+^2p_-^2\sin^2\Theta_+)} \label{pair.24}
\end {eqnarray}
\normalsize
with
\begin {eqnarray}
A=\frac{Z^2\alpha_{fine}^3c^2}{(2\pi)^2\hbar}\frac{|\mathbf{p}_+||\mathbf{p}_-|}
{\omega^3}, \label{pair.25}
\end {eqnarray}
and   with   $\Delta^{(p)}_1,\Delta^{(p)}_2$ defined as
\begin {eqnarray}
\Delta^{(p)}_1&:=&-|\mathbf{p}_+|^2-|\mathbf{p}_-|^2-\left(\frac{\hbar}{c}\omega\right)
+ 2\frac{\hbar}{c}\omega|\mathbf{p}_+|\cos\Theta_+, \label{pair.26}\\
\Delta^{(p)}_2&:=&2\frac{\hbar}{c}\omega|\mathbf{p}_i|-2|\mathbf{p}_+||\mathbf{p}_-|
\cos\Theta_+. \label{pair.27}
\end {eqnarray}

\newpage

\section {Discussion}      \label{disc}

\subsection {Bremsstrahlung}
\subsubsection {Comparison with experiments}
\begin {figure}
\includegraphics [scale=0.56] {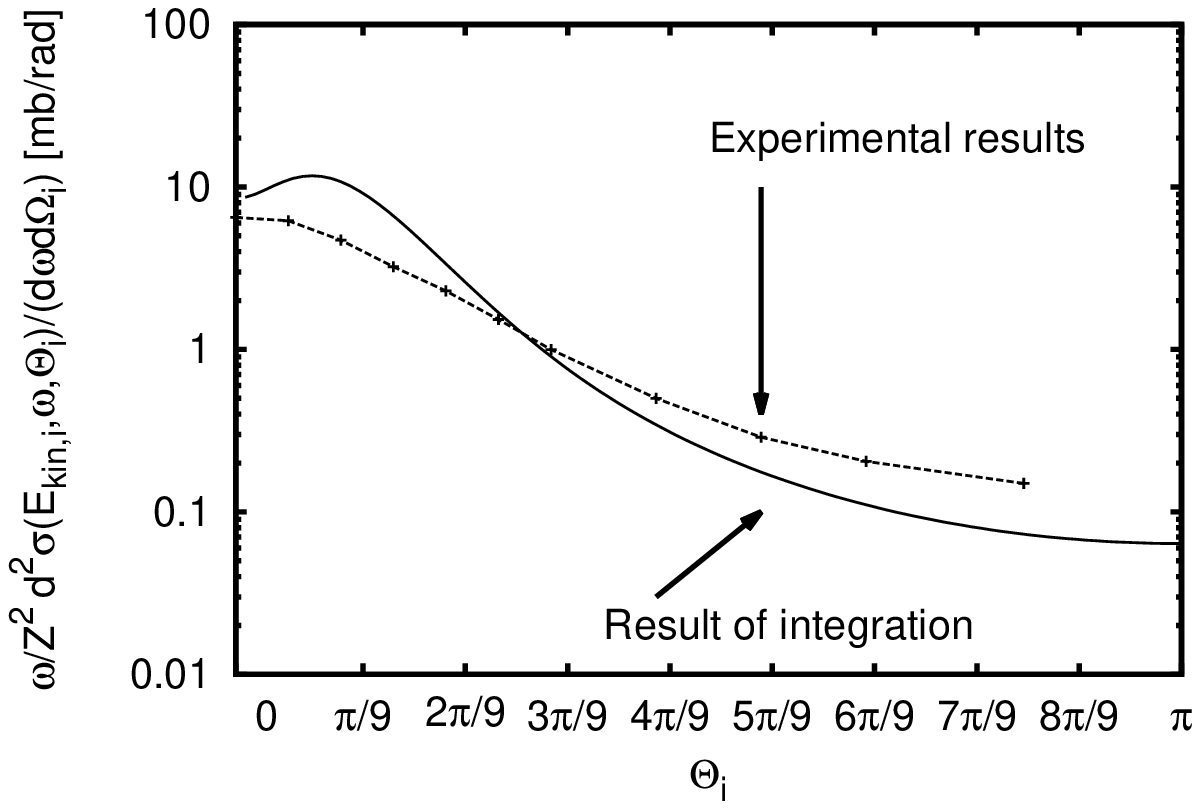}
\includegraphics [scale=0.56] {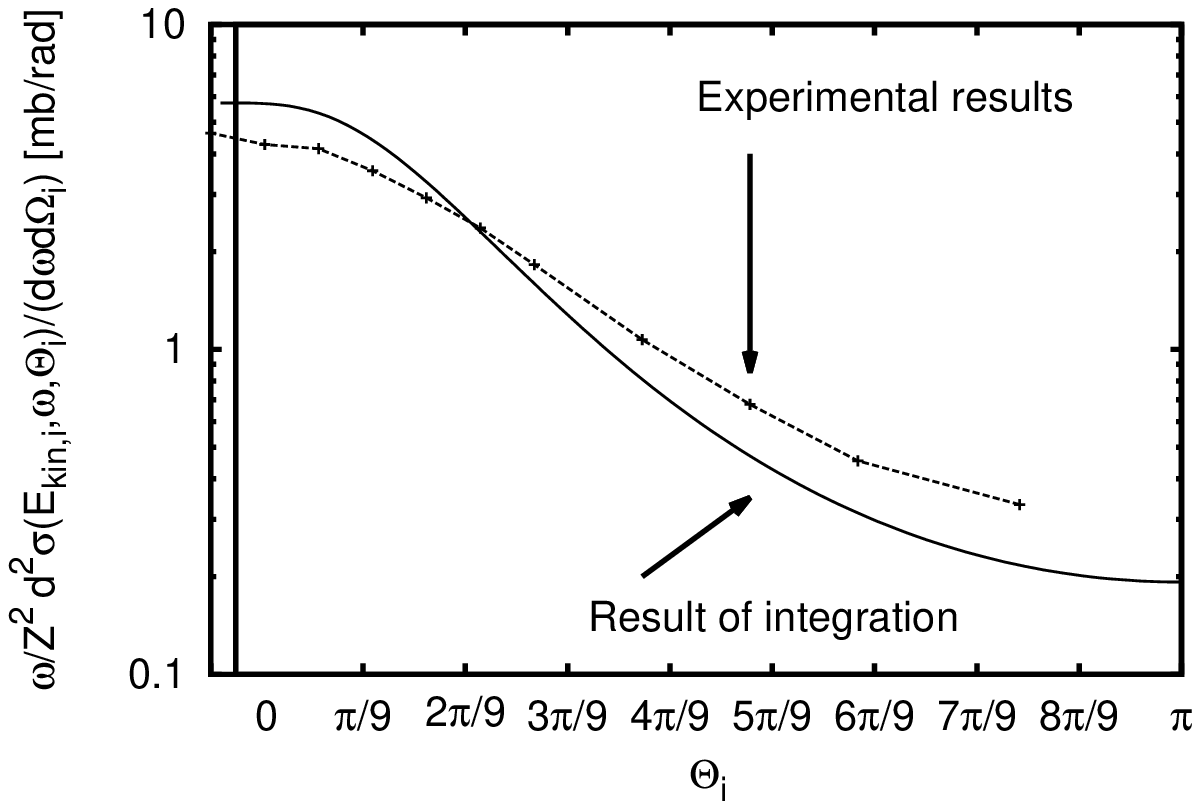}
a) $E_{kin,i}=180$ keV, $\hbar\omega=50$ keV \hspace{1.0cm} b)
$E_{kin,i}=380$ keV, $\hbar\omega=100$ keV\\
\caption {$\omega/Z^2\cdot d^2\sigma/{(d\omega d\Omega_i)}
(E_{kin,i},\omega,\Theta_i)$ for Bremsstrahlung as a function of the scattering angle
$\Theta_i$ between emitted photon and incident electron for gold $Z=79$
where 1 mb = $10^{-31}$ m$^2$. The
energies are a) $E_{kin,i}=180$ keV, $\hbar\omega=50$ keV and b)
$E_{kin,i}=380$ keV, $\hbar\omega=100$ keV. The solid lines shows our result
(\ref{theta.16}); the dotted lines show experimental values (Aiginiger,\
1966).} \label {exp.fig}
\end {figure}

  If electrons are scattered at nuclei, they can produce hard
Brems\-strahlung photons with frequency $\omega$ and direction $\Theta_i$
relative to the direction of the electrons.\\ \indent
Figure \ref{exp.fig} compares our equation (\ref{theta.16}) with experimental results for
gold ($Z=79$) for different electron and photon energies (Aiginger,\ 1966).
For $Z=79$ the minimal electron energy (\ref{born.2}) for the Born
approximation to be valid, is $E_{kin,\{i,f\}}= 115$ keV. Figure \ref{exp.fig}
shows that the cross sections agree overall in size
for $E_{kin,i}=180$ keV, $\hbar\omega=50$ keV and for $E_{kin,i}=380$
keV, $\hbar\omega=100$ keV. However, for the first case, the energy of the
electron in the final state is $E_{kin,f}=130$ keV $\approx 115$ keV, thus
close to the velocity limit. Therefore there is a larger deviation,
especially for small angles, than for the second case where a very good
agreement can be observed.
\subsubsection {Angular distribution of Bremsstrahlung}
\begin {figure}
\includegraphics [scale=0.56] {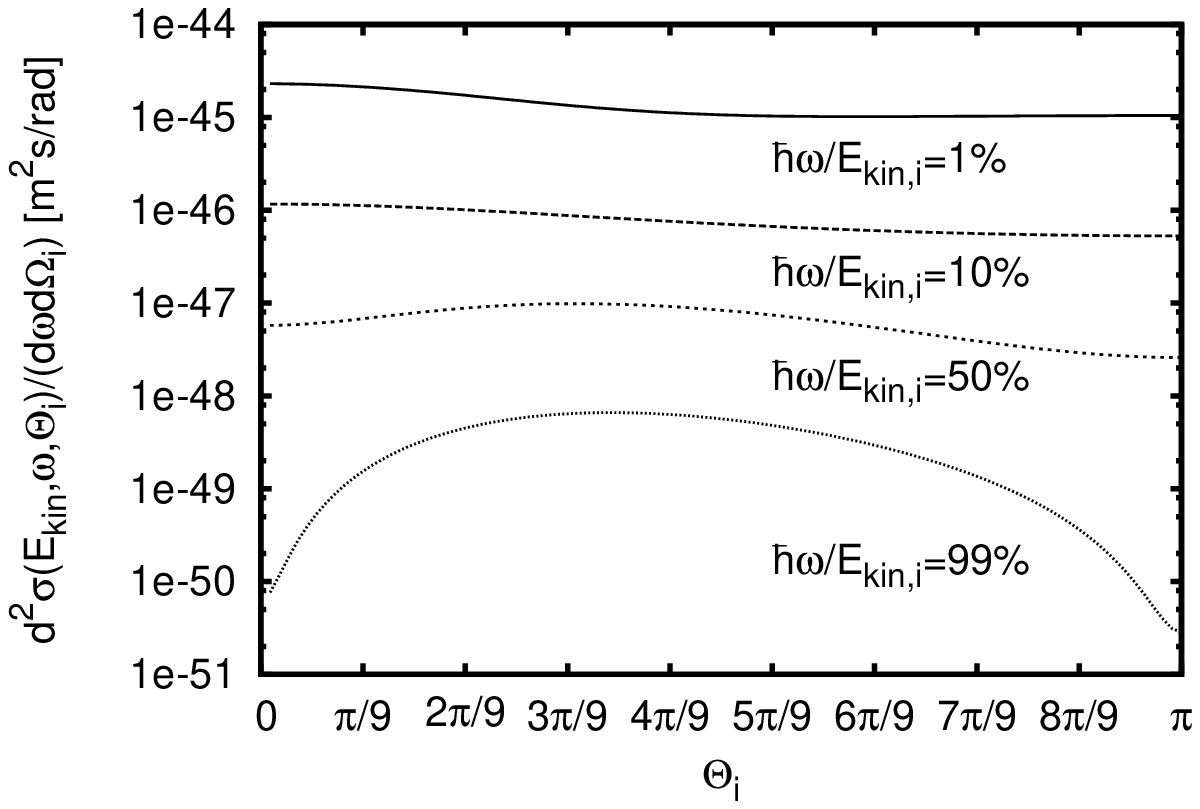}
\includegraphics [scale=0.56] {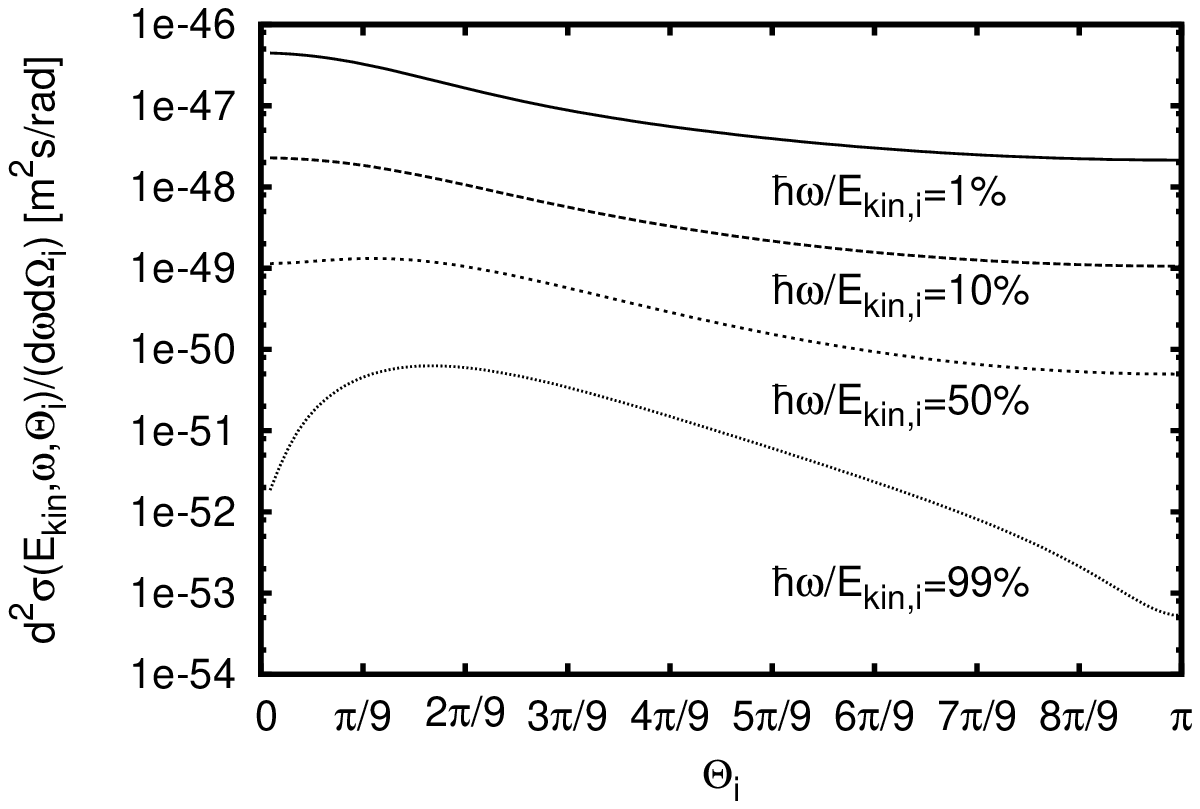}\\
a) $E_{kin,i}=10$ keV \hspace{3.8cm} b) $E_{kin,i}=150$ keV\\
\includegraphics [scale=0.56] {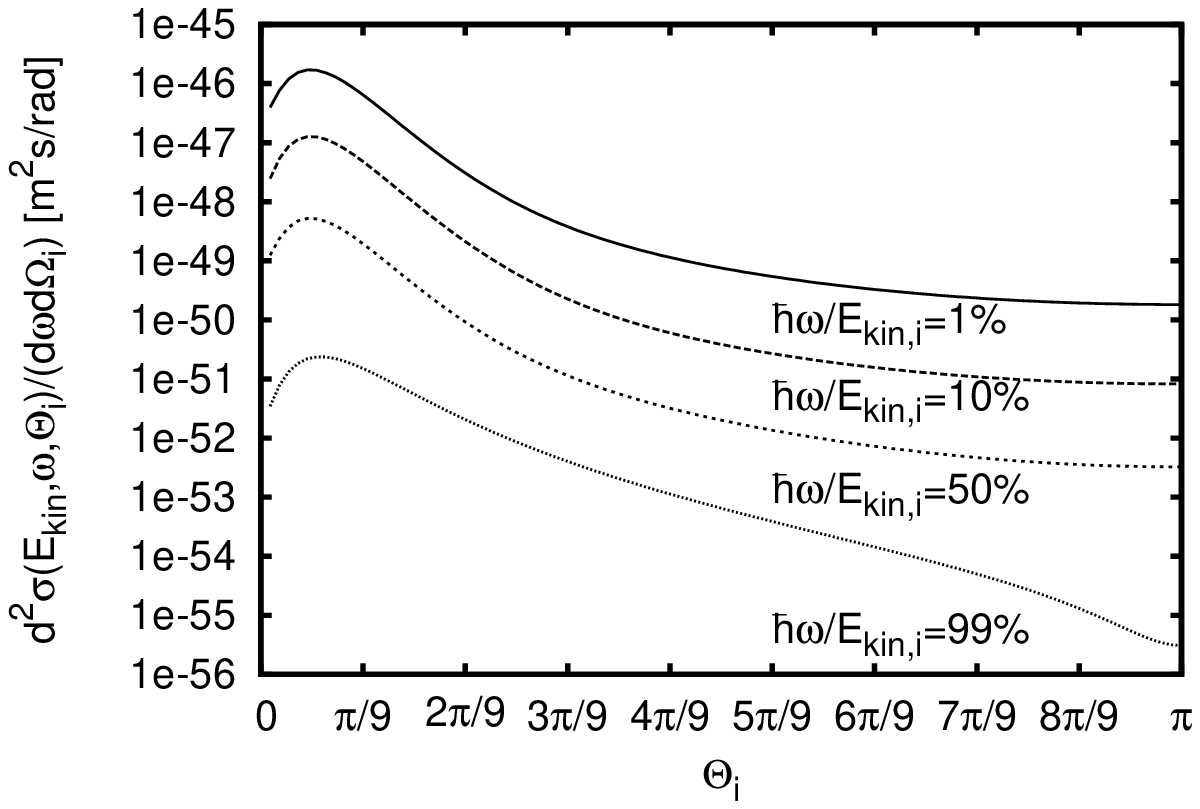}
\includegraphics [scale=0.56] {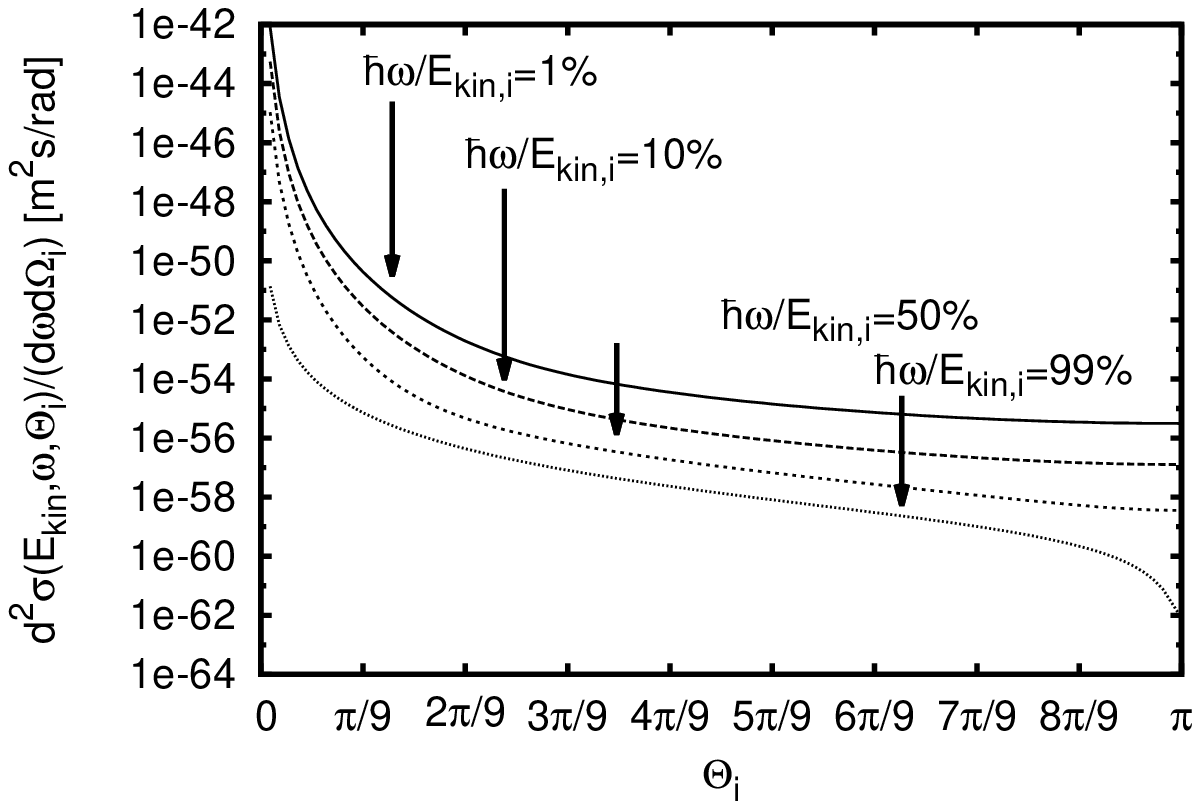}\\
c) $E_{kin,i}=1$ MeV \hspace{3.8cm} d) $E_{kin,i}=100$ MeV\\
\caption {The doubly differential cross section $d^2\sigma/{(d\omega d\Omega_i)}
(E_{kin,i},\omega,\Theta_i)$ for Bremsstrahlung ($Z=7$)   versus the scattering angle
$\Theta_i$ between emitted photon and incident electron. The electron
energies are a) $E_{kin,i}=10$ keV, b) $E_{kin,i}=150$ keV, c) $E_{kin,i}=1$ MeV and d)
$E_{kin,i}=100$ MeV.   In each plot the the photon energy $\hbar\omega$ amounts to $1\%, 10\%, 50\%$ and
$95\%$ of the kinetic energy of the incident electron.
} \label{disc_fig.1}
\end {figure}

Figure \ref{disc_fig.1} shows the doubly differential cross section (\ref{theta.16})
for Brems\-strah\-lung for several electron and photon energies.
At first, the probability for generating photons
decreases with increasing photon energy for fixed electron energy. This can be
understood easily by applying (\ref{limit.7}). As can be seen there, the doubly
differential cross section grows linearly in the momentum of the electron in
the final state which is equivalent to
\begin {eqnarray}
\frac{d^2\sigma}{d\omega d\Omega_i}\sim |\mathbf{p}_f|
\label{disc.1}
\end {eqnarray}
for high photon energies. So, if all kinetic energy is transferred from the electron onto the
photon,   the final momentum $|\mathbf{p}_f|$ vanishes,
  and thus
\begin {eqnarray}
\frac{d^2\sigma}{d\omega d\Omega_i}\rightarrow 0.
\label{disc.2}
\end {eqnarray}
For nonrelativistic electron and photon energies the scattering angle tends
to be mainly equally distributed,\ i.e. the photons do not have a
preference for a particular direction. When the photon energy
increases, photons are  mainly emitted in forward direction, but the ratio between forward and
backward scattering is at least three orders of magnitude
lower than for a
relativistic electron. This case belongs to the classical case where the velocity
is small compared to the speed of light. Namely, it is $v/c|_{E_{kin,i}=10
\textnormal{keV}}\approx 0.20$ and
non-relativistic equations will be enough to describe these phenomena. In the
relativistic case ($v/c|_{E_{kin,i}=1 \textnormal{MeV}}\approx 0.94$ and
$v/c|_{E_{kin,i}=100 \textnormal{MeV}}\approx 0.99999$) the differential cross section
becomes more and more anisotropic. Forward scattering is preferred to
backward scattering although the maximal cross section does not
lie precisely at $\Theta_i=0$ as can be seen in figure \ref{disc_fig.1} c).
But the more the electron energy increases the more the maximum wanders
to smaller angles, for example, it seems in figure
\ref{disc_fig.1} d) that the maximal emission
is indeed for $\Theta_i$=0. As mentioned in section \ref{theta_int} and in
\ref{app_small}, formula (\ref{theta.17})
cannot be evaluated directly at $\Theta_i=0$. However, for this purpose, we derived
(\ref{small.8}) which is valid for $\Theta_i=0$ and $\Theta_i=\pi$.
\begin {figure}
\includegraphics [scale=0.56] {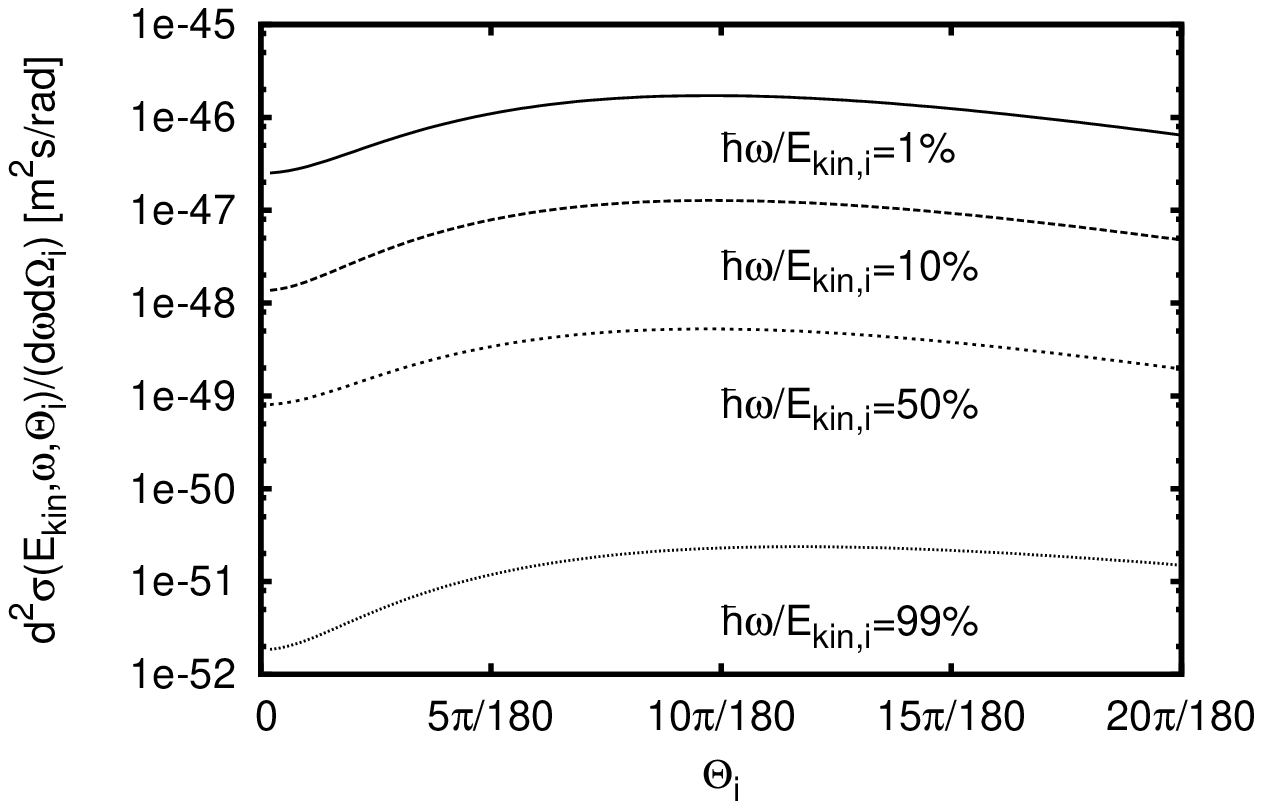}
\includegraphics [scale=0.56] {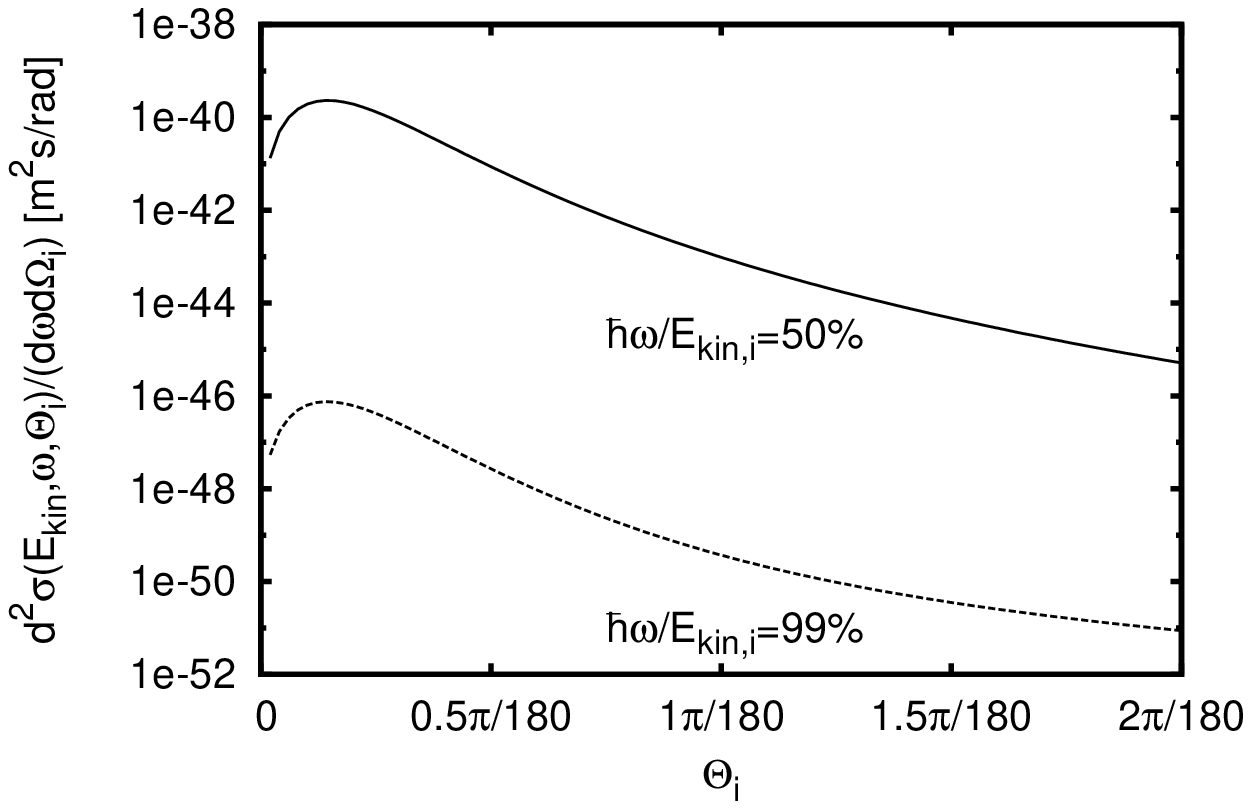}\\
a) $E_{kin,i}=1$ MeV \hspace{3.8cm} b) $E_{kin,i}=100$ MeV
\caption {The doubly differential cross section $d^2\sigma/(d\omega
d\Omega_i)$ as in figure \ref{disc_fig.1} for a smaller angular range} \label{disc_fig.2}
\end {figure}
\begin {figure}
\includegraphics [scale=0.56] {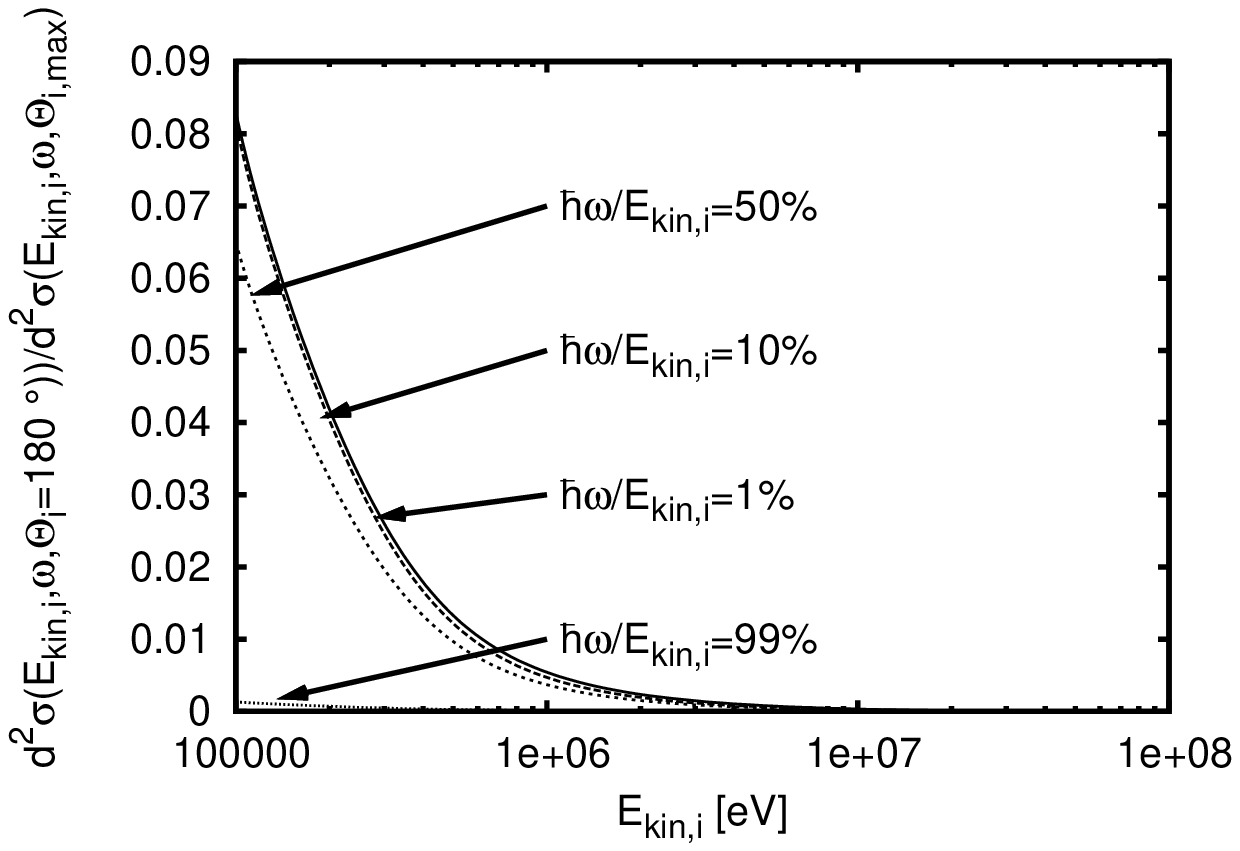}
\includegraphics [scale=0.56] {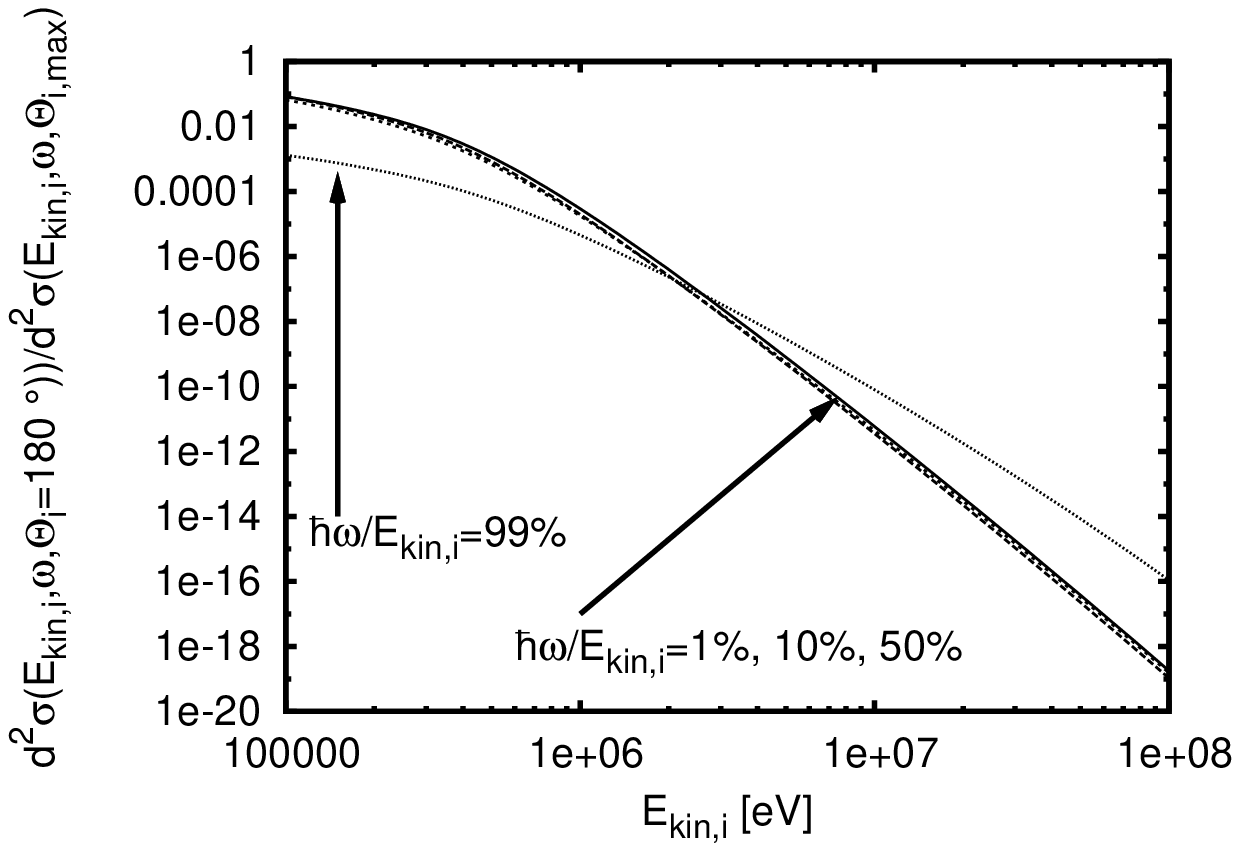}\\
a) \hspace {6.5cm} b)
\caption {The ratio between   the doubly differential cross section for backward
scattering ($\Theta_i=180^{\circ}$)   $d^2\sigma(E_{kin,i},\omega,\Theta_i=180^{\circ})/
(d\omega d\Omega_i)$
and   the maximum of this cross section   $\max\left(d^2\sigma(E_{kin,i},\omega,\Theta_i)/(d\omega d\Omega_i)\right)$
  vs.   the kinetic energy of the incident electron for different
ratios between photon and electron energies   in a a) linear and b)
logarithmic scale for $Z=7$  .} \label{disc.ratio}
\end {figure}
Figure \ref{disc_fig.2} shows again (\ref{theta.16}) for two relativistic electron
and different photon energies but for a smaller range of angles. It shows in
more detail that the angle of maximal scattering is small, but not 0.\\ \indent
Figure \ref{disc.ratio} shows the ratio between the cross section for
backward scattering and forward scattering. It can be clearly seen that the
tendency for backward scattering decreases for increasing electron energy.
The lower the electron energy becomes, the more forward and backward
scattering become similar and in general, the scattering tends to be
isotropic. Only for ratios between photon energies and electron
energies close to 1, forward scattering is preferred for the whole range of energies,
but still decreases with increasing electron energies. \\ \indent
In energetic electron avalanches electrons
scatter frequently which leads to a large velocity dispersion. It depends
on the direction of the applied electric field whether electrons move forward or whether their
directions are distributed arbitrarily. If so, however, this implies that
photons will not necessarily move in a preferred direction, but in the direction of the
incident electron. Their motion and thus change of direction
depend on photon processes, such as Compton scattering.

\subsubsection {Relativistic transformation}
\begin {figure}
\centering
a) \includegraphics [scale=0.25] {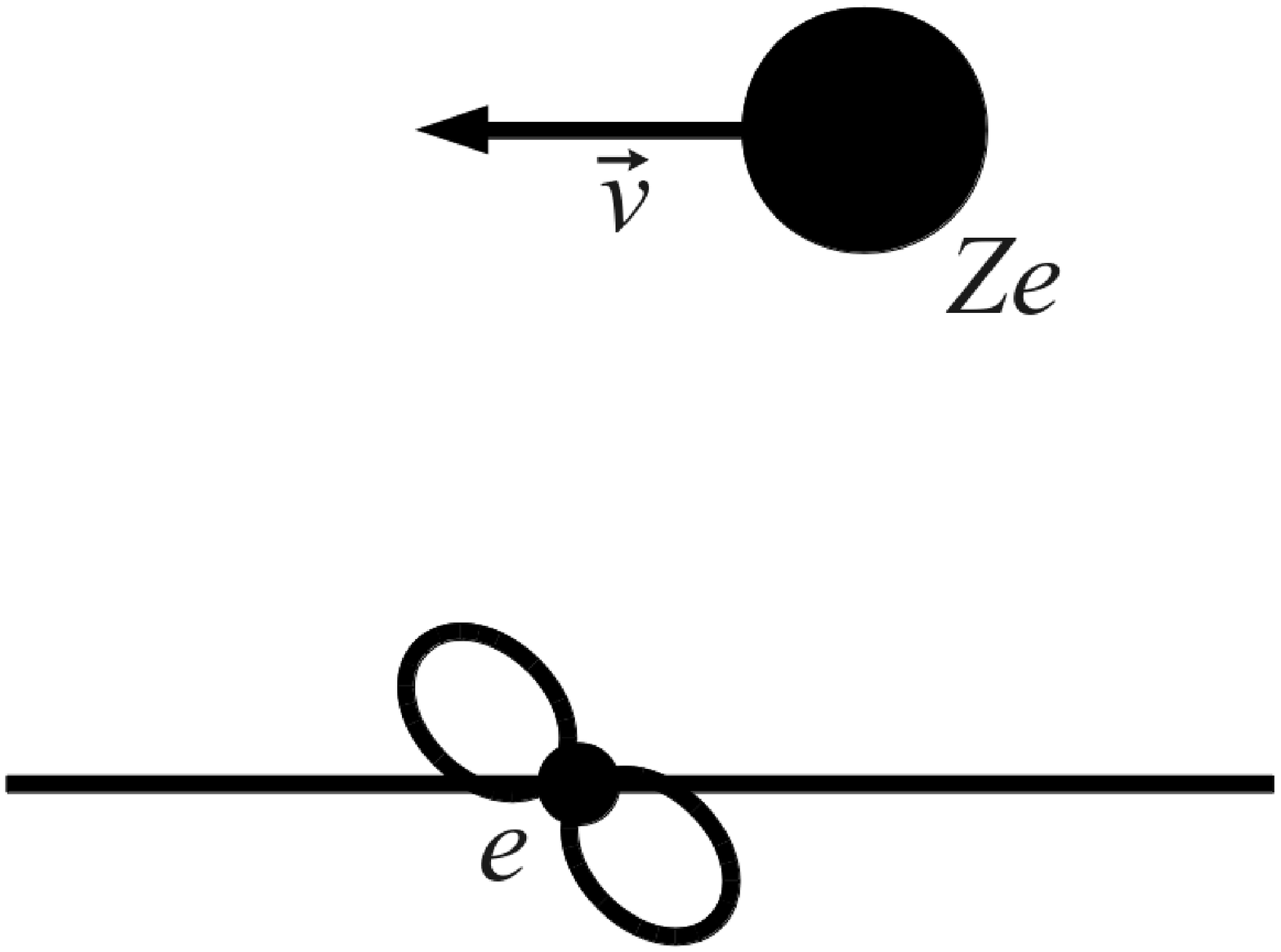}
b) \includegraphics [scale=0.25] {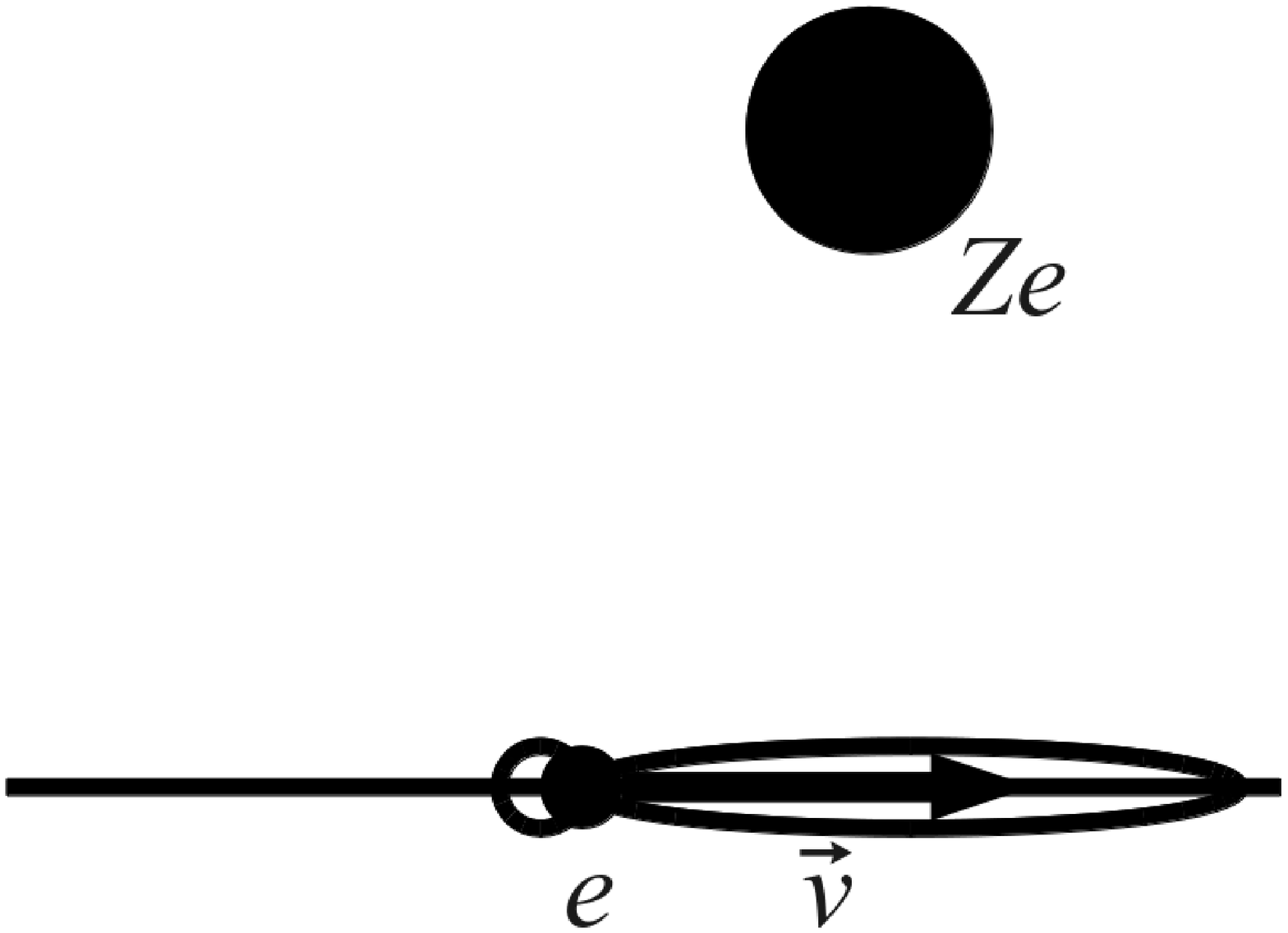}
\caption {a) In the rest frame of the electron where the nucleus is moving
instead radiation is emitted isotropically with a small-angle
deflection. b) If, however, one transforms
the situation into the rest frame of the nucleus where the electron is moving,
most of the radiation is emitted in forward direction relative to the
direction of the electron.} \label{disc_fig.3}
\end {figure}

The tendency of forward scattering in the case of
relativistic incident electrons can be understood by applying the laws
of relativistic transformations. Imagine a non-quantum field theoretical
description of Bremsstrahlung ~\cite{jackson}. If one regards an inertial
system in which the incident particle is at rest
(Fig. \ref{disc_fig.3} a) ), radiation is emitted isotropically with a small-angle
deflection.
If the physical laws for this process are relativistically transformed into
the laboratory system where the nucleus is at rest and the electron moving,
most of the radiation is emitted in forward direction relative to the
electron direction (Fig. \ref{disc_fig.3} b) ). Because this
transformation is valid for a non-quantum field theoretical, relativistic electron, it
must also be true for a relativistic quantum theoretical description,
therefore we see that the forward
scattering of photons can simply be explained as a result of the
relativistic transformation.\\ \indent
\begin {figure}
\includegraphics [scale=0.56] {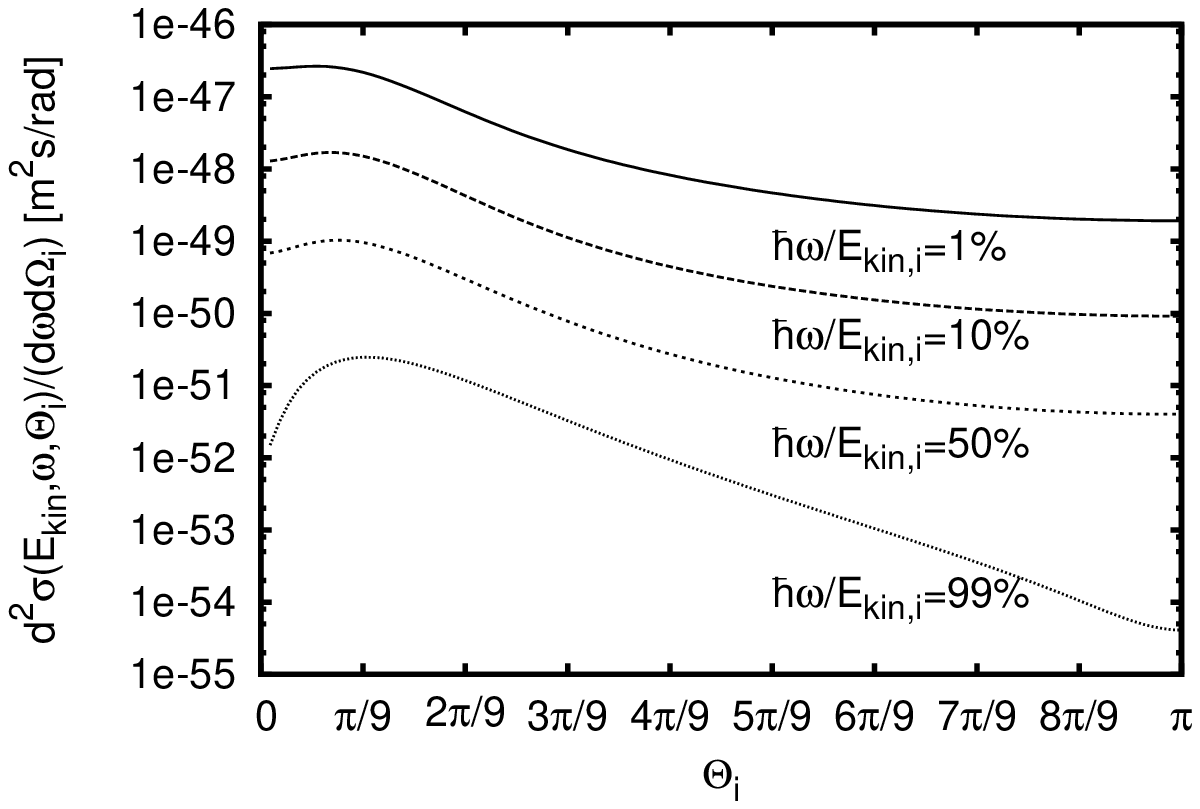}
\includegraphics [scale=0.56] {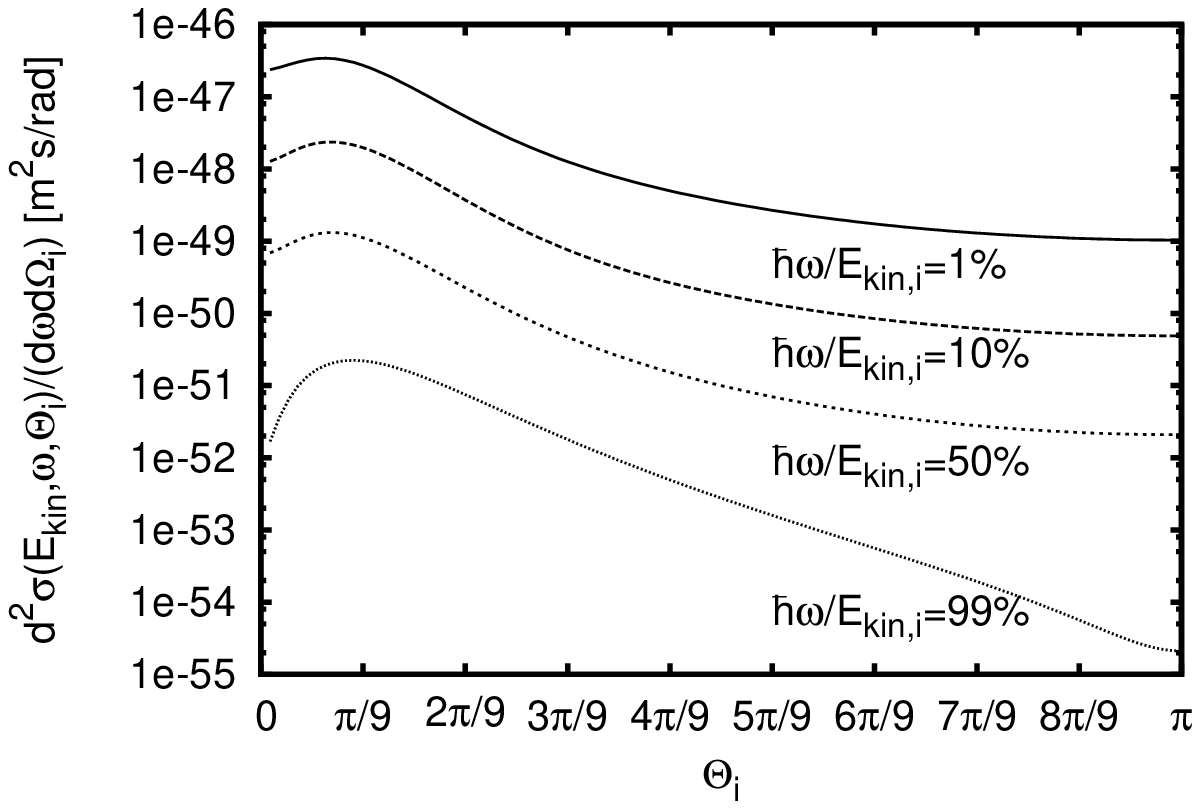}\\
a) $E_{kin,i}=400$ keV \hspace{3.8cm} b) $E_{kin,i}=m_ec^2\approx 511$ keV\\
\caption {  The doubly differential cross section $d^2\sigma/{(d\omega d\Omega_i)}
(E_{kin,i},\omega,\Theta_i)$ for Bremsstrahlung ($Z=7$) versus the scattering angle
$\Theta_i$ between emitted photon and incident electron. The electron
energies are a) $E_{kin,i}=400$ keV, b) $E_{kin,i}\approx 511$ keV.
 In each plot the the photon energy $\hbar\omega$ amounts to $1\%, 10\%, 50\%$ and
$95\%$ of the kinetic energy of the incident electron.
} \label{disc_fig.4}
\end {figure}
The forward scattering can moreover be understood by using the conservation
laws of energy and momentum. They predict that photons have to be
scattered in forward direction if electron and photon energy are high. The
interested reader is referred to \ref{app_cons}.\\ \indent
Although figure \ref{disc_fig.1} shows that the maxima of the doubly
differential cross section form with increasing electron energy,
it is difficult to determine in these plots when these maxima really start to be
generated clearly. Figure \ref{disc_fig.4} shows the doubly differential cross
section in dependence of the incident electron energy for
  $E_{kin,i}=400$ keV and $E_{kin,i}\approx 511$ keV
  when the kinetic energy is
equal to the rest energy. For $400$ keV and for
$\hbar\omega/E_{kin,i}=0.01$ the cross section for forward scattering is
already two orders of magnitude larger than for backward scattering, but a clear maximum cannot be
seen. However, for the same kinetic energy, but for $\hbar\omega/E_{kin,i}=0.95$
there is already a clear maximum formed. But if the kinetic energy
grows up to $511$ keV which is equal to the rest energy of the
electron, there is even a maximum for $\hbar\omega=0.01 E_{kin,i}$.
This can be expected due to the relativistic transformation. If $E_{kin,i}
\ll m_ec^2$, then the photon emission is relatively isotropic. But if the kinetic
energy is approximately equal to the rest energy of the electron, relativistic laws are valid.
Especially for $E_{kin,i}=m_ec^2$
\begin {eqnarray}
\frac{v}{c}=\frac{\sqrt{3}}{2}\approx 87\%, \label{disc.3}
\end {eqnarray}
therefore the electron has to be treated relativistically and clear maxima
close to $\Theta_i=0$ form for
every possible photon energy.

\subsubsection {Dependence on the energy of the emitted photon}

\begin {figure}
\includegraphics [scale=0.56] {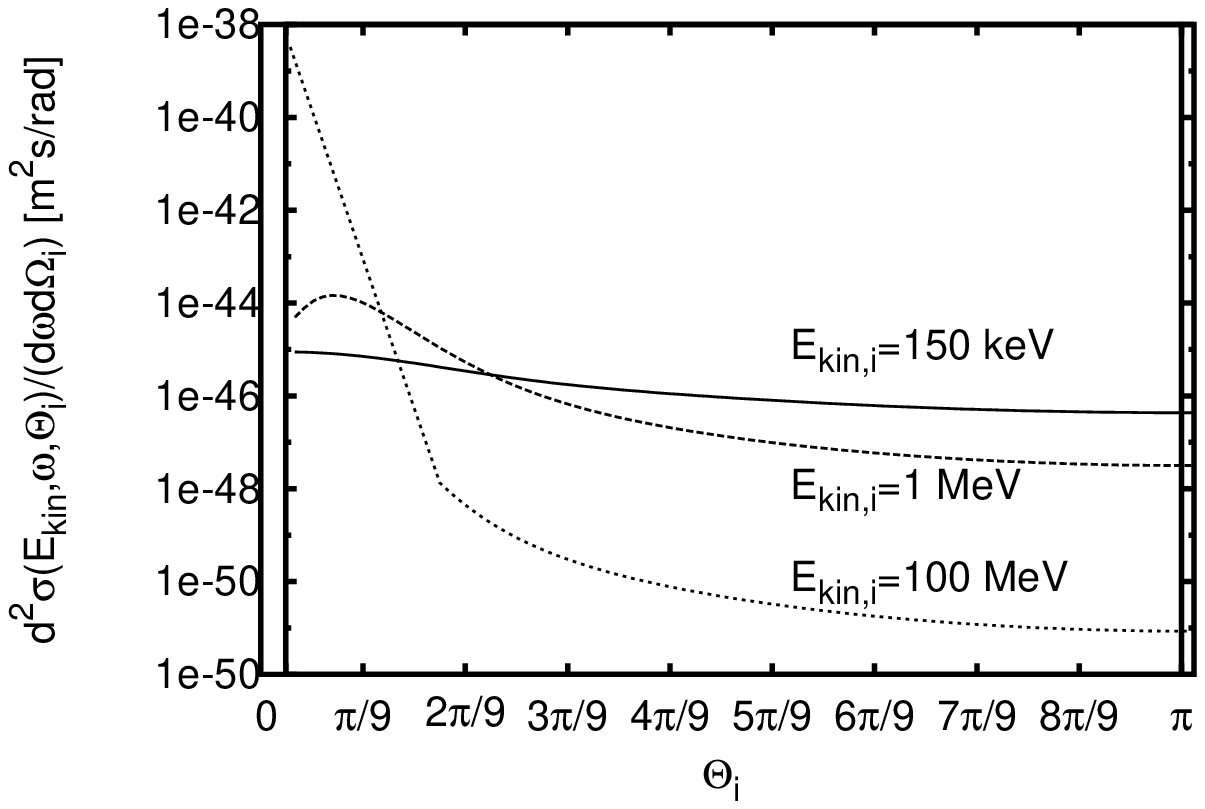}
\includegraphics [scale=0.56] {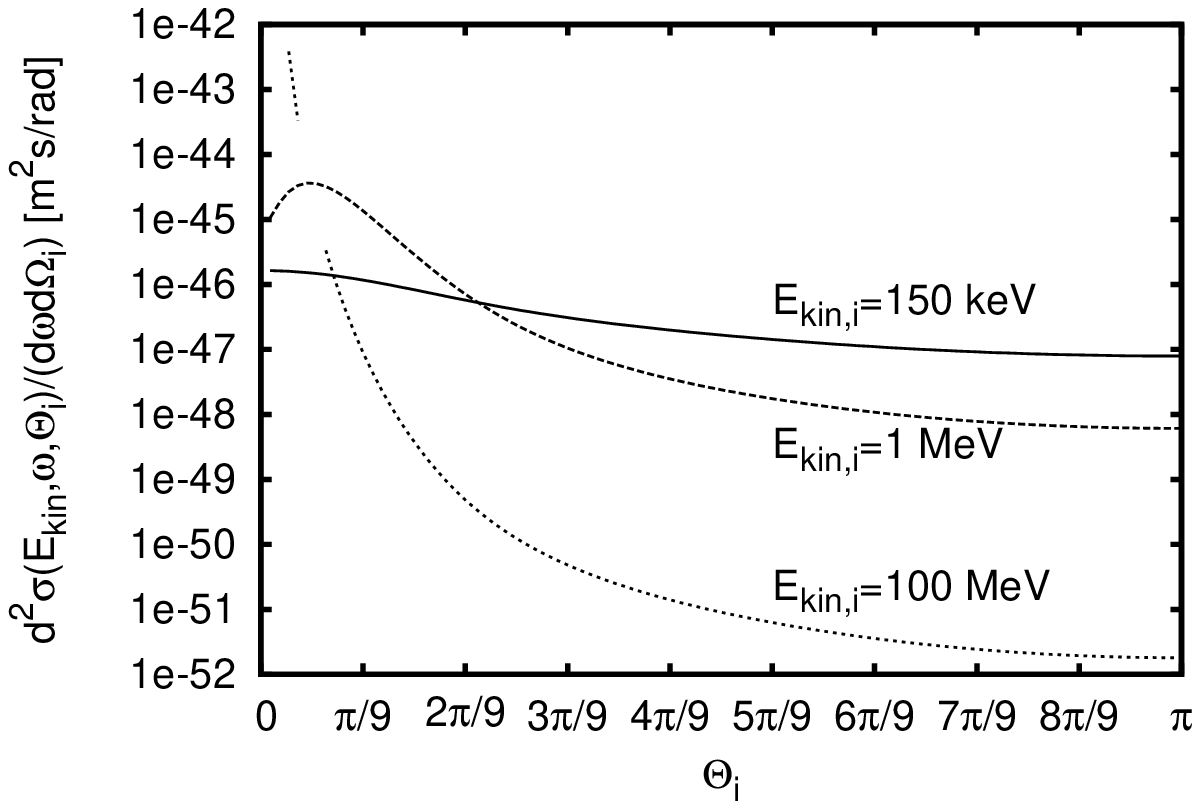}\\
a) $\hbar\omega=50$ eV \hspace{3.8cm} b) $\hbar\omega=500$ eV\\
\includegraphics [scale=0.56] {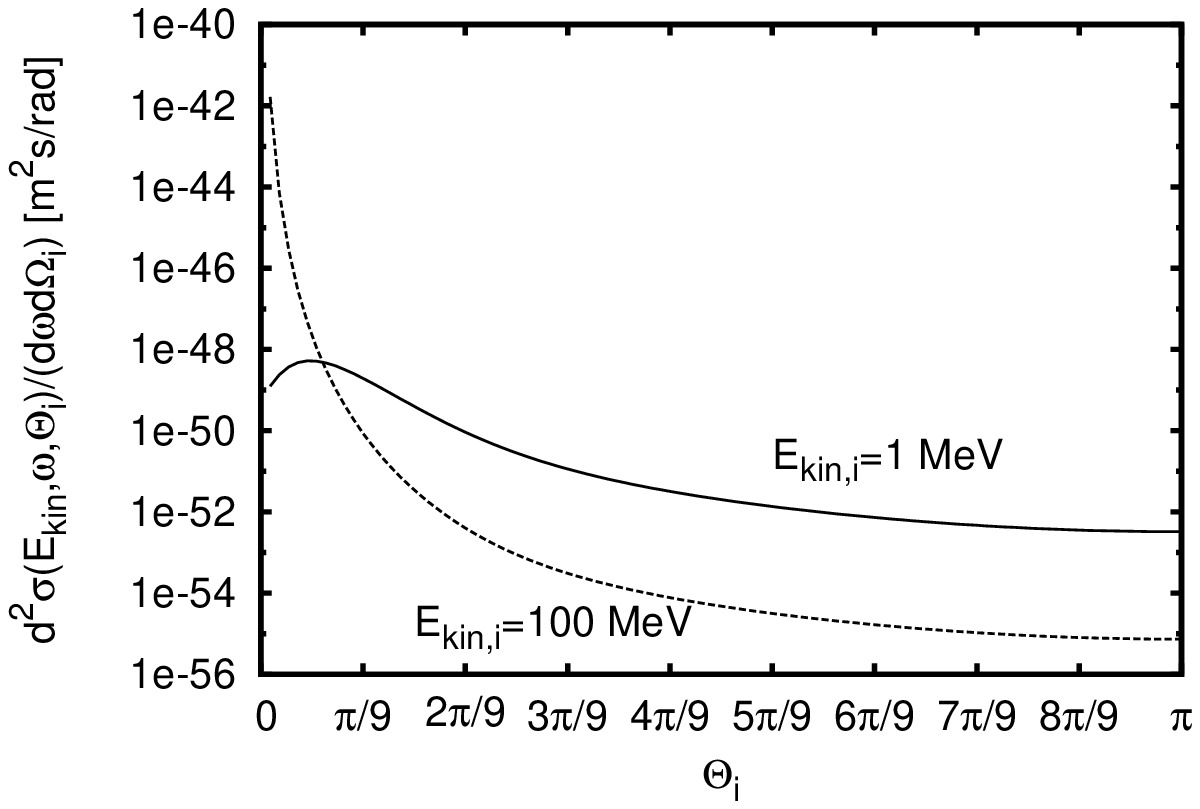}
\includegraphics [scale=0.56] {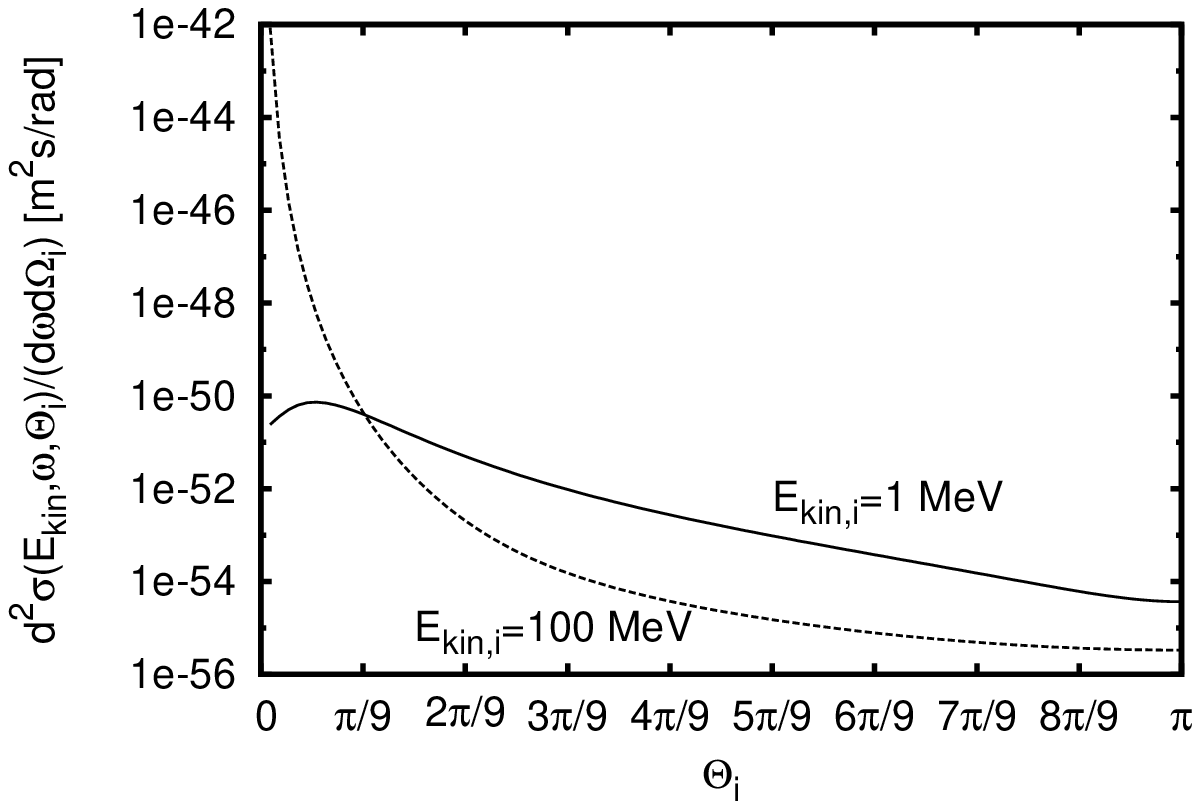}\\
c) $\hbar\omega=500$ keV \hspace{3.8cm} d) $\hbar\omega=950$ keV
\caption {  The doubly differential cross section
$d^2\sigma/{(d\omega d\Omega_i)}
(E_{kin,i},\omega,\Theta_i)$ for Bremsstrahlung ($Z=7$) vs. the scattering angle $\Theta_i$ for several electron and
photon energies. In each panel the photon energy $\hbar\omega$ is fixed and
the cross section is plotted for various kinetic energies of the incident
electron. }\label {disc_fig.6}
\end {figure}

Figure \ref{disc_fig.1} shows that for both slow and
relativistic electrons   the doubly differentical cross section
also varies with   the photon
energy for fixed electron energies. For fixed electron energy,
lower photon   energies are more likely.
Moreover, photons are more likely for certain angles. 
They are more likely for lowly energetic electrons in the limit
$\Theta_i\rightarrow 180^{\circ}$ and for highly energetic electrons
in the limit $\Theta_i\rightarrow 0^{\circ}$.
Figure \ref{disc_fig.6} shows
the doubly differential cross section in another way. Now the photon energy
is fixed and the electron energy differs within one plot.
For all cases  it is more likely that low
energetic electrons create photons than relativistic electrons do, in the limit $\Theta_i\rightarrow
180^{\circ}$. 
But for small angles, i.e. for forward emission of photons, the probability
rapidly increases for relativistic electrons and exceeds the probability at
small electron energies.

\subsubsection {  The most probable scattering angle  }   \label{max_scat.sec}

Figure \ref{disc_fig.1} also shows that the angle for which maximal scattering takes place,
is rather independent of the photon energy. Hence, one can use
(\ref{limit.7}) to
determine a formula for that scattering angle. Actually this derivation
leads to a quartic equation which can, however, be approximated
for small angles, i.e. $\Theta_i\lesssim 20^{\circ}$, through a
quadratic equation. The reader is referred to \ref{app_theta}
for the detailed calculation. The solution of the quadratic equation reads
\begin {eqnarray}
\Theta_i=\sqrt{\frac{-\frac{\delta_0}{\hbar\omega}(4E_f^2+\delta_0c^2)-\frac{2
\delta_0\hbar\omega}{E_f}(E_i-c|\mathbf{p}_i|)}{2\frac{|\mathbf{p}_i|}{c}\left[
4E_f^2+\delta_0c^2+\frac{2\hbar^2\omega^2}{E_f}(E_i-c|\mathbf{p}_i|)\right]
-|\mathbf{p}_i|\delta_0c-\frac{\hbar\omega}{E_f}c|\mathbf{p}_i|\delta_0}}
\label{disc.9}
\end {eqnarray}
with
\begin {eqnarray}
\delta_0:= -|\mathbf{p}_i|^2-\left(\frac{\hbar}{c}\omega\right)^2
+2\frac{\hbar}{c}\omega|\mathbf{p}_i|,
\label{disc.7}
\end {eqnarray}
  $\hbar\omega\rightarrow E_{kin,i}$, e.g. $\hbar\omega=0.9999E_{kin,i}$
  and
\begin {eqnarray}
p_i=\sqrt{E_{kin,i}\left(\frac{E_{kin,i}}{c^2}+2m_e\right)}. \label{disc.11}
\end {eqnarray}
\begin {figure}
\includegraphics [scale=0.5] {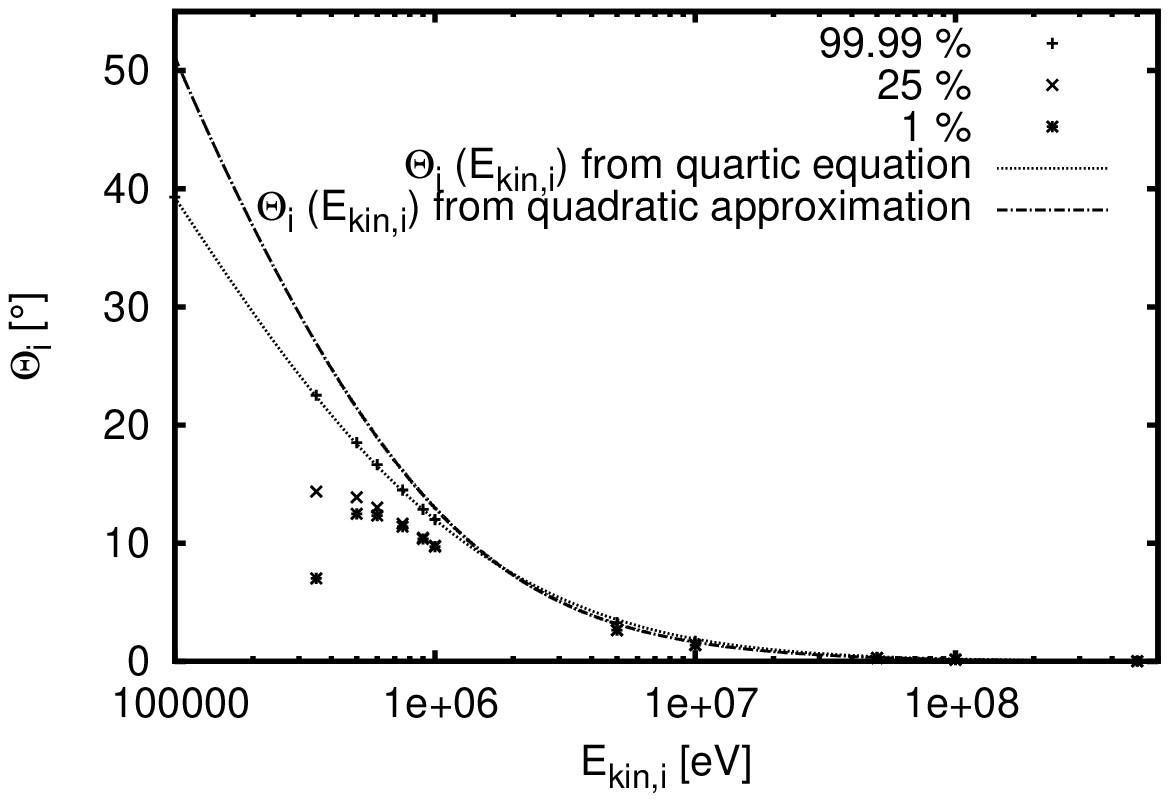}
\includegraphics [scale=0.5] {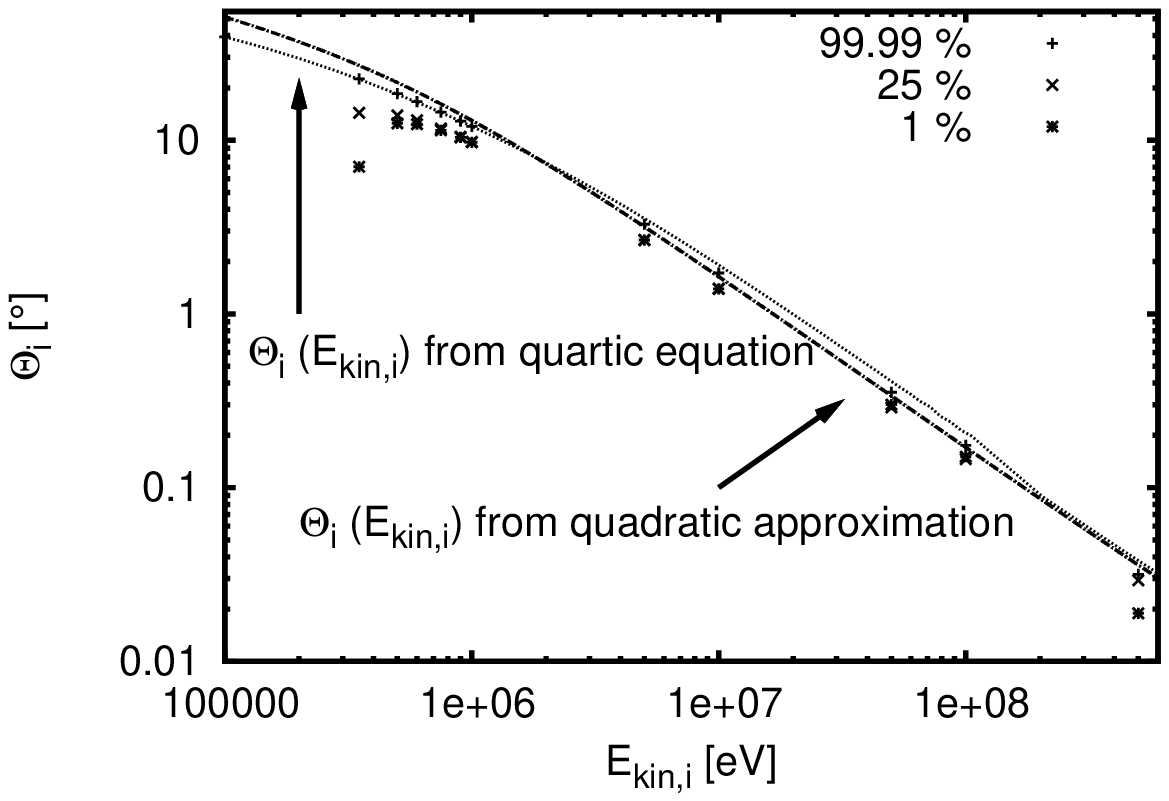}\\
a) \hspace {6.5cm} b)
\caption {$\Theta_i$ for maximal scattering vs. incident electron energy in
a a) semilog and b) loglog plot for   $Z=7$  . Besides (\ref{disc.9}) for
$\hbar\omega=0.9999E_{kin,i}$, the exact solution of the quartic equation and
various data for different $\hbar\omega/E_{kin,i}$ are shown.} \label{disc_fig.7}
\end {figure}
Figure \ref{disc_fig.7} shows   (\ref{disc.9})   and manually extracted values
for $\Theta_i$ for different photon energies. It shows much better than
figure \ref{disc_fig.1} that $\Theta_i$ is rather independent of the photon energy
for relativistic electron energies. Besides (\ref{disc.9}), the
solution of the quartic equation, 
is shown. Moreover, it shows that (\ref{disc.9})
gives a good approximation for those angles $\Theta_i$ for which scattering is maximal.
Actually, we see that the exact solution describes the angle for
maximal scattering better, especially for low energies, but for high
energies both curves fit very well.\\ \indent
By inserting $E_{kin,i}=\hbar\omega/0.9999$ into   (\ref{disc.9}) one
obtains a formula which relates the photon energy to the most probable
scattering angle.   \\ \indent
\subsection{Pair production}
\subsubsection{Basic properties of pair production}
We now proceed from Bremsstrahlung to pair production. One
photon with energy $\hbar\omega$ creates two particles, namely an electron
and a positron, both   with   rest energy
$m_ec^2$.   Therefore
\begin {eqnarray}
E_{kin,-}+E_{kin,+}=\hbar\omega-2m_ec^2 \label{disc.12}
\end {eqnarray}
follows for the kinetic energies of these two particles.
Thus the photon energy has to be $\hbar\omega\ge2m_ec^2\approx
1.022$ MeV   for pair production and   the kinetic energy of the
particles   is bounded as
$E_{kin,\pm}\le\hbar\omega-2m_ec^2$.
\begin {figure}
\includegraphics[scale=0.56]{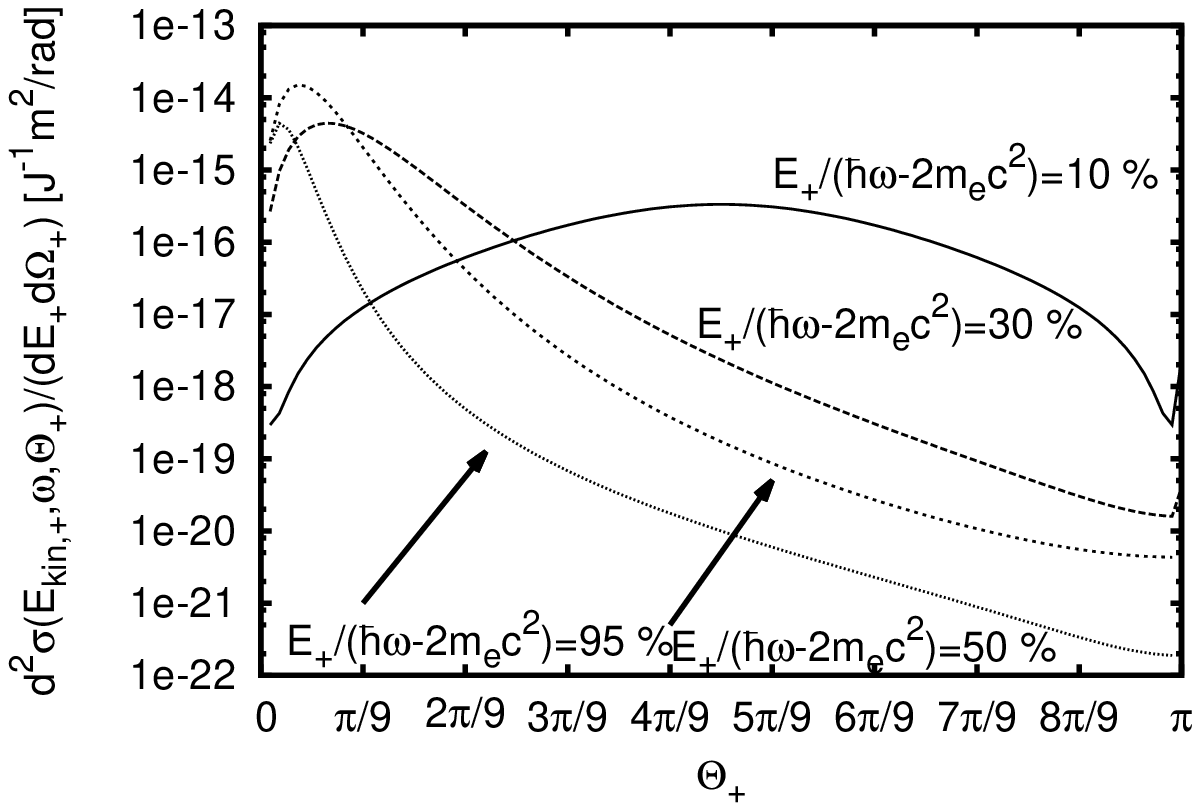}
\includegraphics[scale=0.56]{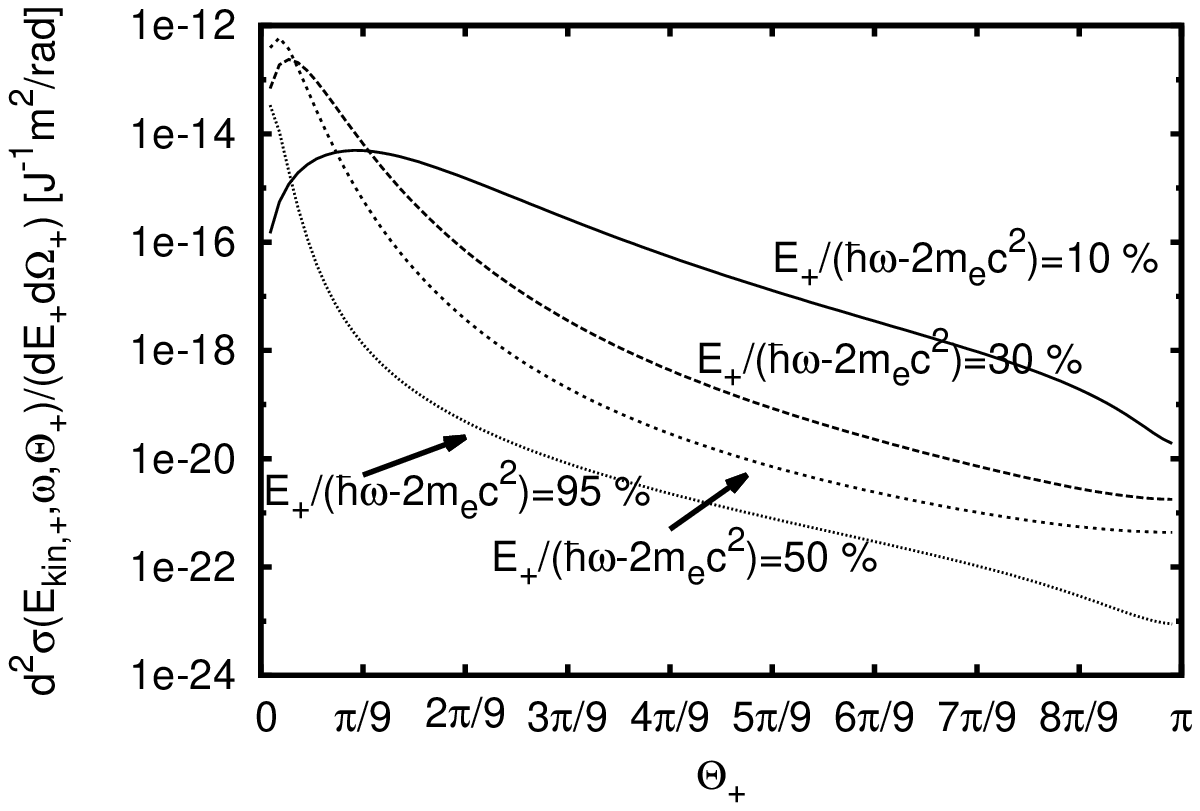}\\
a) $\hbar\omega=5$ MeV \hspace{3.8cm} b) $\hbar\omega=10$ MeV\\
\includegraphics [scale=0.56] {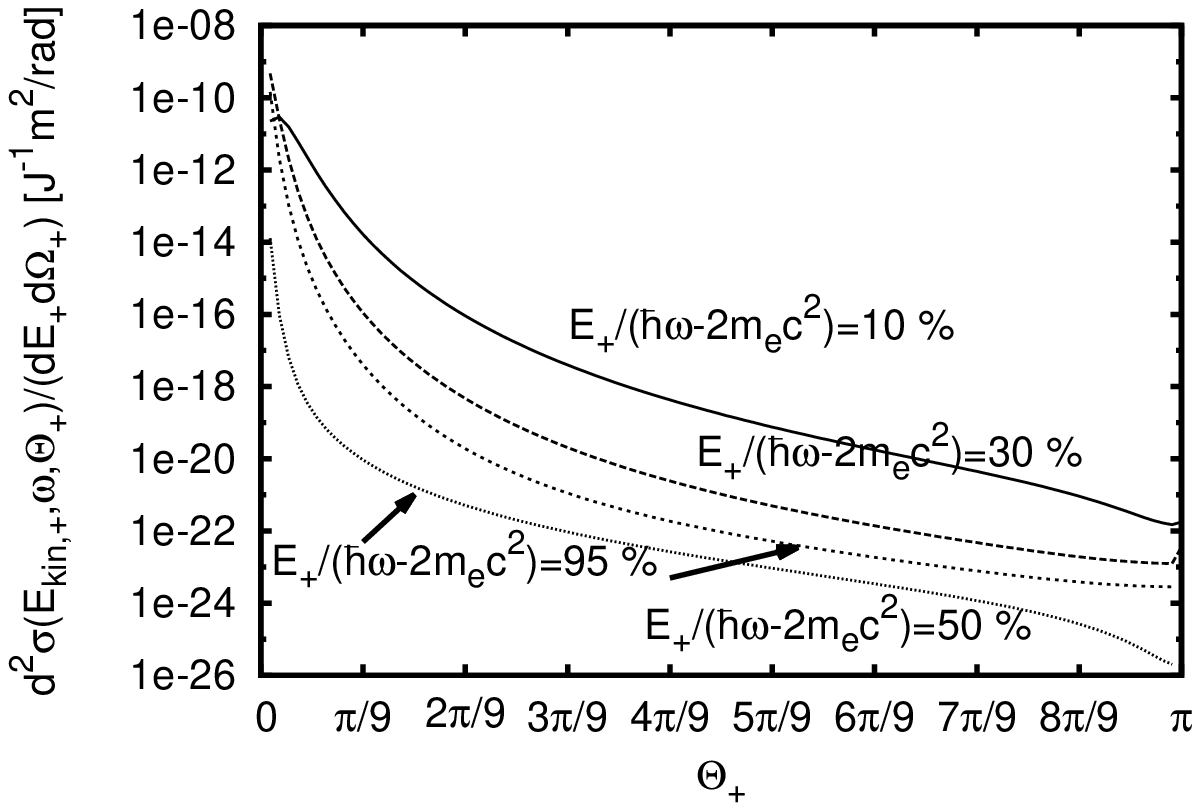}
\includegraphics [scale=0.56] {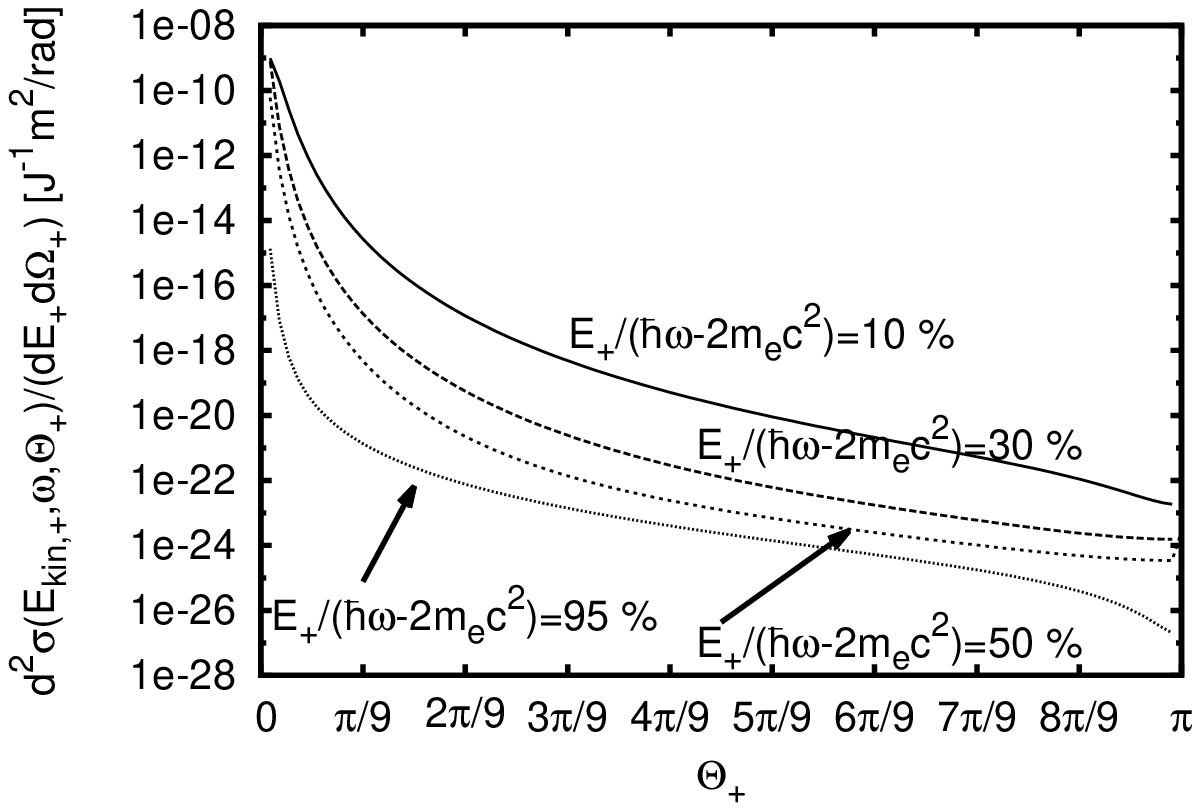}\\
c) $\hbar\omega=50$ MeV \hspace{3.8cm} d) $\hbar\omega=100$ MeV
\caption {Doubly differential cross section ${d^2\sigma(E_+,\omega,
\Theta_+)}/{(dE_+d\Omega_+)}$ as a function of the angle $\Theta_+$ between
incident photon and created positron for $Z=7$: The cross section is shown for
fixed photon energies a) $\hbar\omega=5$ MeV, b) $\hbar\omega=10$ MeV, c) $\hbar\omega=50$ MeV and
d) $\hbar\omega=100$ MeV. In each panel   different positron energies $E_+$ relative to the
available photon energy $\hbar\omega-2m_ec^2$   are plotted.
 } \label{disc_fig.8}
\end {figure}
Figure \ref{disc_fig.8} shows the doubly differential cross section
(\ref{pair.18}) for different photon
and positron energies. 
Forward scattering is dominant, there is almost no case  
now of more isotropic scattering.
This results from the fact that almost all positron energies in Figure
\ref{disc_fig.8} are relativistic. For very highly energetic
photons, e.g. $50$ MeV and $100$ MeV, and thus relativistic
positron energies in Fig. \ref{disc_fig.8} there are clear maxima for
forward scattering. 
For energies   $\hbar\omega<50$ MeV  ,
however, the maxima are   $>5^{\circ}$  , but
forward scattering is still preferred. 
Pair production is symmetric in positron and electron energy. Thus
for the singly differential cross section
\begin {eqnarray}
\frac{d\sigma}{dE_{+}}(E_{+},\omega)=\int\limits_{0}^{\pi}
\frac{d^2\sigma(E_{+},\omega,\Theta_{+})}{dE_{+}d\Omega_{+}} \sin\Theta_{+}
d\Theta_{+}
\end {eqnarray}
the probability of the creation of a positron with a given energy is as
large as the probability to create an electron with this energy:
\begin {eqnarray}
\left.\frac{d\sigma}{dE_{+}}\right|_{\frac{E_{+}}{\hbar\omega-2m_ec^2}} =
\left.\frac{d\sigma}{dE_{+}}\right|_{1-\frac{E_{+}}{\hbar\omega-2m_ec^2}}.
\end {eqnarray}\\ \indent
\begin {figure}
\includegraphics [scale=0.56] {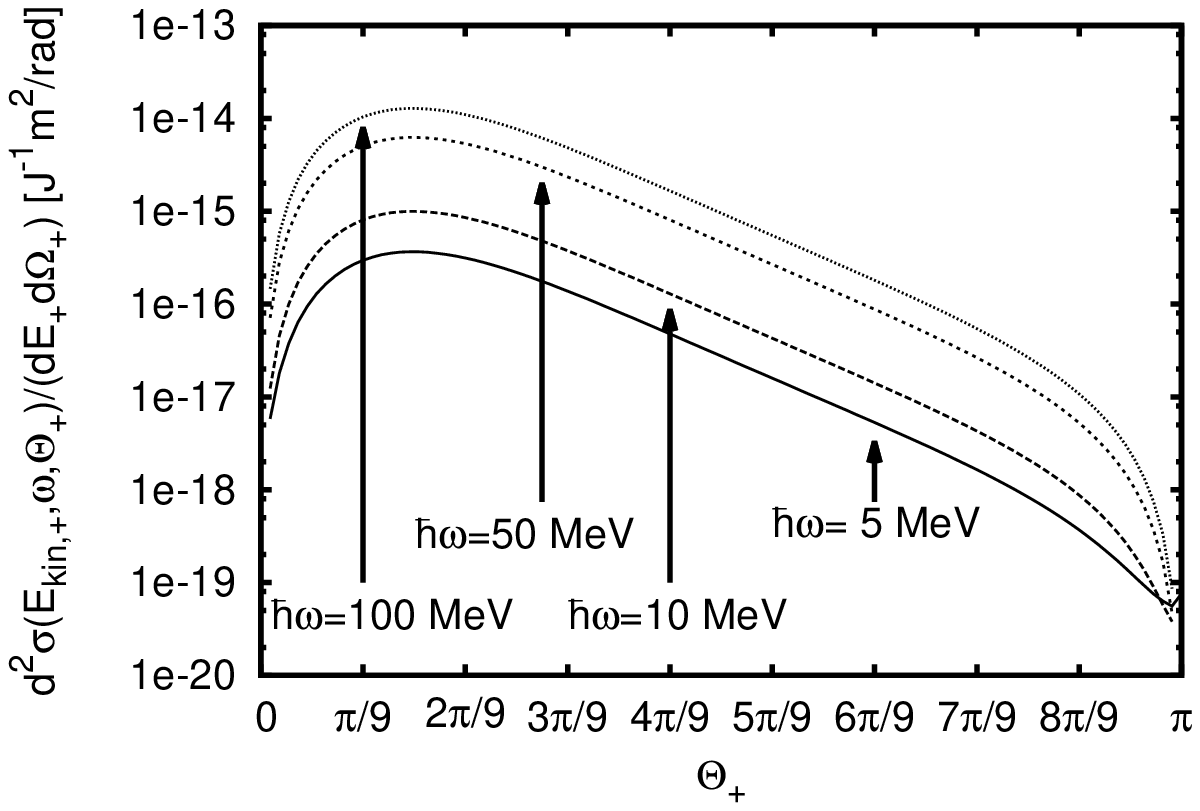}
\includegraphics [scale=0.56] {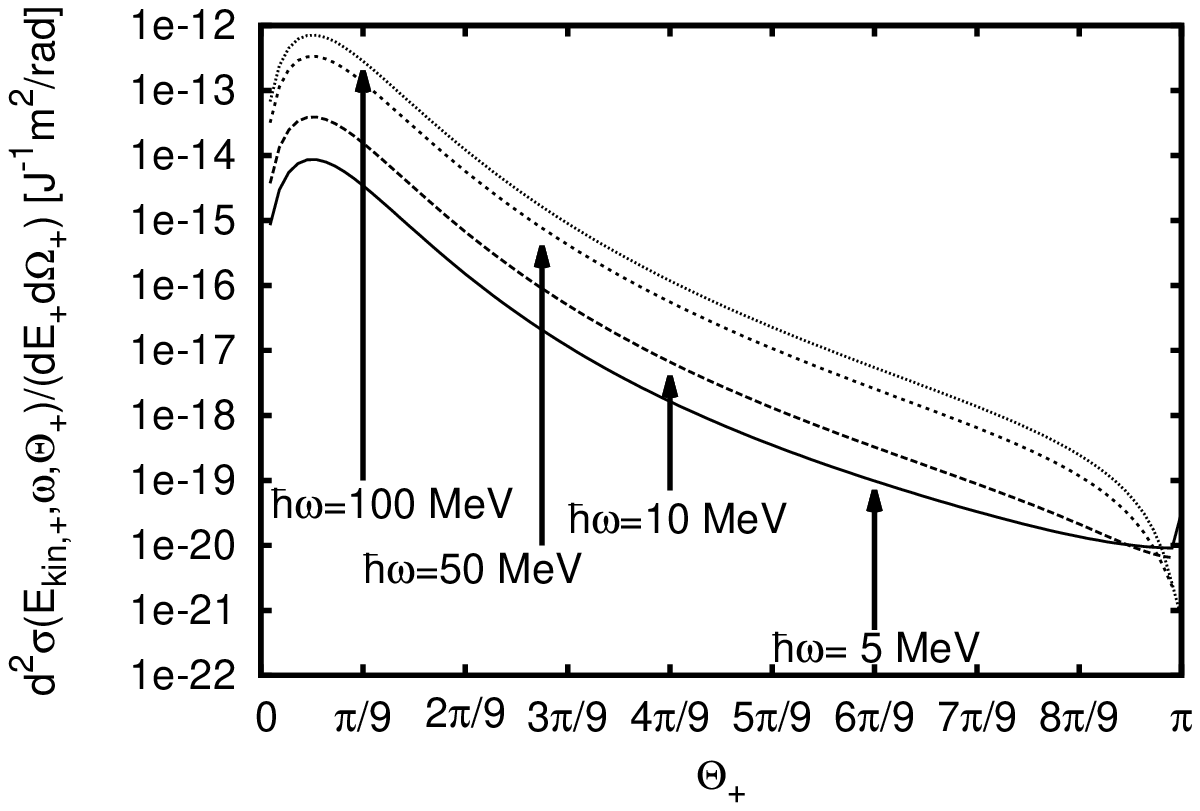}\\
a) $E_{kin,+}=150$ keV \hspace{3.8cm} b) $E_{kin,+}=1$ MeV
\caption { Doubly differential cross section ${d^2\sigma(E_+,\omega,
\Theta_+)}/{(dE_+d\Omega_+)}$ as a function of the angle $\Theta_+$ between
incident photon and created positron for $Z=7$: The cross section is shown for
fixed positron energies a) $E_{kin,+}=150$ keV and b) $E_{kin,+}=1$ MeV. In each panel
curves for the photon energies $\hbar\omega=$ 5 MeV, 10 MeV, 50 MeV and 100 MeV
are included.} \label{disc_fig.9}
\end {figure}
Figure \ref{disc_fig.9} shows the doubly differential cross section (\ref{pair.18}) for fixed positron and
different photon energies. Again positrons which are generated
with high velocities predominantly scatter forward while this tendency
vanishes if the positron energy is very low. This can be traced back to the
relativistic behaviour again. If a positron is very energetic, it has
to be treated relativistically and the relativistic transformation leads to
forward scattering (this is the same explanation as for Bremsstrahlung). We
also see that the creation of positrons is more likely for highly energetic
photons.
\subsubsection {The most probable scattering angle}
\begin {figure}
\includegraphics [scale=0.56] {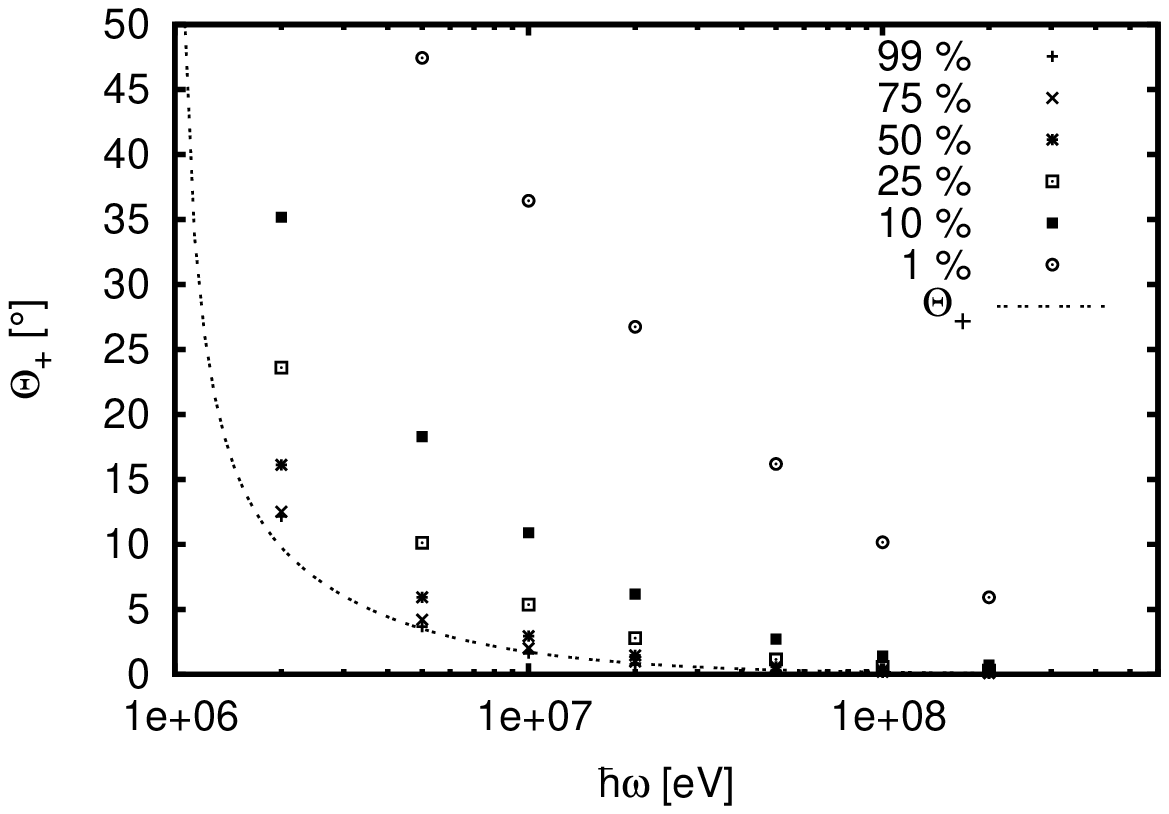}
\includegraphics [scale=0.56] {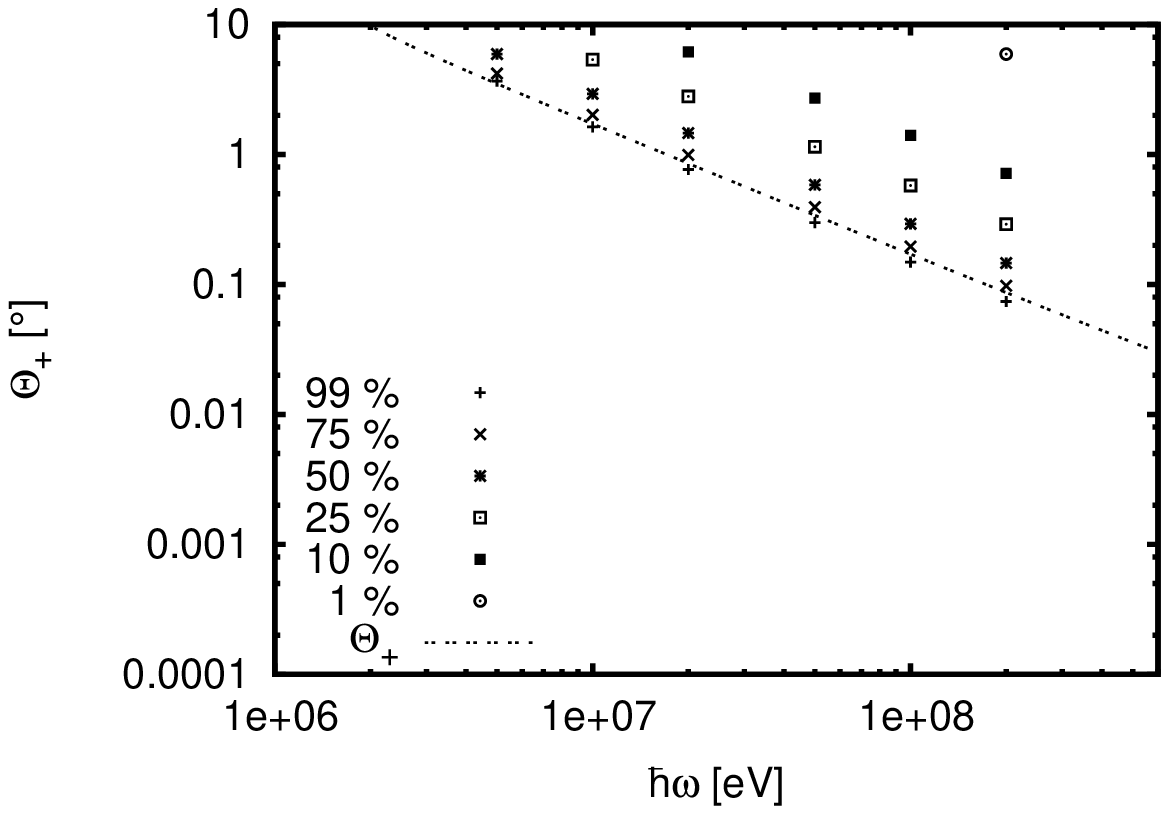}\\
a) \hspace {6.5cm} b)
\caption {$\Theta_+$ for maximal scattering vs. incident photon energy in
a a) semilog and b) loglog plot   for $Z=7$  . Besides (\ref{disc.14}) for
$E_{kin,+}/(\hbar\omega-2m_ec^2)=0.9999$
various data for different $E_{kin,+}/(\hbar\omega-2m_ec^2)$ are shown.} \label {disc_fig.10}
\end {figure}

As for Bremsstrahlung one can get a simple formula for the preferred
direction. Performing the same calculation as for Bremsstrahlung one obtains
\begin {eqnarray}
\Theta_+&=&\Bigg[\Big(-\frac{\delta^{(p)}_0}{\hbar\omega}(-4E_-^2-\delta^{(p)}_0c^2)-\frac{2
\delta^{(p)}_0\hbar\omega}{E_-}(E_+-c|\mathbf{p}_+|)\Big) \nonumber\\
&\times&\Big(2\frac{|\mathbf{p}_+|}{c}\left[
-4E_-^2-\delta^{(p)}_0c^2+\frac{2\hbar^2\omega^2}{E_-}(E_+-c|\mathbf{p}_+|)\right]
-|\mathbf{p}_+|\delta^{(p)}_0c \nonumber\\
&-&\frac{\hbar\omega}{E_-}c|\mathbf{p}_+|\delta^{(p)}_0\Big)^{-1}\Bigg]^{\frac{1}{2}}
 \label{disc.14}
\end {eqnarray}
with
\begin {eqnarray}
\delta^{(p)}:=-|\mathbf{p}_+|^2-\left(\frac{\hbar}{c}\omega\right)^2+2|\mathbf{p}_+|
\frac{\hbar}{c}\omega \label{disc.15}
\end {eqnarray}
and
\begin {eqnarray}
\frac{E_+-m_ec^2}{\hbar\omega-2m_ec^2}\approx 1. \label{disc.16}
\end {eqnarray}
Figure \ref{disc_fig.10}
shows that (\ref{disc.14}) is a good approximation for $\Theta_+$ for high
photon energies and high ratios between photon and positron energy. The
smaller the ratio between photon and positron energy, however, is, the worse
(\ref{disc.14}) becomes for low photon energies. If the photon energy
is larger than 50 MeV, relativistic
positrons are created; therefore forward
scattering takes place and $\Theta_+$ can be calculated with
(\ref{disc.14}).

\newpage

\section {Conclusion} \label{concl}

   We have reviewed literature relevant for Bremsstrahlung in Terrestrial
gamma-ray flashes (TGFs) (Bethe and Heitler, 1934; Heitler, 1944; Elwert and
Haug, 1969; Seltzer and Berger, 1985; Shaffer et al., 1996; Agostinelle et
al., 2003). Focussing on atomic numbers $Z=7$ (nitrogen) and $Z=8$
(oxygen) and an energy range of 1 keV to 1 GeV, no good parametrization of an energy
resolved angular distribution in the form of doubly differential cross section is
available. The theory of Bethe and Heitler covers
this energy range for $Z=7,8$, but it parameterizes the direction of the
scattered electron as well; therefore we integrated their triply
differential cross section to obtain the correct energy resolved angular
distribution for Bremsstrahlung and pair production. Other authors
(Lehtinen, 2000; Dwyer, 2007; Carlson et al., 2009, 2010)
used different approaches, as discussed in the introduction. They use
singly or triply differential cross
sections which do not give a direct relation between the photon energy
and the direction of the photon relative to the motion of the electron.
As positrons are created within a thundercloud as well \cite{fermi}, we
used a symmetry between the production of Bremsstrahlung and the creation of an
electron-positron pair both in the field of a nucleus to obtain a cross
section which relates the energy of the created positron with its direction.

We have seen that emitted Bremsstrahlung photons are mainly released in
forward direction if the electron which interacts with the nucleus has
such a high energy that it has to be treated relativistically. For lower energies
scattering tends to be more isotropic. For the case that almost all
kinetic energy of the incident electron is transformed into photon
energy, we derived an approximation for the most probable photon emission angle as a
function of the incident electron energy and of the photon energy.
The expression is valid for all ratios of photon over electron energy if
the electron motion is relativistic. So, when photons have been created
within a thundercloud or discharge, they are mainly scattered in forward
direction as long as the electrons move relativistically, i.e. if their
kinetic energy is at least as large as their rest energy.

Similar results hold for pair production. Next to the doubly
differential cross section we derived a simple approximation for the most probable positron
emission angle for the case that the photon energy is larger than 10 MeV
(for ratios between the kinetic energy of the positron and available
photon energy down to 25\%)
or than 100 MeV (for ratios lower than 25\%). We have seen
that for very highly energetic photons that almost all positrons are
scattered in forward direction. If, however, the photon energy decreases,
the probability of forward scattering decreases as well. Instead the maximal
cross section can be found at $\Theta_+\approx 90^{\circ}$ for low ratios
between $E_+$ and $\hbar\omega-2m_ec^2$ and is, beyond
that, symmetric to this angle.

Our analytical results for the doubly differential cross-sections for
Brems\-strahlung and pair production are also supplied in the form of two
functions written in C++. In this form the functions can be implemented into Monte
Carlo codes simulating energetic processes like the production of
gamma-rays or electron positron pairs in thunderstorms.\\ \indent
\\
{\bf Acknowledgements:} We dedicate this article to the memory of Davis D.
Sentman. He was a great inspiration and discussion partner in the summer of 2011 when C.K.
worked on this project, and he was an invaluable colleague and
co-organizer for U.E. for many years. 
We acknowledge fruitful and motivating discussions with Brant Carlsson. C.K. acknowledges financial support by STW-project 10757, where Stichting Technische Wetenschappen (STW) is part of The Netherlands' Organization for Scientific Research
NWO.

\newpage

\begin {appendix}
\section {The residual theorem to calculate integrals with trigonometric
functions} \label{app_res}
In this appendix the method how to calculate integrals of the form
\begin {eqnarray}
\int\limits_{0}^{2\pi} R(\cos\Phi,\sin\Phi) d\Phi \label {res.1}
\end {eqnarray}
shall be discussed where $R(x,y):\mathbb{R}^2
\setminus\{x,y\in\mathbb{R}|y=\pm\sqrt{1-x^2}\}
\rightarrow \mathbb{R}$ is a rational function without poles on
the unit circle $x^2+y^2=1$. But before explaining this method let's briefly
review some general facts about residua.
\subsection {The residual theorem}
Let $f:\mathbb{C}\supset I\rightarrow \mathbb{C},z\mapsto f(z)$, be a
holomorphic function and $\Gamma:[a,b]\rightarrow\mathbb{C},t\mapsto
\Gamma(t)$, a closed curve in the complex plane.
Then one can calculate complex curve integrals via
\begin {eqnarray}
\int\limits_{\Gamma}f(z)dz=2\pi i \sum\limits_{j} \Res(f,z_j)
\label{res.2}
\end {eqnarray}
where the sum has to be taken over all poles $z_j$ of $f$ and complex curve
integrals are defined as
\begin {eqnarray}
\int\limits_{\Gamma}f(z)dz:=\int\limits_a^b f(\Gamma(t))\cdot\frac{d\Gamma}{dt}(t)
dt. \label{res.3}
\end {eqnarray}
The residuum of a pole $z_j$ can be calculated via
\begin {eqnarray}
\Res(f,z_j)=\frac{1}{(n-1)!}\lim_{z\rightarrow z_j}\frac{d^{n-1}}{dz^{n-1}}
\Big[(z-z_j)f(z)\Big] \label {res.4}
\end {eqnarray}
where $n$ denotes the order of the pole.

\subsection {Integral with trigonometric functions}
With the help of the (\ref{res.2}) one can simply perform the integration of
(\ref{res.1}). For that purpose define
\begin {eqnarray}
f(z):=\frac{1}{iz} R\left(\frac{1}{2}\left(z+\frac{1}{z}\right),
\frac{1}{2i} \left(z-\frac{1}{z}\right)\right) \label {res.5}
\end {eqnarray}
and choose the unit circle
\begin {eqnarray}
\Gamma(t)=e^{it}, t\in[0,2\pi] \label{res.6}
\end {eqnarray}
as closed curve; hence (\ref{res.3}) becomes
\begin {eqnarray}
\int\limits_{\Gamma}f(z)dz&=& \int\limits_{0}^{2\pi}
\frac{1}{ie^{it}} R\left(\frac{1}{2}\left(e^{it}+e^{-it}\right),\frac{1}{2i}
\left(e^{it}-e^{-it}\right)\right)ie^{it} dt \\
&=& \int\limits_{0}^{2\pi}
R\left(\frac{1}{2}\left(e^{it}+e^{-it}\right),\frac{1}{2i}
\left(e^{it}-e^{-it}\right)\right) dt \\
&=& \int\limits_{0}^{2\pi} R(\cos t,\sin t)dt \label{res.7}
\end {eqnarray}
where the   identities
$\cos t=1/2\left(e^{it}+e^{-it}\right)$   and
$\sin t=1/(2i)\left(e^{it}-e^{-it}\right)$   were   used in the last step.\\ \indent
Finally with (\ref{res.2}) and (\ref{res.7}) one gets a simple formula to
calculate (\ref{res.1}):
\begin {eqnarray}
\int\limits_{0}^{2\pi} R(\cos\Phi,\sin\Phi) d\Phi  = 2\pi i \sum\limits_
{|z|<1} \Res(f,z) \label{res.10}
\end {eqnarray}
with $f$ being defined in (\ref{res.5}).

\section {The doubly differential cross section for $\Theta_i=0$ and
$\Theta_i=\pi$} \label{app_small}
In order to get (\ref{small.8}) from (\ref{theta.16}) it is rather straight
forward to set $\Theta_i=0$ or $\Theta_i=\pi$. Especially it is
\begin {eqnarray}
\Delta_1(\Theta_i=0,\pi)&=&\tilde{\Delta}_1 \label{appC.1}, \\
\Delta_2(\Theta_i=0,\pi)&=&\tilde{\Delta}_2 \label{appC.2}.
\end {eqnarray}
But there is one case which should be considered a bit more thoroughly.\\ \indent
This regards the logarithm in (\ref{theta.17}). For $\Theta_i=\pi$
it is $\Delta_2(\Theta=\pi)=\tilde{\Delta}_2=-2p_f(\hbar/c\ \omega+p_i)<0$;
thus $|\tilde{\Delta}_2|=-\tilde{\Delta}_2$ and
\begin {eqnarray}
&&\left.\frac{\ln\left(
\frac{\Delta_2^2+4p_i^2p_f^2\sin^2\Theta_i-\sqrt{\Delta_2^2+4p_i^2p_f^2\sin^2
\Theta_i}(\Delta_1+\Delta_2)+\Delta_1\Delta_2}{-\Delta_2^2-4p_i^2p_f^2\sin^2\Theta_i
-\sqrt{\Delta_2^2+4p_i^2p_f^2\sin^2 \Theta_i}(\Delta_1-\Delta_2)+\Delta_1\Delta_2
}\right)}{\sqrt{\Delta_2^2+4p_i^2p_f^2\sin^2\Theta_i}}\right|_{\Theta_i=\pi} \label{appC.3}
\\
&=&\frac{1}{|\tilde{\Delta}_2|}\ln\left(
\frac{\tilde{\Delta}_2^2-|\tilde{\Delta}_2|(\tilde{\Delta}_1+\tilde{\Delta}_2)+\tilde{\Delta}_1
\tilde{\Delta_2}}{-\tilde{\Delta_2}^2
-|\tilde{\Delta}_2|(\tilde{\Delta}_1-\tilde{\Delta}_2)+\tilde{\Delta}_1\tilde{\Delta}_2
}\right) \label{appC.4}\\
&=&-\frac{1}{\tilde{\Delta}_2}\ln\left(\frac{\tilde{\Delta}_1+\tilde{\Delta}_2}
{\tilde{\Delta}_1-\tilde{\Delta}_2}\right) \label{appC.5}
\end {eqnarray}
which is a very simple calculation. However, for $\Theta_i=0$ it is
$\Delta_2(\Theta_i=0)=\tilde{\Delta}_2=-2p_f(\hbar/c\ \omega-p_i)$ which can
be both negative or positive depending on   values of
  $p_i$ and $\hbar/c\  \omega$ If $\tilde{\Delta}_2<0$ then
  equations   (\ref{appC.3}) -
(\ref{appC.5}) are valid again. If $\tilde{\Delta}_2>0$, however, it follows for the
argument of the logarithm
\begin {eqnarray}
&&\left.\frac{\Delta_2^2+4p_i^2p_f^2\sin^2\Theta_i-\sqrt{\Delta_2^2+4p_i^2p_f^2\sin^2
\Theta_i}(\Delta_1+\Delta_2)+\Delta_1\Delta_2}{-\Delta_2^2-4p_i^2p_f^2\sin^2\Theta_i
-\sqrt{\Delta_2^2+4p_i^2p_f^2\sin^2 \Theta_i}(\Delta_1-\Delta_2)+\Delta_1\Delta_2
}\right|_{\Theta_i=0} \label{appC.6}\\
&=&\frac{\tilde{\Delta}_2^2-\tilde{\Delta}_2(\tilde{\Delta}_1+\tilde{\Delta}_2)+\tilde{\Delta}_1
\tilde{\Delta}_2}{-\tilde{\Delta}_2^2
-\tilde{\Delta_2}(\tilde{\Delta}_1-\tilde{\Delta}_2)+\tilde{\Delta}_1\tilde{\Delta}_2
}=\frac{0}{0}. \label{appC.7}
\end {eqnarray}
Hence it is necessary to use the rule of L'H\^{o}pital:
\begin {eqnarray}
&&\lim_{\Theta_i\rightarrow0}\frac{\Delta_2^2+4p_i^2p_f^2\sin^2\Theta_i-\sqrt{\Delta_2^2+4p_i^2p_f^2\sin^2
\Theta_i}(\Delta_1+\Delta_2)+\Delta_1\Delta_2}{-\Delta_2^2-4p_i^2p_f^2\sin^2\Theta_i
-\sqrt{\Delta_2^2+4p_i^2p_f^2\sin^2 \Theta_i}(\Delta_1-\Delta_2)+\Delta_1\Delta_2
}\nonumber\\ \label{appC.8} \\
&=&\lim_{\Theta_i\rightarrow0}\frac{8p_i^2p_f^2\sin\Theta_i\cos\Theta_i-
\frac{\Delta_1+\Delta_2}{2\sqrt{\Delta_2^2+4p_i^2p_f^2\sin^2\Theta_i}}\cdot 8p_i^2p_f^2\sin\Theta_i
\cos\Theta_i}{-8p_i^2p_f^2\sin\Theta_i\cos\Theta_i-
\frac{\Delta_1-\Delta_2}{2\sqrt{\Delta_2^2+4p_i^2p_f^2\sin^2\Theta_i}}\cdot 8p_i^2p_f^2\sin\Theta_i
\cos\Theta_i} \nonumber\\ \label{appC.9}\\
&=&\ln\left(\frac{\tilde{\Delta}_1-\tilde{\Delta}_2}{\tilde{\Delta}_1+\tilde{\Delta}_2}
\right). \label {appC.10}
\end {eqnarray}
With (\ref{appC.10}) the whole limit yields
\begin {eqnarray}
&&\left.\frac{\ln\left(
\frac{\Delta_2^2+4p_i^2p_f^2\sin^2\Theta_i-\sqrt{\Delta_2^2+4p_i^2p_f^2\sin^2
\Theta_i}(\Delta_1+\Delta_2)+\Delta_1\Delta_2}{-\Delta_2^2-4p_i^2p_f^2\sin^2\Theta_i
-\sqrt{\Delta_2^2+4p_i^2p_f^2\sin^2 \Theta_i}(\Delta_1-\Delta_2)+\Delta_1\Delta_2
}\right)}{\sqrt{\Delta_2^2+4p_i^2p_f^2\sin^2\Theta_i}}\right|_{\Theta_i=0}
\label{appC.11} \\
&=&\frac{1}{\tilde{\Delta}_2}\ln\left(\frac{\tilde{\Delta}_1-\tilde{\Delta}_2}{\tilde{\Delta}_1+\tilde{\Delta}_2}
\right)=-\frac{1}{\tilde{\Delta}_2}\ln\left(\frac{\tilde{\Delta}_1+\tilde{\Delta}_2}{\tilde{\Delta}_1-\tilde{\Delta}_2}
\right) \label{appC.12}
\end {eqnarray}
which is and has to be identical with (\ref{appC.5}). So in both cases,
$\tilde{\Delta}_2>0$ and $\tilde{\Delta}_2<0$, (\ref{appC.5},\ref{appC.12})
are generated by setting $\Theta_i=0,\pi$; therefore one does not have to distinguish
between the these cases in (\ref{small.8}).\\ \indent
But it is of importance to mention that due to (\ref{appC.7}) one can get
numerical problems if one only implements (\ref{theta.16}) and wants to
calculate the doubly differential cross section for $\Theta_i=0$. Thus it is
useful to distinguish for $\Theta_i\not=0$ and $\Theta_i=0$ and to use
(\ref{small.8}) instead for the latter case.\\ \indent
For the rest of limiting forward and/or backward scattering it is, however,
straight forward to insert $\Theta_i=0,\pi$ and thus can deduce
(\ref{small.8}) from (\ref{theta.16}) with the additional help of (\ref{appC.10}).

\section {The doubly differential cross section for $\hbar\omega \rightarrow
E_{kin,i}$} \label{app_large}
There are three contributions from (\ref{theta.16}) which lead to
(\ref{limit.7}) in the limit
$\hbar\omega\rightarrow E_{kin,i} \Leftrightarrow |\mathbf{p}_f|\rightarrow
0$:
\tiny
\begin {eqnarray}
\iota_1&=&\frac{16\pi E_f^2 p_i^2\sin^2\Theta_i A}{(E_i-cp_i\cos\Theta_i)^2
(\Delta_1^2-\Delta_2^2-4p_i^2p_f^2\sin^2\Theta_i)}, \label{appD.1} \\
\iota_2&=&-\frac{2\pi Ap_i^2c^2\sin^2\Theta_i}{(E_i-cp_i\cos\Theta_i)^2}
\frac{1}{\sqrt{\Delta_2^2+4p_i^2p_f^2\sin^2\Theta_i}} \nonumber\\
&\times&\ln\left(\frac{\Delta_2^2+4p_i^2p_f^2\sin^2\Theta_i-\sqrt{\Delta_2^2+4p_i^2p_f^2\sin^2
\Theta_i}(\Delta_1+\Delta_2)+\Delta_1\Delta_2}{-\Delta_2^2-4p_i^2p_f^2\sin^2\Theta_i
-\sqrt{\Delta_2^2+4p_i^2p_f^2\sin^2 \Theta_i}(\Delta_1-\Delta_2)+\Delta_1\Delta_2
}\right), \label{appD.2} \\
\iota_3&=&-\frac{4\pi\hbar^2\omega^2p_i^2\sin^2\Theta_iA}{E_i-cp_i\cos\Theta_i}
\left[-\frac{2\Delta_1\Delta_2p_fc+2\Delta_2^2E_f+8p_i^2p_f^2\sin^2\Theta_i
E_f}{(-\Delta_2^2+\Delta_1^2-4p_i^2p_f^2\sin^2\Theta_i)((\Delta_2E_f
+\Delta_1 p_f c)^2+4m^2c^4p_i^2p_f^2\sin^2\Theta_i)}\right. \nonumber\\
&+&\frac{p_fc(\Delta_2E_f+\Delta_1p_fc)}{\sqrt{((\Delta_2
E_f+\Delta_1 p_fc)^2+4m^2c^4p_i^2p_f^2\sin^2\Theta_i)^3}} \nonumber \\
&\times&\ln\Bigg(\Big((E_f+p_fc)(4p_i^2p_f^2\sin^2\Theta_i(E_f-p_fc)+(\Delta_1+\Delta_2)
((\Delta_2E_f+\Delta_1p_fc) \nonumber\\
&-&\sqrt{(\Delta_2E_f+\Delta_1p_fc)^2+4m^2c^4p_i^2p_f^2\sin^2\Theta_i}))\Big)\Big((E_f-p_fc)
(4p_i^2p_f^2\sin^2\Theta_i(-E_f-p_fc) \nonumber \\
&+&(\Delta_1-\Delta_2)
((\Delta_2E_f+\Delta_1p_fc)-\sqrt{(\Delta_2E_f+\Delta_1p_fc)^2+4m^2c^4p_i^2p_f^2\sin^2\Theta_i}))\Big)^{-1}
\Bigg)\Bigg] \label{appD.3}
\end {eqnarray}
\normalsize
while all other integrals which appear in (\ref{theta.16}) cancel each other (which will be
shown   in   an example later). It can be verified easily that
\begin {eqnarray}
\lim_{p_f\rightarrow 0} \Delta_1&=&\delta, \label{appD.4} \\
\lim_{p_f\rightarrow 0} \Delta_2&=&0 \ \textnormal{  with
$\Delta_2 \sim |\mathbf{p}_f|$ }\label{appD.5}
\end {eqnarray}
according to definitions (\ref{theta.5}), (\ref{theta.6}) and (\ref{limit.3}). With these limits the
behavior of $\iota_1$ for small $p_f$ can be calculated in a straight forward
way:
\begin {eqnarray}
\lim_{p_f\rightarrow 0} \iota_1 = \frac{16\pi A E_f^2p_i^2\sin^2\Theta_i}
{(E_i-cp_i\cos\Theta_i)^2\delta^2}. \label{appD.6}
\end {eqnarray}
For (\ref{appD.2}) and (\ref{appD.3}), however, there is more effort to be invested. As it is
$\sqrt{\Delta_2^2+4p_i^2p_f^2\sin^2\Theta_i}\rightarrow0$ and $\ln
\Bigg(\Big(\Delta_2^2+4p_i^2p_f^2\sin^2\Theta_i-\sqrt{\Delta_2^2+4p_i^2p_f^2\sin^2
\Theta_i}\linebreak
\times(\Delta_1+\Delta_2)+\Delta_1\Delta_2\Big)/\Big(-\Delta_2^2-4p_i^2p_f^2\sin^2\Theta_i
-\sqrt{\Delta_2^2+4p_i^2p_f^2\sin^2 \Theta_i}(\Delta_1 \linebreak
-\Delta_2)+\Delta_1\Delta_2\Big)\Bigg)\rightarrow0$ for
$p_f\rightarrow 0$, one has to use the rule of L'H\^{o}pital. If one rewrites
\begin {eqnarray}
\Delta_2=\Psi p_f \label{appD.7}
\end {eqnarray}
with
\begin {eqnarray}
\Psi:=-2\frac{\hbar}{c}\omega+2p_i\cos\Theta_i \label{appD.8}
\end {eqnarray}
this rule leads to
\begin {eqnarray}
&&\lim_{p_f\rightarrow 0}\left[\frac{1}{\sqrt{\Delta_2^2+4p_i^2p_f^2\sin^2\Theta_i}}
\right. \times\nonumber\\
&\times&\left.\ln\left(\frac{\Delta_2^2+4p_i^2p_f^2\sin^2\Theta_i-\sqrt{\Delta_2^2+4p_i^2p_f^2\sin^2
\Theta_i}(\Delta_1+\Delta_2)+\Delta_1\Delta_2}{-\Delta_2^2-4p_i^2p_f^2\sin^2\Theta_i
-\sqrt{\Delta_2^2+4p_i^2p_f^2\sin^2 \Theta_i}(\Delta_1-\Delta_2)+\Delta_1\Delta_2
}\right)\right] \nonumber \\
&=&-\frac{2}{\delta}; \label{appD.9}
\end {eqnarray}
thus
\begin {eqnarray}
\lim_{p_f\rightarrow 0}\iota_2=\frac{4\pi Ap_i^2c^2\sin^2\Theta_i}
{(E_i-cp_i\cos\Theta_i)^2\delta}. \label{appD.10}
\end {eqnarray}
The limit of (\ref{appD.3}) can also be calculated by using (\ref{appD.7}).
It is
\tiny
\begin {eqnarray}
\iota_3&=&-\frac{4\pi\hbar^2\omega^2p_i^2\sin^2\Theta_i A}{E_i-cp_i\cos
\Theta_i}\left[-\frac{p_f^2(2\Delta_1\Psi c+2\Psi^2E_f+8p_i^2\sin^2\Theta_i
E_f)}{p_f^2(-\Delta_2^2+\Delta_1^2-4p_i^2\sin^2\Theta_i)((\Psi E_f+\Delta_1
c)^2+4m^2c^4p_i^2\sin^2\Theta_i)} \right. \nonumber\\
&\times& \frac{p_f^2 c(\Psi E_f+\Delta_1c)}{p_f^2((\Psi E_f+
\Delta_1c)^2+4m^2c^4p_i^2\sin^2\Theta_i)} \nonumber\\
&\times&\frac{1}{p_f\sqrt{(\Psi E_f+\Delta_1c)^2+4m^2c^4p_i^2\sin^2
\Theta_i)}} \nonumber\\
&\times&\ln\Bigg(\Big((E_f+p_fc)(4p_i^2p_f^2\sin^2\Theta_i(E_f-p_fc)+(\Delta_1+\Psi p_f)
(p_f(\Psi E_f+\Delta_1c) \nonumber\\
&-&p_f\sqrt{(\Psi E_f+\Delta_1c)^2+4m^2c^4p_i^2\sin^2\Theta_i}))\Big)\Big((E_f-p_fc)
(4p_i^2p_f^2\sin^2\Theta_i(-E_f-p_fc) \nonumber \\
&+&(\Delta_1-\Psi p_f)
(p_f(\Psi E_f+\Delta_1c)-p_f\sqrt{(\Psi E_f+\Delta_1c)^2+4m^2c^4p_i^2\sin^2\Theta_i}))\Big)^{-1}
\Bigg)\Bigg]
\end {eqnarray}
\normalsize
While $p_f$ can simply be reduced in the fractions, one has to use the rule
of L'H\^{o}pital again for the logarithmic part because it is \linebreak
$p_f\sqrt{(\Psi
E_f+\Delta_1c)^2+4m^2c^4p_i^2\sin^2\Theta_i}\rightarrow 0$ and the logarithm
$\rightarrow 0$ for $p_f\rightarrow 0$. Its limit is
\tiny
\begin {eqnarray}
&&\lim_{p_f\rightarrow 0}\frac{1}{p_f\sqrt{(\Psi E_f+\Delta_1c)^2+4m^2c^4p_i^2\sin^2\Theta_i}}
\nonumber\\
&\times&\ln\Bigg(\Big((E_f+p_fc)(4p_i^2p_f^2\sin^2\Theta_i(E_f-p_fc)+(\Delta_1+\Psi p_f)
(p_f(\Psi E_f+\Delta_1c) \nonumber\\
&-&p_f\sqrt{(\Psi E_f+\Delta_1c)^2+4m^2c^4p_i^2\sin^2\Theta_i}))\Big)\Big((E_f-p_fc)
(4p_i^2p_f^2\sin^2\Theta_i(-E_f-p_fc) \nonumber \\
&+&(\Delta_1-\Psi p_f)
(p_f(\Psi E_f+\Delta_1c)-p_f\sqrt{(\Psi E_f+\Delta_1c)^2+4m^2c^4p_i^2\sin^2\Theta_i}))\Big)^{-1}
\Bigg)\Bigg] \nonumber \\
&=&\left.-\frac{2}{E_f\Delta_1}\right|_{p_f
\rightarrow 0}
=-\frac{2}{E_f\delta} \label{appD.12};
\end {eqnarray}
\normalsize
thus the whole limit yields after some further calculations
\begin {eqnarray}
\lim_{p_f\rightarrow 0}\iota_3 =\frac{8\pi\hbar^2\omega^2p_i^2\sin^2\Theta_i A}
{(E_i-cp_i\cos\Theta_i)\delta^2 E_f}. \label{appD.13}
\end {eqnarray}
Finally, if one inserts (\ref{int.3}), the sum of (\ref{appD.6}), (\ref{appD.10})
and (\ref{appD.13}) leads to (\ref{limit.7}).\\ \indent
All other terms which appear in (\ref{theta.16}) vanish. For this purpose one should
regroup all terms according to their origin. As an example let's consider
the three contributions which have arisen from $\int\limits_{\Theta_f=0}^{\pi}
\int\limits_{\Phi=0}^{2\pi}\frac{a_2\cos\Phi}{\alpha\cos\Phi+\beta}d\Phi
d\Omega_f$. For this integral it follows
\tiny
\begin {eqnarray}
&&\int\limits_{\Theta_f=0}^{\pi}
\int\limits_{\Phi=0}^{2\pi}\frac{a_2\cos\Phi}{\alpha\cos\Phi+\beta}d\Phi
d\Omega_f \nonumber\\
&=&-\frac{2\pi Ac}{(E_i-cp_i\cos\Theta_i)p_f}\ln\left(\frac{E_f+p_fc}
{E_f-p_fc}\right) \nonumber \\
&-&\frac{2\pi A c^2}{E_i-cp_i\cos\Theta_i}\left[-\frac{\Delta_2}{\sqrt
{\Delta_2^2+4p_i^2p_f^2\sin^2\Theta_i}p_fc}\right.\nonumber \\
&\times&\left.\ln\left(\frac{\Delta_2^2+4p_i^2p_f^2\sin^2\Theta_i-\sqrt{\Delta_2^2+4p_i^2p_f^2\sin^2
\Theta_i}(\Delta_1+\Delta_2)+\Delta_1\Delta_2}{-\Delta_2^2-4p_i^2p_f^2\sin^2\Theta_i
-\sqrt{\Delta_2^2+4p_i^2p_f^2\sin^2 \Theta_i}(\Delta_1-\Delta_2)+\Delta_1\Delta_2
}\right)
\right.\nonumber\\
&-&\frac{\Delta_2 E_f+\Delta_1 p_fc}{p_fc\sqrt{(\Delta_2 E_f +\Delta_1
p_fc)^2+4m^2c^4p_i^2p_f^2\sin^2\Theta_i^2}}\nonumber \\
&\times&\ln\Bigg(\Big((E_f+p_fc)(4p_i^2p_f^2\sin^2\Theta_i(E_f-p_fc)+(\Delta_1+\Delta_2)
((\Delta_2E_f+\Delta_1p_fc) \nonumber\\
&-&\sqrt{(\Delta_2E_f+\Delta_1p_fc)^2+4m^2c^4p_i^2p_f^2\sin^2\Theta_i}))\Big)\Big((E_f-p_fc)
(4p_i^2p_f^2\sin^2\Theta_i(-E_f-p_fc) \nonumber \\
&+&(\Delta_1-\Delta_2)
((\Delta_2E_f+\Delta_1p_fc)-\sqrt{(\Delta_2E_f+\Delta_1p_fc)^2+4m^2c^4p_i^2p_f^2\sin^2\Theta_i}))\Big)^{-1}
\Bigg)\Bigg] \nonumber \label{appD.14} \\
&\xrightarrow{\hbar\omega\rightarrow
E_{kin,i}}& \frac{2\pi Ac}{E_i-cp_i\cos\Theta_i}\left[-\frac{2c}{E_f}
+\Psi\left(-\frac{2}{\delta}\right)+\left(\Psi E_f+\delta c\right)
\frac{2}{E_f\delta}\right] \label{appD.15} \nonumber \\
&=& 0 \label{appD.16}
\end {eqnarray}
\normalsize
where we have used (\ref{appD.9}), (\ref{appD.12}) and
\begin {eqnarray}
\lim_{p_f\rightarrow 0} \frac{1}{p_f}\ln\left(\frac{E_f+p_fc}{E_f-p_fc}
\right)=\frac{2c}{E_f} \label{appD.17}
\end {eqnarray}
in the limiting step. Of course, this term has to vanish because
$a_2\sim p_f$, but the concrete calculation after having integrated over
$\Phi$ and $\Theta_f$ is much more complicated.   Therefore
  we have   just   given an
example here. Similarly, all other terms cancel so
that only the limits of $\iota_i, \ i \in\{1,2,3\}$, stay.

\section {Discussion of Geant 4} \label{app_geant4}
As mentioned in the introduction, preimplemented cross sections for
Brems\-strahlung
can be found in the Geant 4 software library 
(Agostinelli et al., 2003; geant4.cern.ch). 
Geant 4 contains data for the total cross
section $\sigma$, the singly differential cross section $d\sigma/d\omega$ and a
singly   differential cross section   $d\sigma/ d\Omega_i$
depending on $\Theta_i$, but not on $\omega$.\\ \indent
The singly differential cross section $d\sigma/d\omega$ by Bethe and Heitler
is appropriate for small $Z$; it is (Bethe and Heitler, 1934; Heitler, 1944)
\begin {eqnarray}
\frac{d\sigma}{d\omega}(E_i,\omega)=\chi_0\frac{1}{\omega}\frac{p_f}{p_i}\left(\chi_1+L\chi_2\right)
\label{bethe.1}
\end {eqnarray}
with
\begin {eqnarray}
\chi_0&=&\frac{Z^2r_0^2}{137}, \label {bethe.2} \\
L&=&\ln\left(\frac{p_i^2+p_ip_f-\frac{E_i\cdot\hbar\omega}{c^2}}{p_i^2-p_ip_f-\frac{E_i\cdot\hbar\omega}{c^2}}\right)
,\label{bethe.3}\\
\epsilon_0&=&2\ln\left(\frac{E_i+cp_i}{m_ec^2}\right), \label{bethe.4} \\
\epsilon&=&2\ln\left(\frac{E_f+cp_f}{m_ec^2}\right), \label{bethe.5}\\
\chi_1&=&\frac{4}{3}-2\frac{E_iE_f}{c^2}\frac{p_f^2+p_i^2}{p_i^2p_f^2}+m_e^2c^2\left(\frac{\epsilon_0
E_f}{cp_i^3}+\frac{\epsilon E_i}{cp_f^3}-\frac{\epsilon_0\epsilon}{p_ip_f}
\right), \label{bethe.6}\\
\chi_2&=&\frac{8}{3}\frac{E_iE_f}{c^2p_ip_f}+\left(\frac{\hbar}{c}\omega\right)^2
\frac{1}{p_i^3p_f^3}\left(\frac{E_i^2E_f^2}{c^4}+p_i^2p_f^2\right) \nonumber \\
&+&\frac{m_e^2c\hbar\omega}{2p_ip_f}\left(\frac{\frac{E_iE_f}{c^2}+p_i^2}{p_i^3}\epsilon_0-
\frac{\frac{E_iE_f}{c^2}+p_f^2}{p_f^3}\epsilon+2\frac{\hbar}{c^3}\frac{\omega
E_i E_f}{p_f^2 p_i^2}\right)
\end {eqnarray}
with the quantities as described in section \ref{brems_def}.\\ \indent
Geant 4 uses a fit formula which is appropriate for large $Z$; it is
(Agostinelli et al., 2003)
\begin {eqnarray}
\frac{d\sigma}{d\omega}(E_i,\omega)=\frac{S\left(\frac{\hbar\omega}{E_{kin,i}},\omega\right)
}{C\omega}
\label{geant.1}
\end {eqnarray}
where $C$ is a constant which is not specified in the Geant 4
documentation, nor in the source code. $S$ is defined as
\begin {eqnarray}
S\left(\frac{\hbar\omega}{E_{kin,i}},\omega\right)=\left\{\begin{array}{l}
1+a_l\frac{\hbar\omega}{E_{kin,i}}+b_l\left(\frac{\hbar\omega}{E_{kin,i}}\right)^2, E_{kin,i}<1
\ \textnormal{MeV} \\
1-a_h\frac{\hbar\omega}{E_i}F_1+b_h\left(\frac{\hbar\omega}{E_i}\right)^2F_2,
E_{kin,i}\ge1\ \textnormal{MeV} \end{array}\right. \label{geant.2}
\end {eqnarray}
where $E_{kin,i}=E_i-m_ec^2$ is the kinetic energy of the incident electron.
$F_1$ and $F_2$ are defined as
\begin {eqnarray}
F_1&=&\left\{\begin{array}{l} F_0(42.392-7.796\delta+1.961\delta^2-F),\
\delta\le 1\\
F_0(42.24-8.368\ln(\delta+0.952)-F),\ \delta>1\end{array}\right. , \label{geant.3}\\
F_2&=&\left\{\begin{array}{l}F_0(41.734-6.484\delta+1.250\delta^2-F),\
\delta\le 1\\
F_0(42.24-8.368\ln(\delta+0.952)-F),\ \delta >1\end{array}\right. \label{geant.4}
\end {eqnarray}
with $F=4\ln(Z)-0.55(\ln(Z))^2$, $F_0=1/(42.392-F)$ and
$\delta=136m_ec^2\epsilon/(Z^{1/3}E_i(1-\epsilon))$ where
$\epsilon=\hbar\omega/E_i$ is the ratio between the photon energy and the
total energy of the incident electron.\\ \indent
$a_h,b_h,a_l$ and $b_l$ in (\ref{geant.2}) are defined as
\begin {eqnarray}
a_h&=&1+\frac{a_{h1}}{u}+\frac{a_{h2}}{u^2}+\frac{a_{h3}}{u^3}
, \label{geant.5}\\
b_h&=&0.75+\frac{b_{h1}}{u}+\frac{b_{h2}}{u^2}+\frac{b_{h3}}{u^3}
, \label{geant.6}\\
a_l&=&a_{l0}+a_{l1}u+a_{l2}u^2, \label{geant.7}\\
b_l&=&b_{l0}+b_{l1}u+b_{l2}u^2, \label{geant.8}
\end {eqnarray}
with $u=\ln\left(E_{kin,i}/(m_ec^2)\right)$. The $a_{hi},b_{hi},a_{li},b_{li}$
are directly defined in the Geant 4 source code as
\begin {eqnarray}
a_{hj}&=&a_{hj,0}[Z(Z+1)]^{\frac{1}{3}}\left(a_{hj,1}+[Z(Z+1)]^{\frac{1}{3}}a_{hj,2}\right),\
j\in\{1,2,3\}, \nonumber \\  \label{geant.9} \\
b_{hj}&=&b_{hj,0}[Z(Z+1)]^{\frac{1}{3}}\left(b_{hj,1}+[Z(Z+1)]^{\frac{1}{3}}b_{hj,2}\right),\
j\in\{1,2,3\}, \nonumber \\ \label{geant.10} \\
a_{lj}&=&a_{lj,0}[Z(Z+1)]^{\frac{1}{3}}\left(a_{lj,1}+[Z(Z+1)]^{\frac{1}{3}}a_{lj,2}\right),\
j\in\{1,2,3\}, \label{geant.11} \\
b_{lj}&=&b_{lj,0}[Z(Z+1)]^{\frac{1}{3}}\left(b_{lj,1}+[Z(Z+1)]^{\frac{1}{3}}b_{lj,2}\right),\
j\in\{1,2,3\}, \label{geant.12} 
\end {eqnarray}
where all the coefficients are also defined in the source code:\\
\begin {eqnarray}
(a_h)_{i,j}&=&\left(\begin{array}{ccc}4.67733 & -0.619012 & -0.020225\\
-7.34101 & 1.00462 & -0.0320985\\
2.93119 & -0.403761 & 0.0125153 \end{array}\right), \label{geant.13}\\
(b_h)_{i,j}&=&\left(\begin{array}{ccc}4.23071 & -6.10995 & -0.0195531\\
-7.12527 & 0.969160 & -0.0274255\\
2.69925 & -0.363283 & -0.00955316\end{array}\right), \label{geant.14}\\
(a_l)_{i,j}&=&\left(\begin{array}{ccc}-2.05398 & 0.0238815 & 0.000525483 \\
-0.0769748 & -0.0691499 & 0.00222453\\
0.0406463 & -0.0101281 & 0.000340919\end{array}\right), \label{geant.15}\\
(b_l)_{i,j}&=&\left(\begin{array}{ccc}1.04133 & -0.00943291 & -0.000454758\\
0.119253 & 0.0407467 & -0.00130718\\
-0.0159391 & 0.00727752 & -0.000194405\end{array}\right), \label{geant.16}
\end {eqnarray}
with $i\in\{1,2,3\}$ and $j\in\{0,1,2\}$.
In (\ref{geant.13}) - (\ref{geant.16}) the first index denotes columns, the
second one denotes rows.\\ \indent
\begin {figure}
\includegraphics [scale=0.56] {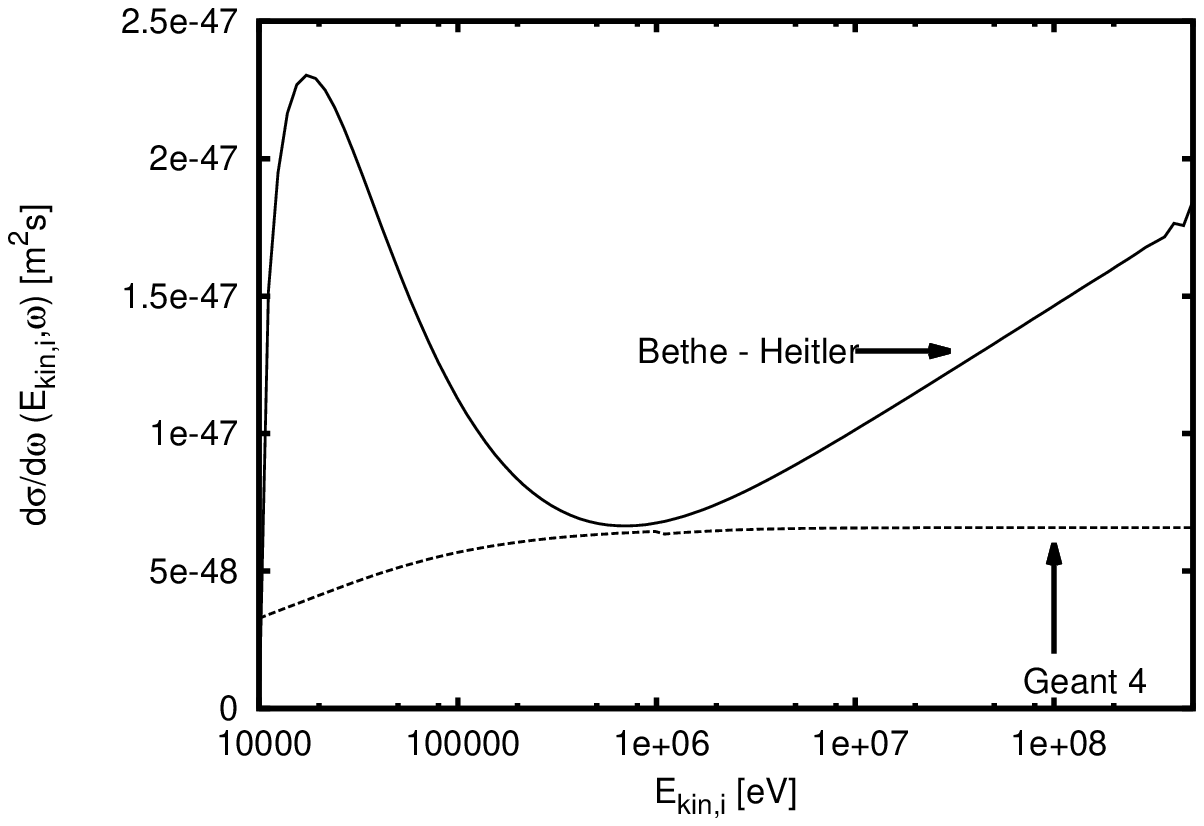}
\includegraphics [scale=0.56] {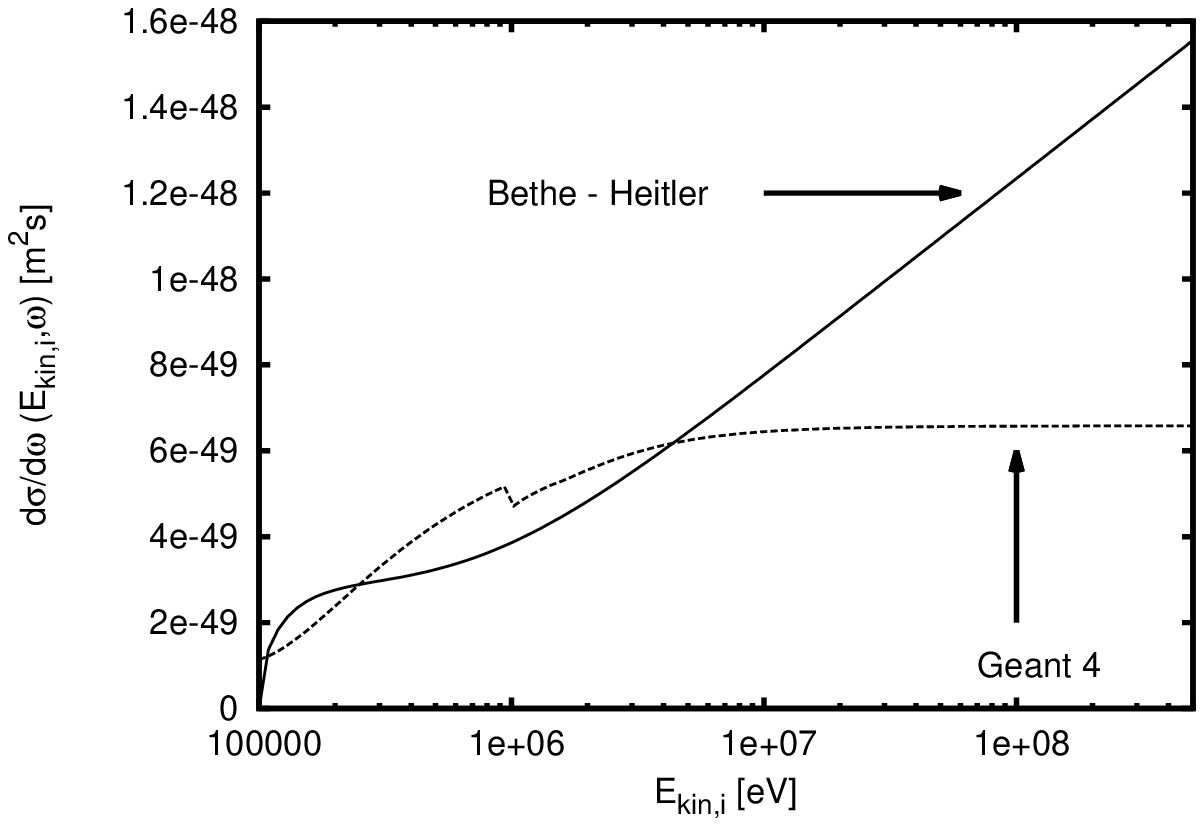}\\
a) $\hbar\omega=10$ keV \hspace{4.1cm} b) $\hbar\omega=100$ keV\\
\includegraphics [scale=0.56] {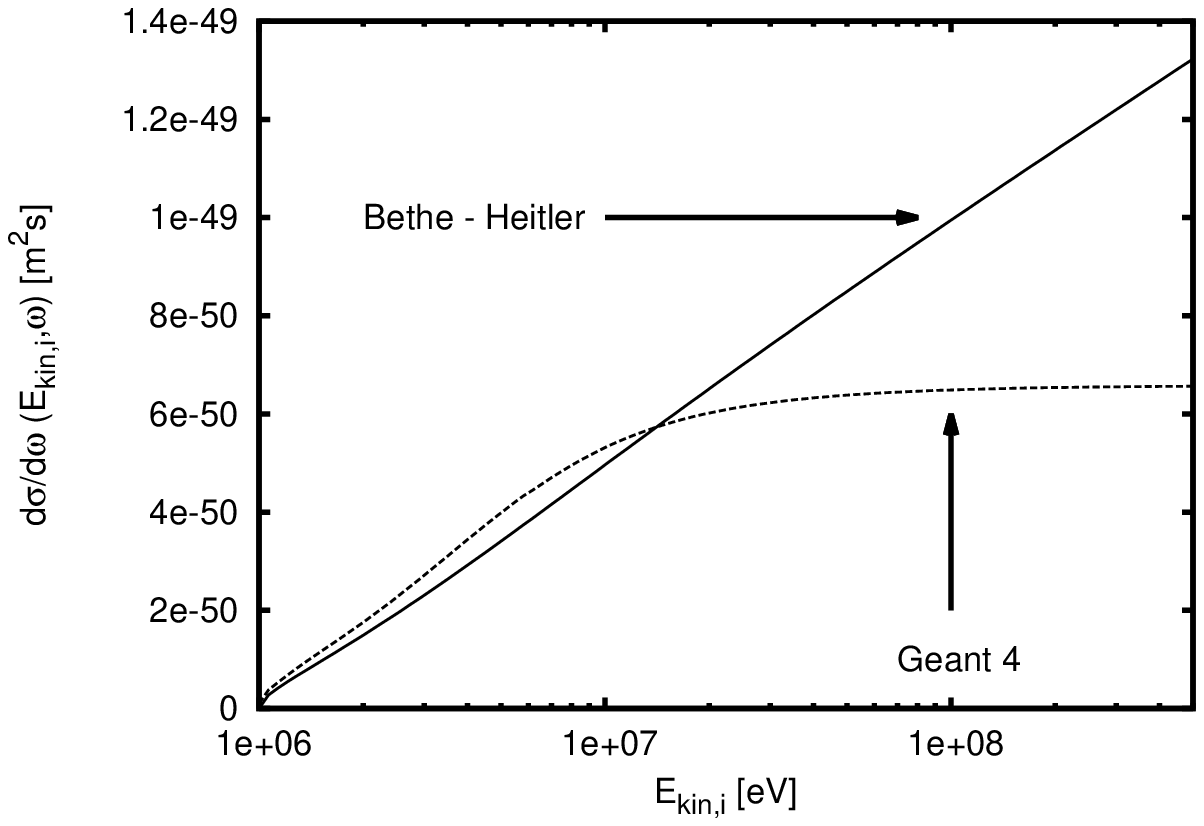}\\
c) $\hbar\omega=1$ MeV
\caption {The singly differential cross sections (\ref{bethe.1}) and
(\ref{geant.1}) as a function of the kinetic energy $E_{kin,i}$ of the
incident electron for $Z=7$ (nitrogen) and for fixed photon energy a)
$\hbar\omega=10$ keV, b) $\hbar\omega=100$ keV and c) $\hbar\omega=1$ MeV.} \label {comp.fig}
\end {figure}
Figure \ref{comp.fig} compares the Bethe Heitler cross section
(\ref{bethe.1}) with that of Geant 4 (\ref{geant.1}) where we have chosen
$C=10^{28}$ for all energies
in such a way that the orders of magnitude of (\ref{bethe.1}) and
(\ref{geant.1}) agree with each other. It shows that (using exactly the values
provided in the source code of Geant 4) that there is a good quantitative and
qualitative agreement for electron energies of $\approx 1$ MeV and $\approx 10$
MeV. But above and below that, both cross sections certainly differ.\\
\indent
That is because Geant 4 was developed for high energy energy physics in
particle accelerators and thus for high atomic numbers. Thus the cross
sections used in Geant 4 are not appropriate to describe the production of Bremsstrahlung
photons in air. The Bethe - Heitler 
theory for the energy range we consider, is used for small atomic numbers.\\
\indent
Geant 4 also includes dielectric suppression, i.e. the suppression of the emission of lowly
energetic photons because of their interaction with the electrons of the background medium
(Ter-Mikaelian,\ 1954), and the Landau-Pomeranchuk-Migdal (LPM)
effect (Landau and Pomeranchuk,\ 1953), i.e. the suppression of photon
production due to the multiple scattering of electrons.\\ \indent
The influence of the dielectric effect can be estimated by
\begin {eqnarray}
S(\hbar\omega)=\frac{(\hbar\omega)^2}{(\hbar\omega)^2+\frac{\hbar^2 E_i^2
n_e e^2}{m_e^3c^2\epsilon_0}} \label
{dielec.1}
\end {eqnarray}
where $n_e$ is the density of free electrons. For densities between
$10^{20}$ m$^{-3}$ and $10^{25}$ m$^{-3}$, $S$ is almost $1$.
\begin {figure}
\includegraphics [scale=0.56] {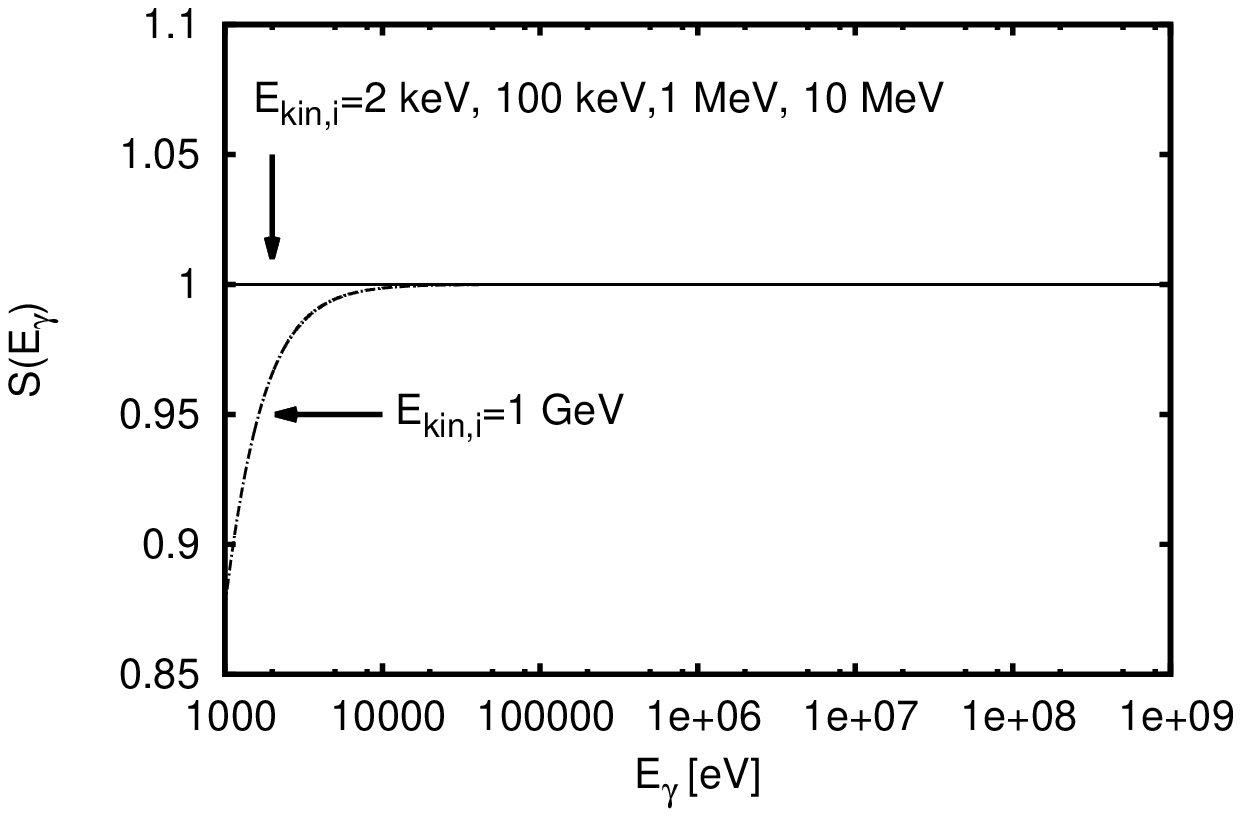}
\includegraphics [scale=0.56] {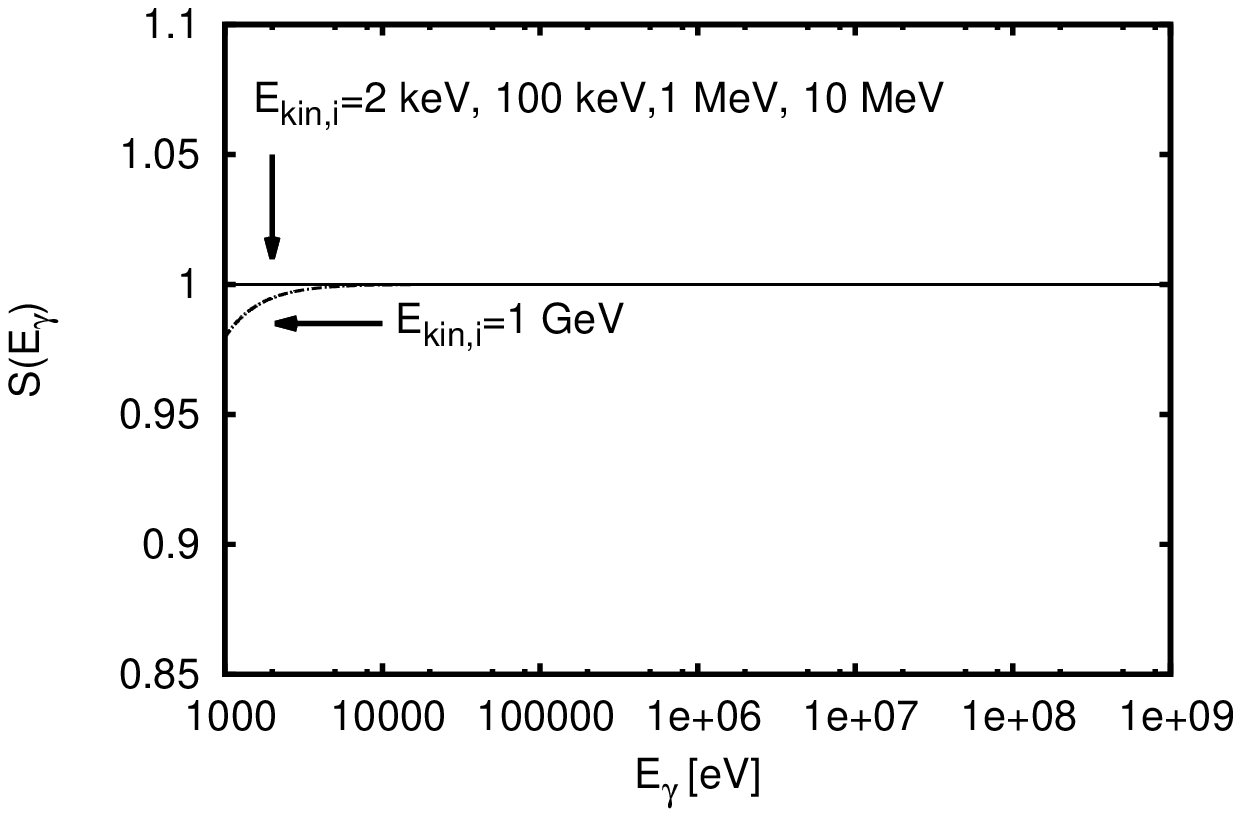}
a) $n_e\approx 2\cdot 10^{25}$ m$^{-3}$  \hspace{3.5cm} b)
$n\approx10^{24}$ m$^{-3}$
\caption {The dielectric factor $S$ (\ref{dielec.1}) vs. the photon energy for different
electron energies for a) $n_e\approx 2\cdot 10^{25}$ m$^{-3}$ and b) $n_e\approx
10^{24}$ m$^{-3}$} \label{dielec_fig}
\end {figure}
Figure \ref{dielec_fig} shows (\ref{dielec.1}) for different photon
energies, electron energies and densities. Dielectric suppression has a
very small effect when $E_{kin,i}\approx 1$ GeV;
thus it can be neglected.\\
\indent
The LPM effect is not important, either. The LPM threshold
energy is $\approx 10^{19}$ eV (Bertou et al.,\ 2000); this is much higher
than typical energies of electrons in the atmosphere.\\ \indent
The preimplemented cross sections used in Geant 4 are supposed to be used
for high electron energies $\gtrsim 1$ MeV and high atomic numbers $Z$. In the
case of TGFs it is necessary to treat electron energies in the keV and MeV
range and small atomic numbers where the LPM effect and dieletric suppression are not
significant.

\section {Comparison with Lehtinen (2000)} \label {app_leht}
\begin {figure}
\includegraphics [scale=0.56] {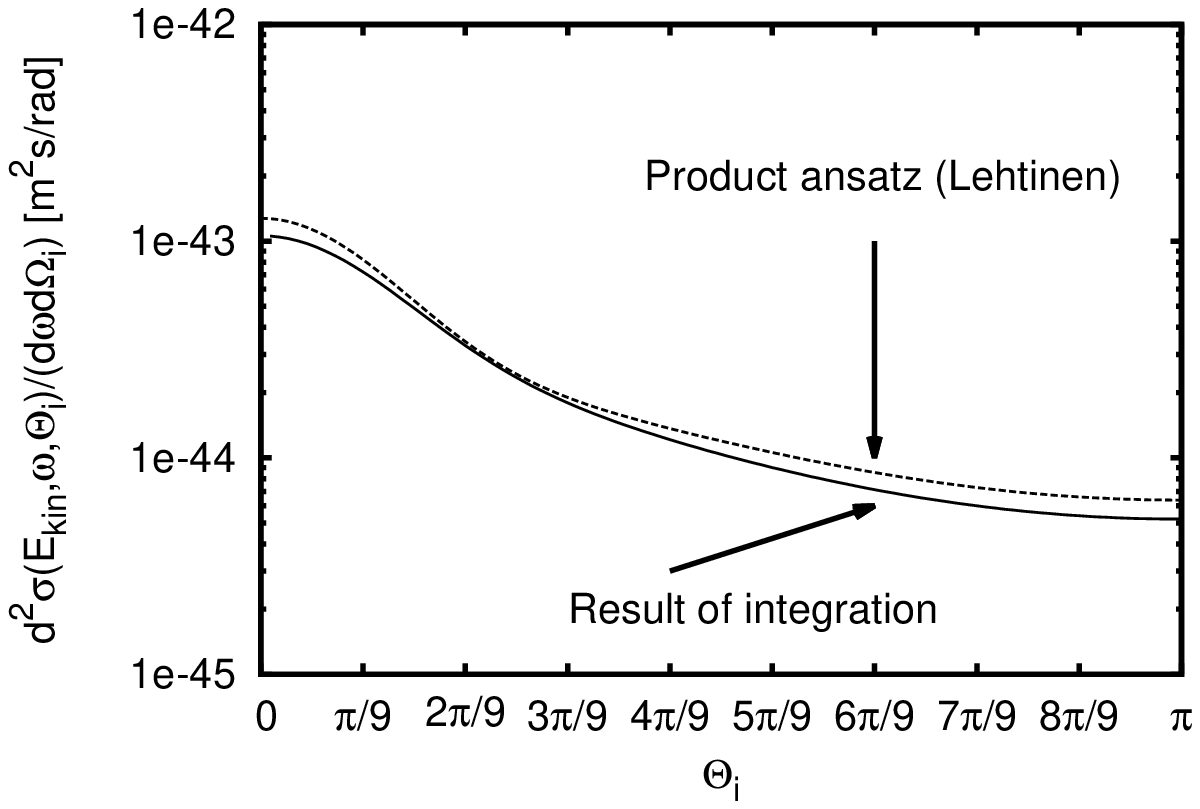}
\includegraphics [scale=0.56] {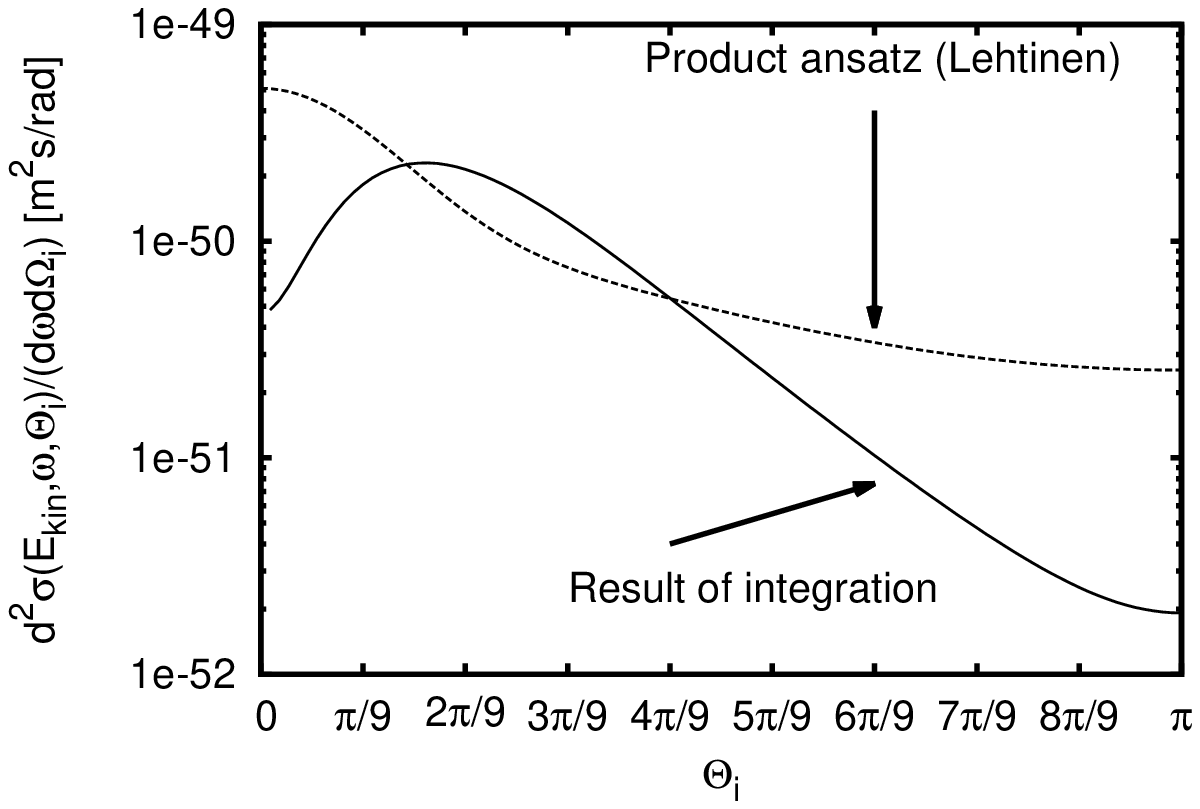}\\
a) $E_{kin,i}=150$ keV, $\frac{\hbar\omega}{E_{kin,i}}=10^{-5}$
\hspace{1.0cm} b) $E_{kin,i}=150$ keV, $\frac{\hbar\omega}{E_{kin,i}}=0.9$\\
\includegraphics [scale=0.56] {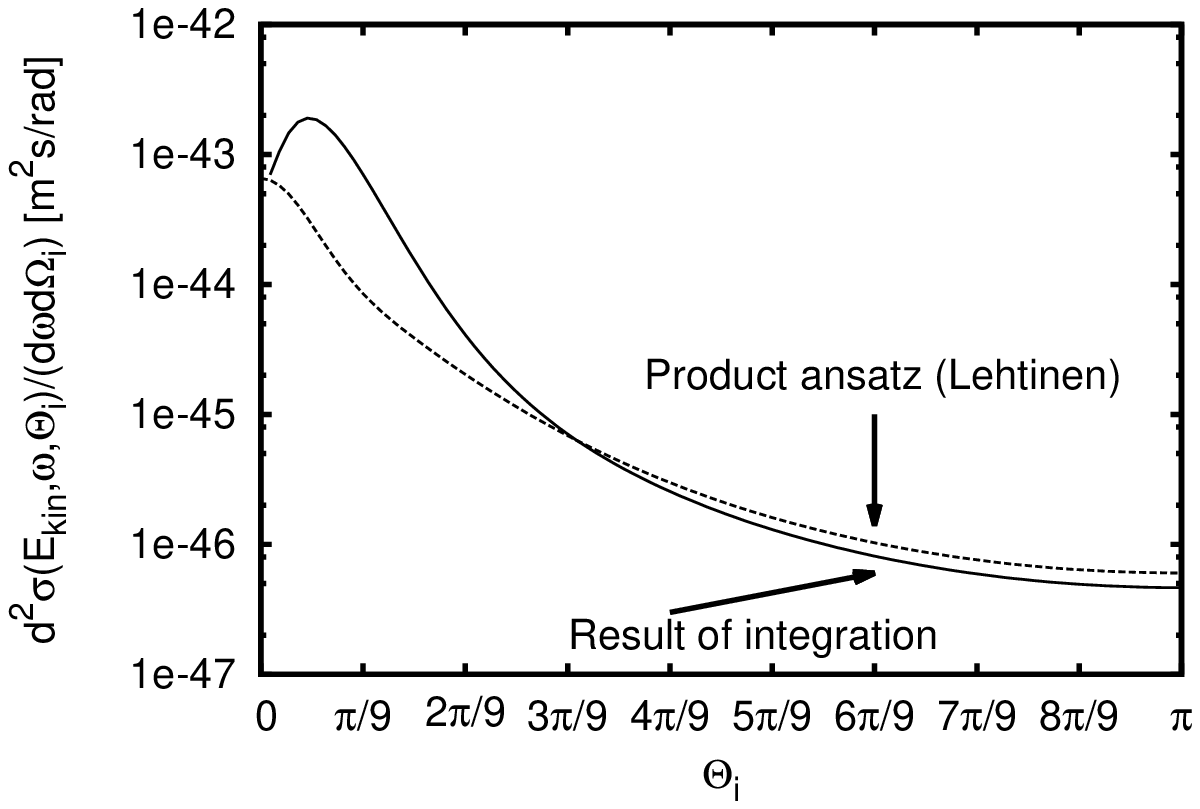}
\includegraphics [scale=0.56] {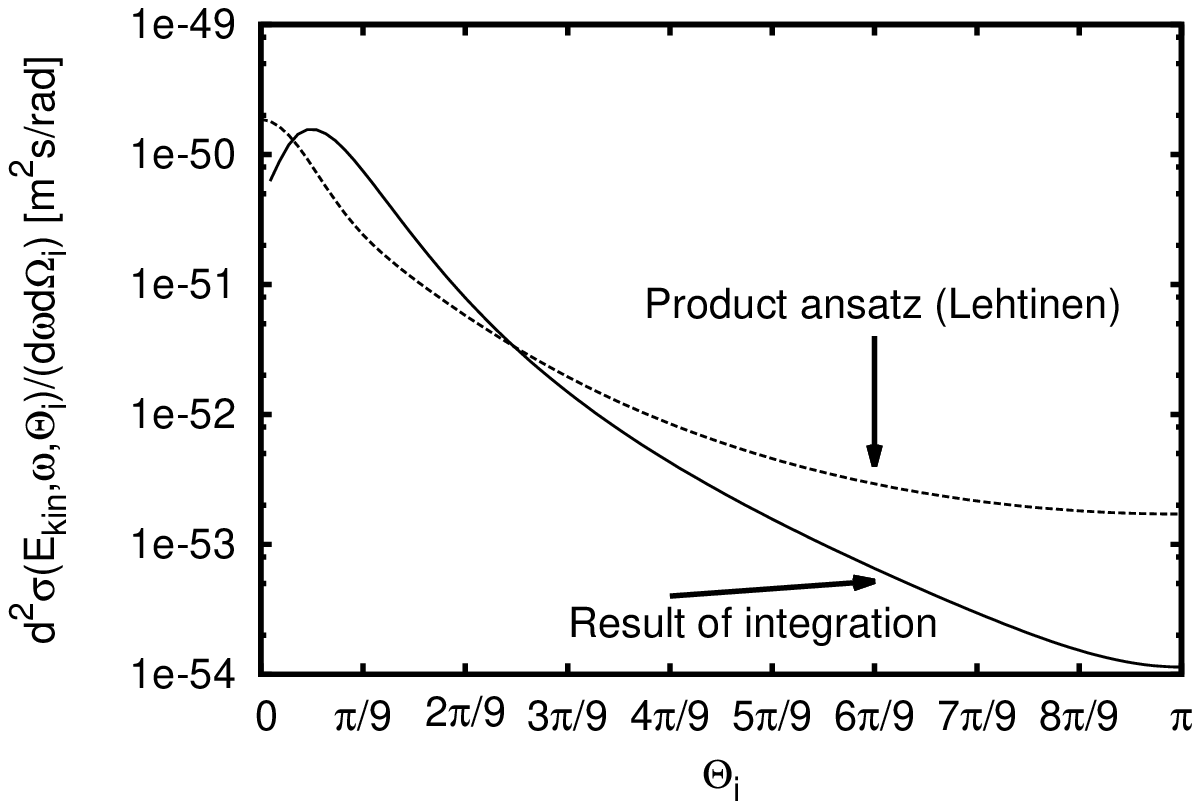}\\
a) $E_{kin,i}=1$ MeV, $\frac{\hbar\omega}{E_{kin,i}}=10^{-5}$
\hspace{1.3cm} b) $E_{kin,i}=1$ MeV, $\frac{\hbar\omega}{E_{kin,i}}=0.9$\\
\caption {Comparison of the product ansatz from Lehtinen (2000) with our
result (\ref{theta.16}) of the integration of (\ref{bcs.1}) for different
electron energies ($Z=7$): doubly differential cross section versus the
scattering angle $\Theta_i$ between incident electron and emitted photon.
The ratio between the photon energy $\hbar\omega$ and the
kinetic electron energy $E_{kin,i}$ is fixed to 0.001\% and 90\%.
The Born approximation (\ref{born.2}) is valid in all cases.
}  \label{leht.fig}
\end {figure}

Figure \ref{leht.fig} shows the comparison of (\ref{theta.16}) and
the doubly differential cross section used by Lehtinen. 
Lehtinen uses a product ansatz for the angular and the
frequency part; here the angular part is a non-quantum mechanical expression
taken from \cite{jackson}.
This cross section is only valid if $\hbar\omega \ll E_{i}$. 
There is a good agreement for low ratios between photon and electron energy, 
but a large deviation for larger ratios. 
Therefore this cross section is not appropriate for high ratios
needed to obtain photons with energies up to several tens of MeV to
determine the high energy tail of the TGF spectrum where almost all electron
energy is converted into photon energy.

\section {Contribution of the atomic form factor} \label{app_form}
Dwyer (2007) uses the triply differential cross section by Bethe and Heitler (1934),
but with an additional form factor $F(\mathbf{q})$ parameterizing the structure of the nucleus~\cite{brems_theory}.
$F$ is defined as
\begin {eqnarray}
F(\mathbf{q}):=-\frac{1}{Ze}\int d^3\mathbf{r}\varrho(\mathbf{r})e^{-\frac{i}{\hbar}
\mathbf{q}\cdot\mathbf{r}}  \label{form.1}
\end {eqnarray}
where $Z$ is the atomic number and $\varrho$ the charge density
\begin {eqnarray}
\varrho(\mathbf{r})=Ze\delta(\mathbf{r})-\frac{Ze}{4\pi a^2r}e^{-\frac{r}{a}}
\label {form.2}
\end {eqnarray}
with $a=111{\lambda\!\!\!/} Z^{-1/3}$ where ${\lambda\!\!\!/}=\lambda/(2\pi)$ is the
reduced Compton wave length of the electron. The
delta function describes the nucleus itself and the Debye term describes the
electrons of the shell. Performing the Fourier transformation in
(\ref{form.1}) gives
\begin {eqnarray}
F(\mathbf{q})=\frac{\mathbf{q}^2}{\mathbf{q}^2+\frac{\hbar^2}{a^2}}
\label{form.3}
\end {eqnarray}
with $\mathbf{q}$ as in Eq. (\ref{bcs.7}).
We calculated the value of $F(\mathbf{q})$ for different angles, electron and photon
energies [a) $E_{kin,i}=100$ keV,
$\hbar\omega=10$ keV, $\Theta_f=37^{\circ},\Phi=87^{\circ}$; b) $E_{kin,i}=100$ keV,
$\hbar\omega=80$ keV, $\Theta_f=62^{\circ},\Phi=43^{\circ}$; c)
$E_{kin,i}=10$ MeV, $\hbar\omega=1$ MeV, $\Theta_f=12^{\circ},\Phi=31^{\circ}$
and d) $E_{kin,i}=50$ MeV, $\hbar\omega=10$ MeV,
$\Theta_f=52^{\circ},\Phi=90^{\circ}$].
In all these cases the atomic form factor is 1. Hence, it can be neglected. As it
makes the integration over $\Phi$ and $\Theta_f$ more complicated, it is
useful not to use $F(\mathbf{q})$.

\section {Contribution of the integrals} \label{app_contr}
As   equation   (\ref{theta.16}) is rather complicated, it is interesting to see which
  terms   have the most important contribution.
\begin {figure}
\includegraphics [scale=0.5] {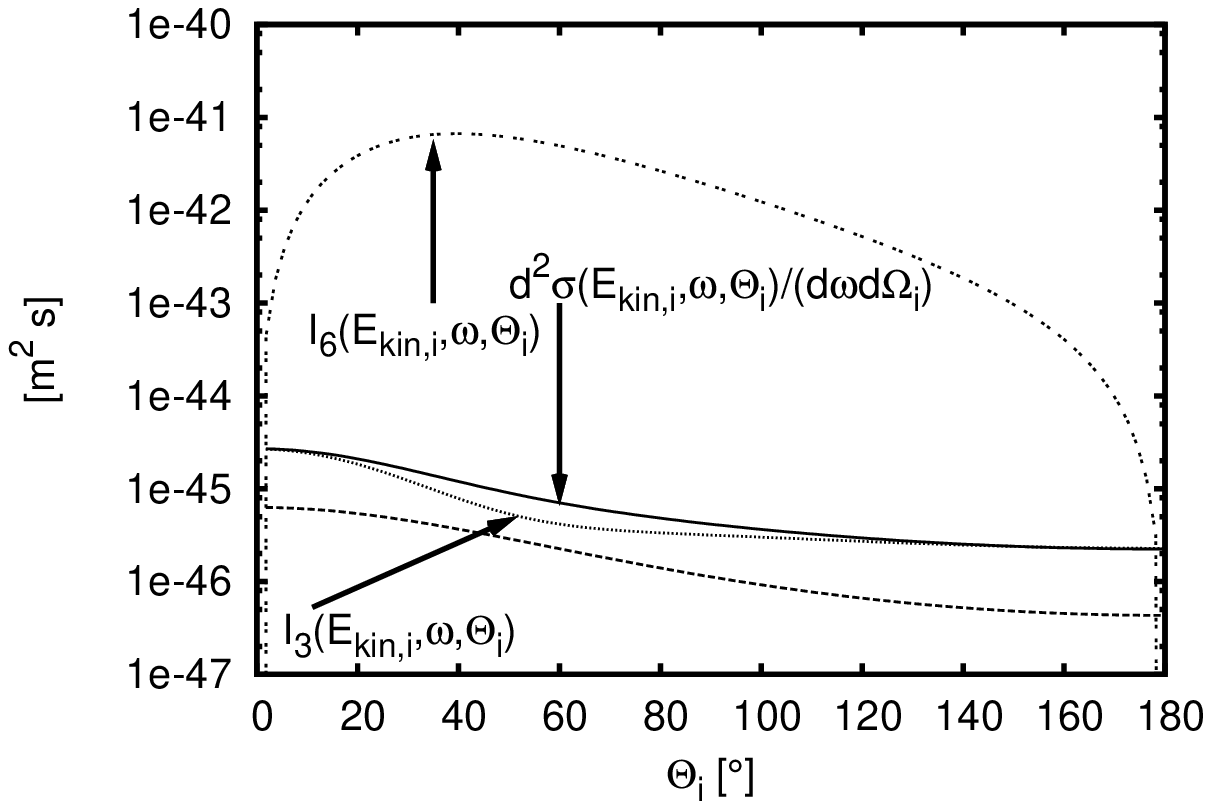}
\includegraphics [scale=0.5] {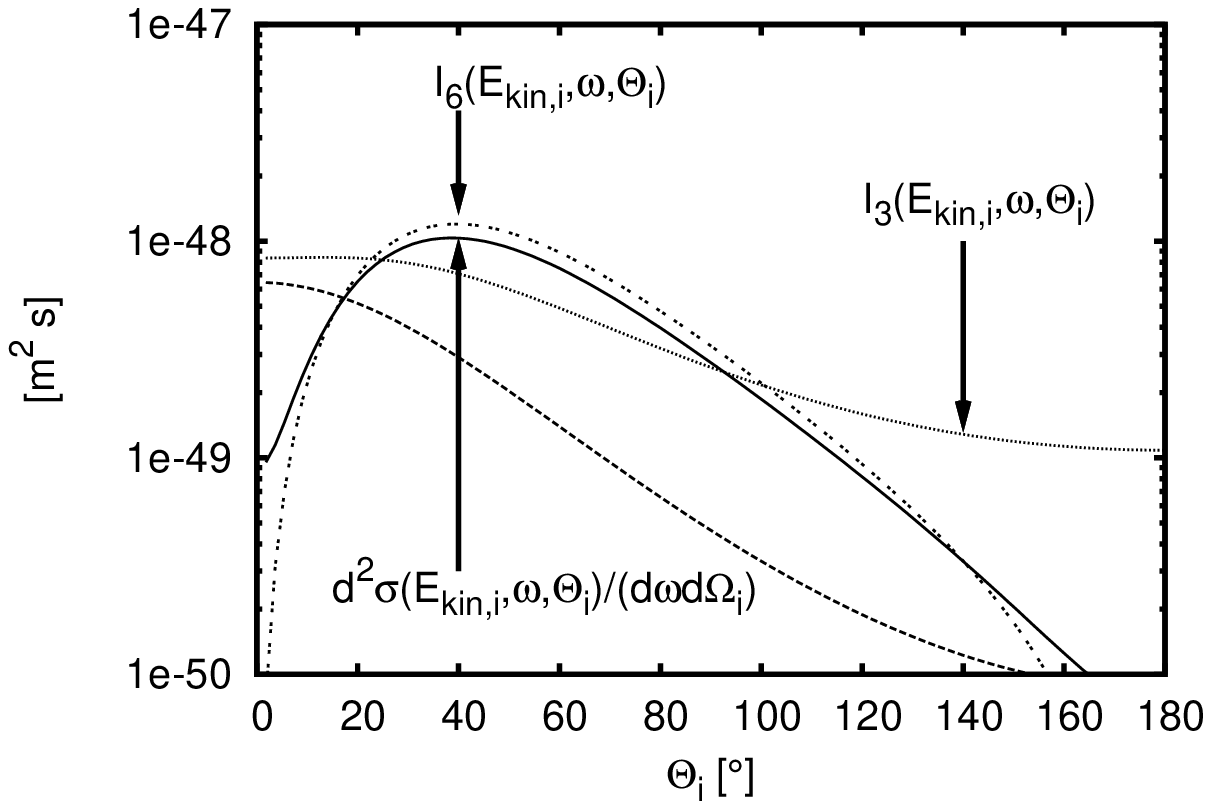}\\
a) $E_{kin,i}=100$ keV, $\hbar\omega=1$ keV \hspace{0.6cm} b) $E_{kin,i}=100$
keV, $\hbar\omega=95$ keV\\
\includegraphics [scale=0.5] {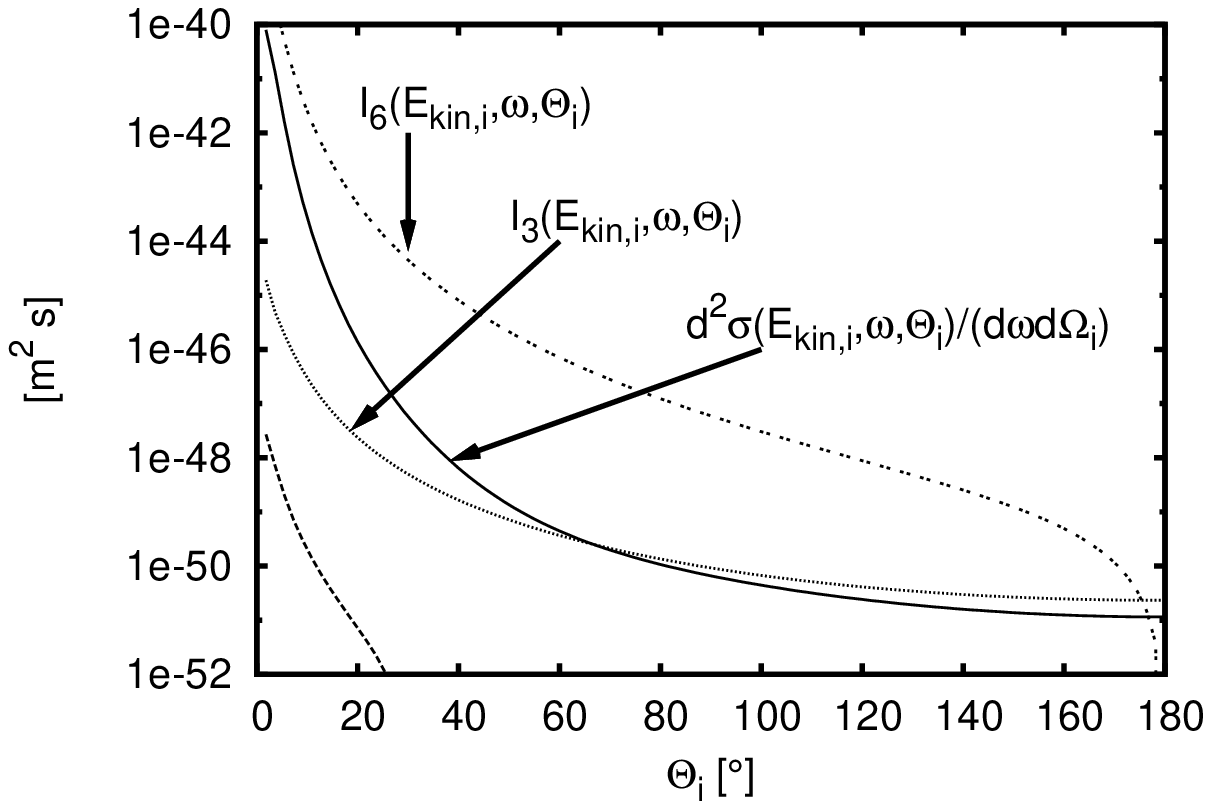}
\includegraphics [scale=0.5] {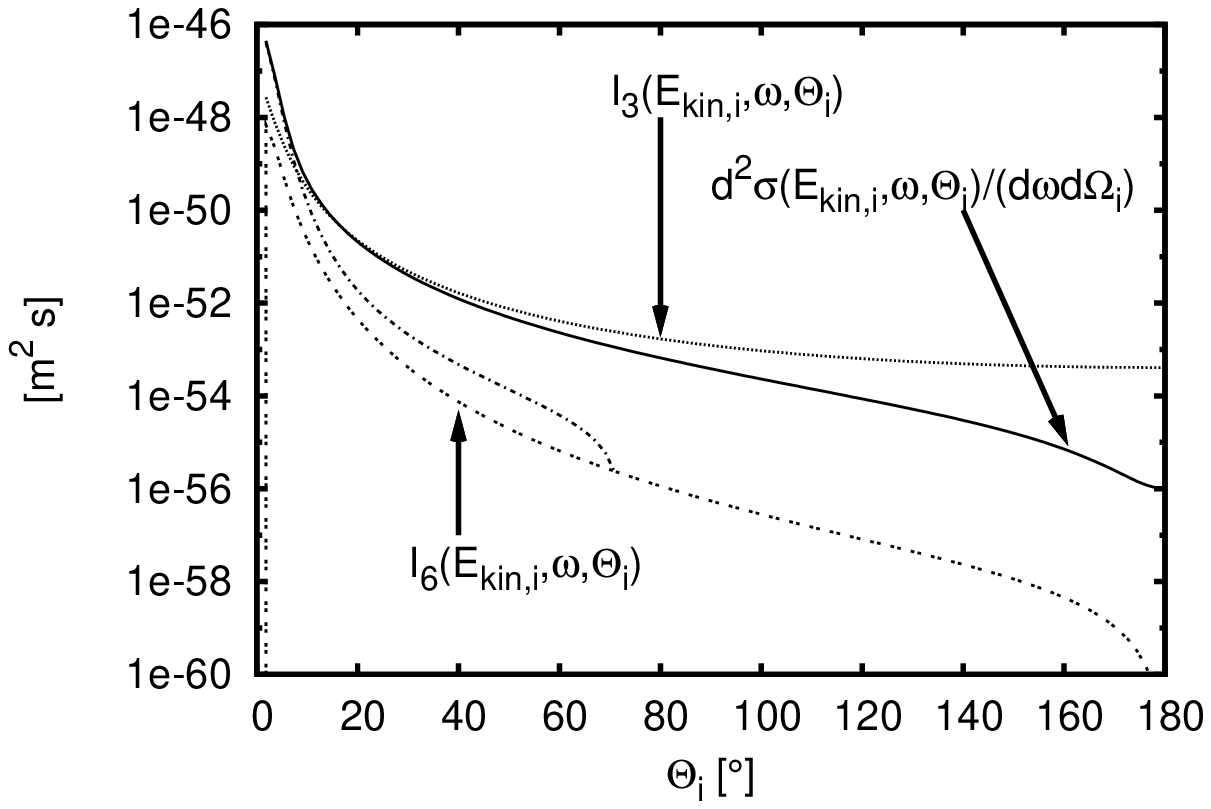}\\
c) $E_{kin,i}=10$ MeV, $\hbar\omega=100$ keV \hspace{0.5cm} d) $E_{kin,i}=10$
MeV, $\hbar\omega=9.5$ MeV
\caption {Contribution of (\ref{theta.17}) - (\ref{theta.22}) to
(\ref{theta.16}) in a semilog plot for different electron and photon
energies ($Z=7$).} \label{contr_fig.1}
\end {figure}
\begin {figure}
\includegraphics [scale=0.5] {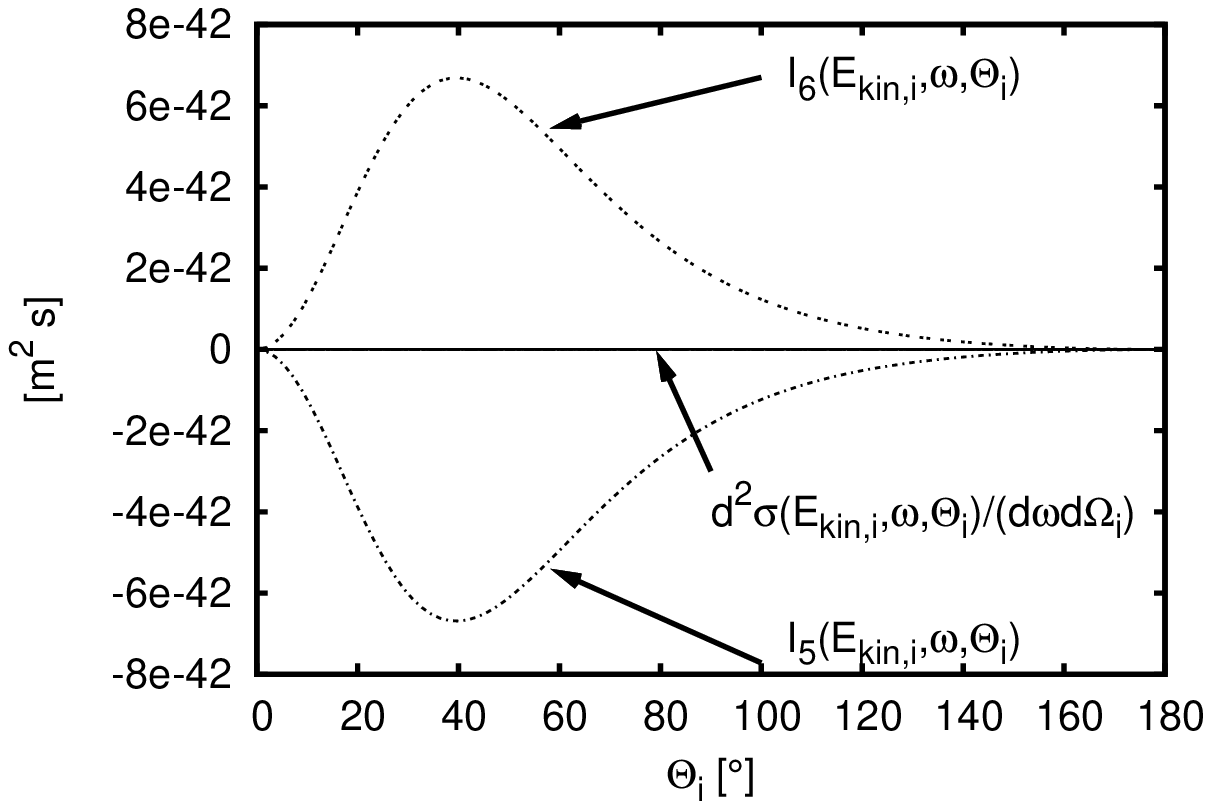}
\includegraphics [scale=0.5] {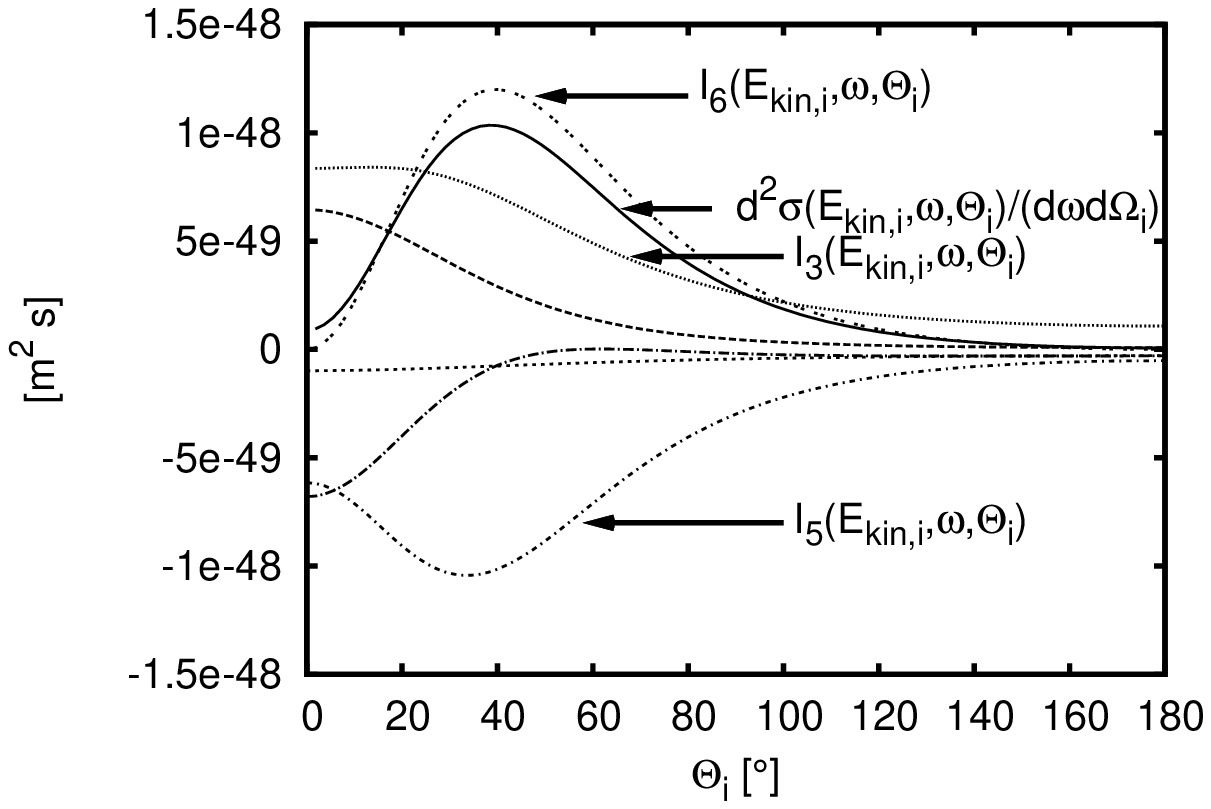}\\
a) $E_{kin,i}=100$ keV, $\hbar\omega=1$ keV \hspace{0.6cm} b) $E_{kin,i}=100$
keV, $\hbar\omega=95$ keV\\
\includegraphics [scale=0.5] {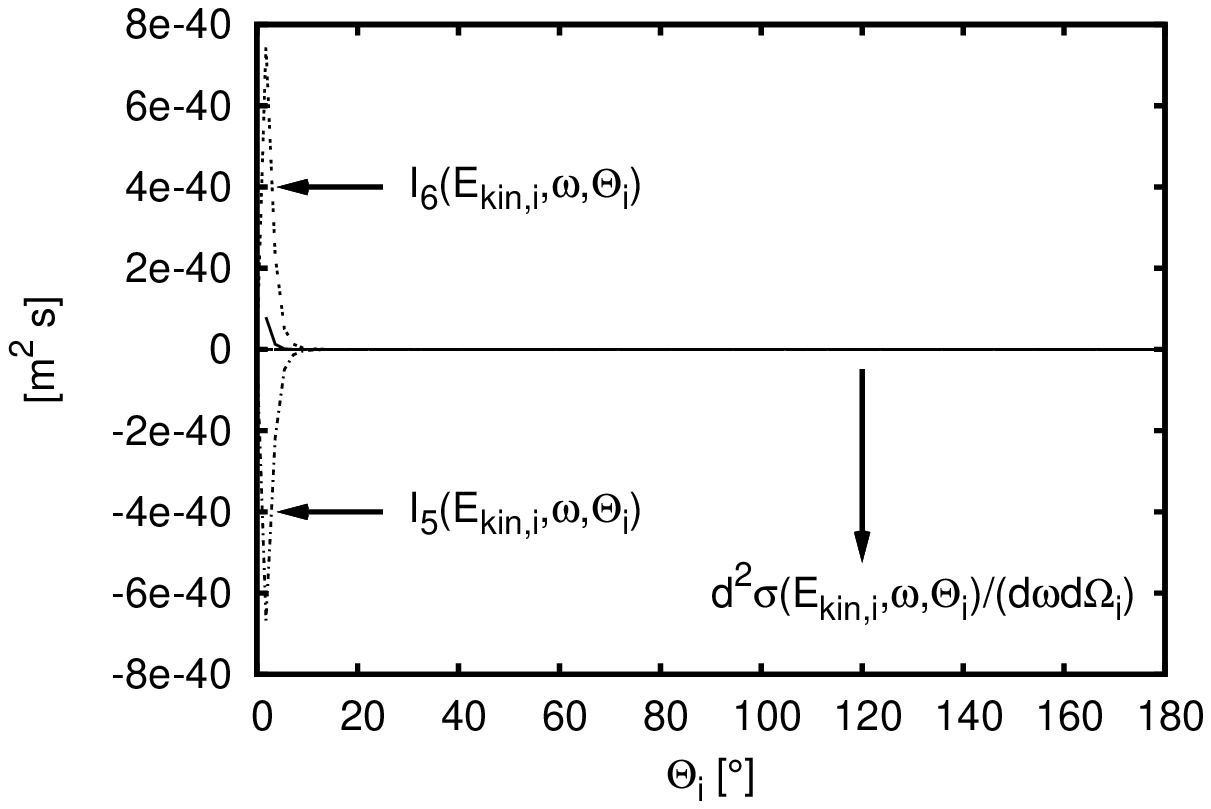}\\
c) $E_{kin,i}=10$ MeV, $\hbar\omega=100$ keV
\caption {Contribution of (\ref{theta.17}) - (\ref{theta.22}) to
(\ref{theta.16}) in a linear plot for different electron and photon
energies ($Z=7$).} \label{contr_fig.2}
\end {figure}
Figure \ref{contr_fig.1} shows the contribution of all parts to the final
result in a logarithmic scale while Fig. \ref{contr_fig.2} shows the same in a linear
scale. In all cases, i.e. low and high electron energies and low and high
ratios between $\hbar\omega$ and $E_{kin,i}$, the main contribution
comes from (\ref{theta.19}). It is important to state that not all
contributions can be seen in figure \ref{contr_fig.1} because some of the
terms have negative values which, however, are shown in figure \ref{contr_fig.2}. So
one might think that for $E_{kin,i}=100$ keV and $\hbar\omega=1$ keV,
   equation   (\ref{theta.22}) has the largest contribution, but as figure
\ref{contr_fig.2} shows, (\ref{theta.21}) has nearly the same absolute
value, but   opposite   sign;   therefore
  they cancel. Thus the third integral
(\ref{theta.19}) is the most important one. The same holds for other
electron  energies and ratios between $\hbar\omega$ and $E_{kin,i}$.
We   conclude   that (\ref{theta.19}) is the
dominant   contribution for all   relevant parameter
values.

\section {Conservation of energy and momentum}   \label{app_cons}
One can also gain information on
the scattering angle $\Theta_i$ for
high electron energies from the conservation of energy and
momentum,
\begin {eqnarray}
E_i+E_q&=&E_f+\hbar\omega, \label{cons.1} \\
\mathbf{p}_i+\mathbf{q}&=& \mathbf{p}_f+\hbar\mathbf{k} \label{cons.2}
\end {eqnarray}
where   $E_{i,f}$ and $p_{i,f}$ are the energy and the momentum of the electron in the initial and final
state.   $\hbar\mathbf{k}$ is the momentum of the photon which is
related   to its energy $\hbar\omega$   through

\begin {eqnarray}
\hbar|\mathbf{k}|=\frac{\hbar}{c}\omega, \label{cons.3}
\end {eqnarray}
and $E_q$ and $\mathbf{q}$ are the energy and the momentum of the virtual photon between electron and
nucleus. $\mathbf{q}$ changes the momentum of the nucleus. But the
contribution to the kinetic energy can be neglected; thus $E_q\equiv 0$ and
\begin {eqnarray}
E_i&=&E_f+\hbar\omega, \label{cons.5} \\
\mathbf{p}_i-\hbar\mathbf{k}&=&\mathbf{p}_f-\mathbf{q}. \label{cons.6}
\end {eqnarray}
Squaring (\ref{cons.6}) and using
$\mathbf{p}_i\cdot\mathbf{k}=|\mathbf{p}_i||\mathbf{k}| \cos\sphericalangle(\mathbf{p}_i,\mathbf{k})
=p_i\ k\cos\Theta_i$, the angle $\Theta_i$ is:
\begin {eqnarray}
\cos\Theta_i=\frac{\left(\mathbf{p}_f-\mathbf{q}\right)^2-p_i^2-\hbar^2k^2}{-2\hbar
p_ik}. \label {cons.7}
\end {eqnarray}
By using (\ref{cons.5}) and the relativistic energy-momentum relation
(\ref{bcs.2}) we get an expression for the momentum of the electron in
the final state
\begin {eqnarray}
p_f=\sqrt{p_i^2+\hbar^2k^2-2\hbar k\sqrt{p_i^2+m_e^2c^2}} \label{cons.8}
\end {eqnarray}
which leads to
\begin {eqnarray}
\cos\Theta_i=\frac{2\hbar\omega\frac{E_i}{c}+2q\sqrt{(E_i-\hbar\omega)^2-m_e^2c^4}
\cos\sphericalangle(\mathbf{p}_f,\mathbf{q})-cq^2}{2\hbar\omega p_i}. \label{cons.9}
\end {eqnarray}
Although this is an analytical expression for the scattering angle
$\Theta_i$ one should   take   into account that it depends on the vector
$\mathbf{q}$ of the virtual photon which is not known in forehand. Thus,
depending on $\mathbf{q}$, only a statistical statement can be made about
$\Theta_i$.\\ \indent
It is, however, possible to investigate the limit of (\ref{cons.9}) for high electron energies
which yields
\begin {eqnarray}
\lim_{E_i\rightarrow\infty}\cos\Theta_i=1+\frac{cq}{\hbar\omega}\cos
\sphericalangle(\mathbf{p}_f,\mathbf{q}) \label{cons.10}
\end {eqnarray}
As $\Theta_i\in\mathbb{R} \Leftrightarrow \cos\Theta_i\in[-1,+1]$ and
$c,q,\hbar\omega>0$ we can conclude that
\begin {eqnarray}
\cos\sphericalangle(\mathbf{p}_f,\mathbf{q})\le 0 \label{cons.11}
\end {eqnarray}
Especially   for   $|\frac{cq}{\hbar\omega}\cos\sphericalangle(\mathbf{p}_f,
\mathbf{q})|\ll 1$,$\Theta_i\approx 0$,   i.e.,
the photon is mainly emitted in forward direction.\\ \indent
If, additionally, the photon energy $\hbar\omega$ also increases more and more (for high
electron energies) it is
\begin {eqnarray}
\lim_{\hbar\omega\rightarrow\infty}\left(1+\frac{cq}{\hbar\omega}\cos
\sphericalangle(\mathbf{p}_f,\mathbf{q})\right)=1; \label{cons.13}
\end {eqnarray}
thus
\begin {eqnarray}
\lim_{E_i,\hbar\omega\rightarrow\infty}\Theta_i = 0.
\label{cons.14}
\end {eqnarray}
Hence, we conclude from simple considerations about energy and momentum
conservation that the photon is mainly scattered in forward direction if
  the   energies of electron and photon are
  both   very high.

\section {Approximation for $\Theta_i$} \label{app_theta}
In order to obtain (\ref{disc.9}) we calculate the derivative of
(\ref{limit.7}) after $\Theta_i$:
\begin {eqnarray}
&&\frac{\partial}{\partial\Theta_i}\left(\frac{d^2\sigma}{d\omega\sin\Theta_i
d\Theta_i}\right)=\frac{Z^2\alpha_{fine}^3\hbar^2}{\pi}\frac{|\mathbf{p}_f||\mathbf{p}_i|}
{\omega}\left[\frac{4\frac{\hbar}{c}\omega|\mathbf{p}_i|\sin\Theta_i}{\delta^3
(\Theta_i)}\right.\times\nonumber\\
&\times&\left(\frac{\sin^2\Theta_i}{(E_i-c|\mathbf{p}_i|\cos\Theta_i)^2}
(4E_f^2+\delta(\Theta_i)c^2)+\frac{2\hbar^2\omega^2}{E_f}\frac{\sin^2\Theta_i}
{E_i-c|\mathbf{p}_i|\cos\Theta_i}\right)\nonumber\\
&+&\frac{1}{\delta^2(\Theta_i)}\left(\frac{2\sin\Theta_i\cos\Theta_i(E_i-c
|\mathbf{p}_i|\cos\Theta_i)-2c|\mathbf{p}_i|\sin^3\Theta_i}{(E_i-c|\mathbf{p}_i|\cos\Theta_i)^3}
\right.\nonumber\\
&\times&(4E_f^2+\delta(\Theta_i)c^2)
-\frac{2\hbar
c\omega|\mathbf{p}_i|\sin^3\Theta_i}{(E_i-c|\mathbf{p}_i|\cos\Theta_i)^2}
\nonumber\\
&+&\left.\left.\frac{2\hbar^2\omega^2}{E_f}\frac{2\sin\Theta_i\cos\Theta_i(E_i-c|\mathbf{p}_i|
\cos\Theta_i)-c|\mathbf{p}_i|\sin^3\Theta_i}{(E_i-c|\mathbf{p}_i|\cos\Theta_i)^2}
\right)\right] \nonumber\\ \label{disc.5}
\end {eqnarray}
  with definition (\ref{limit.3}) for $\delta$.
In order to calculate the extrema one has to set equation (\ref{disc.5})
equal to zero:
\begin {eqnarray}
0&=&4\frac{\hbar}{c}\omega|\mathbf{p}_i|\left[\frac{\sin^2\Theta_i}{(E_i-c|\mathbf{p}_i|\cos\Theta_i)^2}
(4E_f^2+\delta(\Theta_i)c^2) \right.\nonumber\\
&+&\left.\frac{2\hbar^2\omega^2}{E_f}\frac{\sin^2\Theta_i}
{E_i-c|\mathbf{p}_i|\cos\Theta_i}\right]
(E_i-c|\mathbf{p}_i|\cos\Theta_i)^3 \nonumber\\
&+&\delta(\Theta_i)\left[2(E_i\cos\Theta_i
-c|\mathbf{p}_i|)(4E_f^2+\delta(\Theta_i)c^2) \right. \nonumber \\
&-&2\hbar c\omega|\mathbf{p}_i|\sin^2\Theta_i (E_i-c|\mathbf{p}_i|\cos\Theta_i)
\nonumber \\
&+&\frac{2\hbar^2\omega^2}{E_f}\left(2\cos\Theta_i (E_i-c|\mathbf{p}_i|\cos\Theta_i)^2
\right. \nonumber\\
&-&\left.\left.c|\mathbf{p}_i|\sin^2\Theta_i(E_i-c|\mathbf{p}_i|\cos\Theta_i)\right)\right].
\label{disc.6}
\end {eqnarray}
  As   $\delta(\Theta_i)\sim\cos\Theta_i$,
  expression   (\ref{disc.6}) is
quartic in $\cos\Theta_i$;   therefore   (\ref{disc.6}) can be solved analytically
in principle, but the solution will be long and complicated. Figure \ref{disc_fig.1} also shows
that the angles for maximal scattering are very small for relativistic
electrons,   therefore   one can
approximate $\cos\Theta_i\approx 1$ and $\sin\Theta_i\approx\Theta_i$. This
leads to
\begin {eqnarray}
\delta(\Theta_i)\approx -|\mathbf{p}_i|^2-\left(\frac{\hbar}{c}\omega\right)^2
+2\frac{\hbar}{c}\omega|\mathbf{p}_i|=\delta(\Theta_i=0)=:\delta_0
\label{disc_app.7}
\end {eqnarray}
and
\begin {eqnarray}
0&=&4\frac{\hbar}{c}\omega|\mathbf{p}_i|\left[\Theta_i^2(4E_f^2+\delta_0c^2)
+\frac{2\hbar^2\omega^2}{E_f}\Theta_i^2(E_i-c|\mathbf{p}_i|)\right] \nonumber\\
&+&\delta_0
\left[2(4E_f^2+\delta_0c^2)-2\hbar c\omega|\mathbf{p}_i|\Theta_i^2+
\frac{2\hbar^2\omega^2}{E_f}\left(2(E_i-c|\mathbf{p}_i|)-c|\mathbf{p}_i|
\Theta_i^2\right)\right] \nonumber \\ \label{disc_app.8}
\end {eqnarray}
with solution
\begin {eqnarray}
\Theta_i=\sqrt{\frac{-\frac{\delta_0}{\hbar\omega}(4E_f^2+\delta_0c^2)-\frac{2
\delta_0\hbar\omega}{E_f}(E_i-c|\mathbf{p}_i|)}{2\frac{|\mathbf{p}_i|}{c}\left[
4E_f^2+\delta_0c^2+\frac{2\hbar^2\omega^2}{E_f}(E_i-c|\mathbf{p}_i|)\right]
-|\mathbf{p}_i|\delta_0c-\frac{\hbar\omega}{E_f}c|\mathbf{p}_i|\delta_0}}
\label{disc_app.9}
\end {eqnarray}

\end {appendix}

\newpage

{\bf Vitae:}

{\bf Christoph K\"ohn}   studied physics in Kiel
and Hamburg, Germany from 2005 till 2010. After having finished his diploma thesis
on six-dimensional supergravity, he started his PhD studies at CWI Amsterdam, The Netherlands.
\\

{\bf Ute Ebert} studied physics at the University of Heidelberg,
Germany, and she wrote her PhD thesis at the University of Essen, Germany. After a
postdoc period at the University of Leiden, Netherlands, she became
staff member at CWI Amsterdam, Netherlands. Since 2002 she leads a
research group at CWI and is part time professor at Eindhoven University
of Technology. The research of her group concentrates on transient
electrical discharges, both in plasma technology and in atmospheric
electricity.
\\

\begin {itemize}
\item Analytical results on doubly differential cross-sections for typical
TGF parameters
\item Distribution of emission angles and energies for Bremsstrahlung
photons
\item Distribution of emission angles and energies for positrons in pair
production
\item C++ code with the analytical cross section results provided
\end {itemize}
 

\newpage

\begin {thebibliography}{XXX99}
\bibitem [Agostinelli et al.,\ 2003] {geant4} S. Agostinelli et al., 2003. G4-a simulation
toolkit. Nucl. Instrum. Methods Phys. Res., Sect. A, vol. 506, pp. 250-303
\bibitem [Aiginger,\ 1966] {brems_exp_1} H. Aiginger, 1966.
Elektron-Bremsstrahlungwirkungsquerschnitte von 180 und 380 keV-Elektronen.
Zeitschrift fuer Physik , vol. 197, pp. 8-25
\bibitem [Babich,\ 2003] {babich} L. P. Babich, 2003. High-energy phenomena in
electric discharges in dense gases. Theory, experiment and natural
phenomena. ISTC Science and technology series, vol. 2 Futurepast
\bibitem [Bertou et al.,\ 2000] {LPM} Xavier Bertou et al., 2000. LPM effect and
pair production in the geomagnetic field: a signature of ultra-high energy
photons in the Pierre Auger Observatory, Astro. Phys., vol. 14, pp. 121-130
\bibitem [Bethe and Heitler,\ 1934] {bethe} H. A. Bethe and W. Heitler,
1934. On the stopping of fast
particles and on the creation of positive electrons. Proc. Phys. Soc. London,
vol. 146, pp. 83-112
\bibitem [Briggs et al.,\ 2010] {fermi} M. Briggs et al., 2010. First results on terrestrial gamma ray
flashes from the Fermi Gamma-ray Burst Monitor. J. Geophys.
Res., vol. 115, A07323
\bibitem [Briggs et al.,\ 2011] {positron} M. S. Briggs et al., 2011. Electron-positron beams from
terrestrial lightning observed with Fermi GBM. Geophys. Res. Lett., vol. 38,
L02808
\bibitem [Carlson et al.,\ 2009] {stepping} B. E. Carlson, N. G. Lehtinen and U. S. Inan,
2009. Terrestrial gamma ray flashes production by lightning current pulses.
J. Geophys. Res., vol. 114, A00E08
\bibitem [Carlson et al.,\ 2010] {carlson} B. E. Carlson, N. G. Lehtinen and U. S. Inan,
2010. Terrestrial gamma ray flash production by active lightning
leader channels. J. Geophys. Res., vol. 115, A10324
\bibitem [Chanrion and Neubert,\ 2008] {chanrion_2} O. Chanrion and T.
Neubert, 2008. A PIC-MCC code for simulation of streamer propagation in air.
J. Comput. Phys., vol. 227, pp. 7222-7245
\bibitem [Chanrion and Neubert,\ 2010] {chanrion_1} O. Chanrion and T.
Neubert, 2010. Production of runaway electrons by negative streamer
discharges. J. Geophys. Res., vol. 115, A00E32
\bibitem [Cullen et al.,\ 1991] {EEDL} D.E. Cullen, S.T. Perkins and S.M. Seltzer,
1991. Tables and
Graphs of Electron Interaction Cross 10 eV to 100 GeV Derived from the LLNL
Evaluated Electron Data Library (EEDL), Z = 1 - 100. Lawrence Livermore
National Laboratory, UCRL-50400, vol. 31
\bibitem [Cummer et al.,\ 2005] {RHESSI_4} S. A. Cummer, Y. Zhai, W. Hu, D. M. Smith, L. I. Lopez,
M. A. Stanley, 2005. Measurements and implications of the relationship between
lightning and terrestrial gamma ray flashes. Geophys. Res. Lett.,
vol. 32, L08811
\bibitem [Dwyer,\ 2003] {dwyer_2003} J. R. Dwyer, 2003. A fundamental limit on electric fields
in air. Geophys. Res. Lett., vol. 30, no. 2055
\bibitem [Dwyer and Smith,\ 2005] {dwyer_smith} J. R. Dwyer and D. M. Smith,
2005. A comparison between Monte
Carlo simulations of runaway breakdown and terrestrial gamma-ray flashes.
Geophys. Res. Lett., vol. 32, L22804
\bibitem [Dwyer et al.,\ 2005a] {dwyer_2} J. R. Dwyer et al.,
2005a. X-ray bursts
associated with leader steps in cloud-to-ground lightning. Geophys.
Res. Lett., vol. 32, L01803
\bibitem [Dwyer et al.,\ 2005b] {dwyer_2005} J. R. Dwyer et
al., 2005b. X-ray bursts produced by laboratory sparks in air.
Geophys. Res. Lett., vol. 32, L20809
\bibitem [Dwyer,\ 2007] {dwyer_1} J. R. Dwyer, 2007. Relativistic breakdown in
planetary atmospheres, Physics of plasmas. vol. 14, no. 042901
\bibitem [Dwyer et al.,\ 2008a] {dwyer_3} J. R. Dwyer et
al., 2008a. A study of
X-ray emission from laboratory sparks in air at atmospheric pressure.
J. Geophys. Res., vol. 113, D23207
\bibitem [Dwyer et al.,\ 2008b] {electron} J. R. Dwyer et
al., 2008b. High-energy electron beams launched into space by thunderstorms.
Geophys. Res. Lett., vol. 35, L02815
\bibitem [Dweyer, \ 2012] {feedback} J. R. Dwyer, 2012. The relativistic feedback
discharge model of terrestrial gamma ray flashes. Journal of Geophy. Res. -
Space Phys., vol. 117, A02308
\bibitem [Elwert and Haug,\ 1969] {elwert} G. Elwert and E. Haug, 1969. Calculation of Bremsstrahlung cross
sections with Sommerfeld-Maue eigenfunctions. Phys. Rev., vol. 183, pp. 90-105
\bibitem [Fink and Pratt,\ 1973] {fink_pratt} J. K. Fink and R. H. Pratt,
1973. Use of
Furry-Sommerfeld-Maue wave functions in pair production and Bremsstrahlung.
Phys. Rev. A, vol. 7, pp. 392-403
\bibitem [Fishman et al.,\ 1994]{fishman} G. J. Fishman et al., 1994. Discovery of intense gamma-ray
flashes of atmospheric origin. Science, vol. 264, pp. 1313-1316
\bibitem [Galimberti et al.,\ 2002] {galimberti} I. Gallimberti et al.,
2002. Fundamental processes in long air gap discharges. C. R. Physique, vol. 3, pp.
1335-1359
\bibitem [Gjesteland et al.,\ 2011]{ang_brems} T. Gjesteland, N. Ostgaard, A.B. Collier, B.E. Carlson,
M.B. Cohen, N.G. Lehtinen, 2011. Confining the angular distribution of
terrestrial-gamma ray flash emission. J. Geophys. Res., vol. 116,
A11313
\bibitem [Grefenstette et al.,\ 2009]{RHESSI_2} B. W. Grefenstette, D. M. smith, B. J. Hazelton and L. I.
Lopez, 2009. First RHESSI terrestrial gamma ray flash catalog. J.
Geophys. Res., vol. 114, A02314
\bibitem [Greiner and Reinhardt,\ 1995] {QED} W. Greiner and J. Reinhardt,
1995. Quantenelektrodynamik, Verlag Harri Deutsch
\bibitem [Gurevich,\ 1961] {gurevich} A. V. Gurevich, 1961. On the theory of
runaway electrons. Soviet Physics Jetp-USSR, vol. 12, pp. 904-912
\bibitem [Gurevich et al.,\ 1992] {runaway} A. V. Gurevich, G. Milikh, R.
Roussel-Dupr\'{e}, 1992. Runaway electron
mechanism of air breakdown and preconditioning during a thunderstorm.
Phys. Lett. A, vol. 165, pp. 463-468
\bibitem [Gurevich and Zybin,\ 2001] {gurevich_2001} A. V. Gurevich and K.
P. Zybin, 2001. Runaway breakdown and electric discharges in thunderstorms.
Physics-Uspekhi, vol. 44, pp. 1119-1140
\bibitem [Heitler,\ 1944] {heitler} W. Heitler, 1944. The quantum theory of
radiation. Oxford University Press
\bibitem [Hough,\ 1948] {brems} P.V.C. Hough, 1948. The angular distribution of pair-produced
electrons and Bremsstrahlung. Phys. Rev., vol. 74, pp. 80-86
\bibitem [Inan and Lehtinen, 2005] {TGF_inan} U. S. Inan and N. G. Lehtinen,
2005. Production of terrestrial gamma-ray flashes by an electromagnetic pulse from a lightning return
stroke. Geophys. Res. Lett., vol. 32, L19818
\bibitem [Jackson,\ 1975,\ p. 712 et seq.] {jackson} J. D. Jackson, 1975. Classical
electrodynamics. John Wiley \& Sons
\bibitem [Koch and Motz,\ 1959] {brems_theory} H. W. Koch and J. W. Motz,
1959. Bremsstrahlung Cross-Section
Formulas and Related Data. Rev. Mod. Phys., vol. 31, pp. 920-956
\bibitem [Kostyrya et al.,\ 2006] {kos_2006} I. D. Kostyrya, V. F.
Tarasenko, A. N. Tkachev and S. I. Yakovlenko, 2006. X-ray radiation due to
nanosecond volume discharges in air under atmospheric pressure. Techn.
Phys., vol. 51, pp. 356-361
\bibitem [Landau and Pomeranchuk,\ 1953] {landau} L. Landau and I.
Pomeranchuk, 1953. Dokl. Akad. Nauk SSSR, vol. 92, pp. 535-536
\bibitem [Lehtinen,\ 2000] {lehtinen} N.G. Lehtinen, 2000. Relativistic runaway
electrons above thunderstorms. Ph.D. thesis, Stanford University, Stanford, CA
\bibitem [Li et al.,\ 2007] {li_2007} C. Li, W. J. M. Brok, U. Ebert and J. J. A.
M. van der Mullen, 2007. Deviations from the local field approximation in negative
streamer heads. J. Appl. Phys., vol. 101, no. 123305
\bibitem [Li et al.,\ 2009] {li_2009} C. Li, U. Ebert and W. Hundsdorfer,
2009. 3D hybrid computations for streamer discharges and production of runaway
electrons. J. Phys. D-Appl. Phys., vol. 42, no. 202003
\bibitem [Li et al.,\ 2010] {li_2010} C. Li, U. Ebert and W. Hundsdorfer,
2010. Spatially hybrid computations for streamer discharges with generic features
of pulled fronts: I. Planar fronts. J. Comput. Phys., vol. 229, pp. 200-220
\bibitem [Lu et al.,\ 2011] {RHESSI_5} G. Lu, S. A. Cummer, J. Li, F. Han, D. M. Smith and B.
Grefenstette, 2011. Characteristics of broadband magnetic lightning emissions
associated with terrestrial gamma-ray flashes. J. Geophys.
Res., vol. 116, A03316
\bibitem [March and Montany\`{a},\ 2010] {X-Ray_1} V. March and J. Montany\`{a},
2010. Influence of the
voltage-time derivative in X-ray emission from laboratory sparks.
Geophys. Res. Lett., vol. 37, L19801
\bibitem [Marisaldi et al.,\ 2010] {AGILE_1} M. Marisaldi et. al., 2010. Detection of terrestrial gamma ray flashes up to 40 MeV
by the AGILE satellite. J. Geophys. Res., vol. 115, A00E13
\bibitem [Milikh and Roussel-Dupr\'{e},\ 2010] {milikh_1} G. Milikh and R.
Roussel-Dupr\'{e}, 2010. Runaway breakdown and electrical discharges in
thunderstorms. J. Geophys. Res., vol. 115, A00E60
\bibitem [Moore et al.,\ 2001] {moore} C. B. Moore, K. B. Eack, G. D. Aulich
and W. Rison, 2001. Energetic radiation associated with lightning
stepped-leaders. Geophys. Res. Lett., vol. 28, pp. 2141-2144
\bibitem [Moss et al.,\ 2006] {moss} G. D. Moss, V. P. Pasko, N. Y. Liu and
G. Veronis, 2006. Monte Carlo model for analysis of thermal runaway electrons in
streamer tips in transient luminous events and streamer zones of lightning
leaders. J. Geophys. Res., vol. 111, A02307
\bibitem [Nagel,\ 1994] {nackel} W. Nackel, 1994. The elementary process of
Bremsstrahlung. Phys. Rep., vol. 243, pp. 317-353
\bibitem [Nguyen et al.,\ 2008] {nguyen_1} C. V. Nguyen, A. P. J. van Deursen and Ute
Ebert, 2008. Multiple x-ray bursts from long discharges in air. J. Phys.
D-Appl. Phys., vol. 41, no. 234012
\bibitem [Nguyen et al.,\ 2010] {nguyen_2} C. V. Nguyen, A. P. J. van
Deursen E. J. M. van Heesch and G. J. J. Winands, 2010. X-ray emission in
streamer-corona plasma. J. Phys. D-Appl. Phys., vol. 43,
no. 025202
\bibitem [Penczynski and Wehner,\ 1970] {brems_exp_2} P. E. Penczynski and H. L. Wehner,
1970. Measurement of the
energetic and angular dependence of the external Bremsstrahlung asymmetry.
Z. Physik, vol. 237, pp. 75-85
\bibitem [Peskin and Schroeder,\ 1995] {QFT} M. E. Peskin and D. V. Schroeder,
1995. An introduction to quantum field theory, Westview Press
\bibitem [Phelps et al.,\ 1987] {phelps} A. V. Phelps, B. M.
Jelenkovic and L. C. Pitchford, 1987. Simplified models of electron excitation and ionizattion at very
high E/n. Phys. Rev. A, vol. 36, no. 5327
\bibitem [Rahman et al.,\ 2008] {rahman_2008} M. Rahman, V. Cooray, N. A.
Ahmad, J. Nyberg, V. A. Rakov and S. Sharma, 2008. X-rays from 80 cm long sparks
in air. Geophys. Res. Lett., vol. 35, L06805
\bibitem [Rep'ev and Repin,\ 2008] {repev_2008} A. G. Rep'ev and P. B.
Repin, 2008. Spatiotemporal parameters of the X-ray radiation from a diffuse
atmospheric-pressure discharge. Techn. Phys., vol. 53, pp. 73-80
\bibitem [Seltzer and Berger,\ 1985] {seltzer} S.M.Seltzer and M.J.Berger,
1985. Bremsstrahlung spectra from electron interactions with screened atomic
nuclei and orbital electrons. Nucl. Inst. Meth. B12, pp. 95-134
\bibitem [Shaffer et al.,\ 1996] {shaffer_tong} C. D. Shaffer, X. M. Tong and R. H. Pratt,
1996. Triply differential cross section and polarization correlations in electron
Bremsstrahlung emission. Phys. Rev. A, vol. 53, pp. 4158-4163
\bibitem [Shaffer and Pratt,\ 1997] {shaffer_pratt} C. D. Shaffer and R. H. Pratt,
1997. Comparison of relativistic partial-wave calculations of triply differential electron-atom
bremsstrahlung with simpler theories. Phys. Rev. A, vol. 56, pp. 3653-3658
\bibitem [Shao et al.,\ 2011] {shao_2011} T. Shao, C. Zhang, Z. Niu, P. Yan,
V. F. Tarasenko, E. K. Baksht, A. G. Burahenko and Y. B. Shut'ko, 2011. Diffuse
discharge, runaway electron, and x-ray in atmospheric pressure air in a
inhomogeneous electrical field in repetitive pulsed modes. Appl. Phys.
Lett., vol. 98, no. 021506
\bibitem [Smith et al.,\ 2005]{RHESSI_1} D. M. Smith, L. I. Lopez, R. P. Lin and C. P.
Barrington-Leigh, 2005. Terrestrial gamma-ray flashes observed up to 20 Mev.
Science, vol. 307, pp. 1085-1088
\bibitem [Smith et al.,\ 2010]{RHESSI_3} D. M. Smith et al., 2010. Terrestrial gamma ray flashes
correlated to storm phase and tropopause height. J. Geophys. Res.,
vol. 115, A00E49
\bibitem [Smith et al.,\ 2011] {AGILE_3} D. M. Smith, J. Dwyer, B. Hazelton, B. Greffenstette, G.
F. M. Martinez-McKinney, Z. Zhang, A. Lowell, N. Kelley, M. Splitt, S.
Lazarus, W. Ulrich, M. Schaal, Z. Saleh, E. Cramer, H. Rassoul, S.A. Cummer,
G. Lu and R. Blakeslee, 2011. The rarity of terrestrial gamma-ray flashes.
Geophys. Res. Lett., vol. 38, L08807
\bibitem [Stankevich and Kalinin,\ 1967] {stan_67} Y. L. Stankevich and V.
G. Kalinin, 1967. Fast electrons and X-ray radiation during the initial stage of
growth of a pulsed spark discharge in air. Sov. Phys. Dokl., vol. 12, pp.
1042-1043
\bibitem [Tavani et al.,\ 2011] {AGILE_2} M. Tavani et al., 2011. Terrestrial gamma-ray flashes as powerful
particle accelerators. Phys. Rev. Lett., vol. 106, no. 018501
\bibitem [Ter-Mikaelian,\ 1954] {ter-mil} M. L. Ter-Mikaelian, 1954. Dokl. Akad.
Nauk SSSR, vol. 94, p 1033
\bibitem [Torii et al.,\ 2004]{emission} T. Torii, T. Nishijima, Zl. Kawasaki, T. Sugita,
2004. Downward emission of runaway electrons and bremsstrahlung photons in
thunderstorm electric fields. Geophys. Res. Lett., vol. 31, L05113
\bibitem [Tsai,\ 1974] {Tsai_1} Y-S. Tsai, 1974. Rev. Mod. Phys, vol. 46, p. 815
\bibitem [Tsai,\ 1977] {Tsai_2} Y-S. Tsai, 1977. Rev. Mod. Phys, vol. 49, p. 421
\bibitem [Tseng and Pratt,\ 1971] {tseng_pratt} H. K. Tseng and R. H. Pratt,
1971. Exact screened calculations of atomic-field Bremsstrahlung. Phys. Rev. A,
vol. 3, pp. 100-115
\bibitem [Tsuchiya et al.,\ 2011] {GROWTH} H. Tsuchiya, T. Enoto, S. Yamada, T. Yuasa, K.
Nakazawa, T. Kitaguchi, M. Kawaharada, M. Kokubun, H. Kato, M. Okano, K.
Makishima, 2011. Long-duration gamma ray emissions from 2007 and 2008 winter
thunderstorms. J. Geophys. Res., vol. 116, D09113
\bibitem [Wilson,\ 1925] {wilson} C. Wilson, 1925. The electric field of a
thundercloud and some of its effects. Proc. Phys. Soc. London, vol. 37A, pp.
32D-37D
\bibitem [Xu et al.,\ 2012] {xu} W. Xu, S. Celestin and V. P. Pasko, 2012.
Source altitudes of Terrestrial Gamma-ray Flashes produced by lightning
leaders, Geophys. Res. Lett., vol. 39, L08801

\end {thebibliography}
\end {document}